\documentclass[twocolumn]{aastex631}

\newcommand{\Lya}{Ly$\alpha$}
\newcommand{\kms}{km s$^{-1}$}

\shortauthors{Saccardi et al.}
\graphicspath{{./}{figures/}}
\usepackage{float}
\usepackage{appendix}
\begin{document}

\title{Evidence of first stars-enriched gas in high-redshift absorbers}
\thanks{Based on data products from the XQ-100 survey made with \\ ESO Telescopes at the La Silla Paranal Observatory under \\ programme ID 189.A-0424.}

\correspondingauthor{Andrea Saccardi}
\email{andrea.saccardi@obspm.fr}

\author[0000-0002-6950-4587]{Andrea Saccardi}
\affiliation{GEPI, Observatoire de Paris, Université PSL, CNRS, 5 Place Jules Janssen, 92190 Meudon, France}
\affiliation{Dipartimento di Fisica e Astronomia, Università degli Studi di Firenze, Via G. Sansone 1, I-50019 Sesto Fiorentino, Italy}

\author[0000-0001-7298-2478]{Stefania Salvadori}
\affiliation{Dipartimento di Fisica e Astronomia, Università degli Studi di Firenze, Via G. Sansone 1, I-50019 Sesto Fiorentino, Italy}
\affiliation{INAF - Osservatorio Astrofisico di Arcetri, Largo E. Fermi 5, I-50125 Firenze, Italy}

\author[0000-0003-3693-3091]{Valentina D'Odorico}
\affiliation{Scuola Normale Superiore, Piazza dei Cavalieri 7, I-56126 Pisa, Italy}
\affiliation{INAF - Osservatorio Astronomico di Trieste, via G.B. Tiepolo, 11 I-34143 Trieste, Italy}
\affiliation{IFPU - Institute for fundamental physics of the Universe, Via Beirut 2, 34014 Trieste, Italy}

\author[0000-0002-6830-9093]{Guido Cupani}
\affiliation{INAF - Osservatorio Astronomico di Trieste, via G.B. Tiepolo, 11 I-34143 Trieste, Italy}
\affiliation{IFPU - Institute for fundamental physics of the Universe, Via Beirut 2, 34014 Trieste, Italy}

\author[0000-0001-6676-3842]{Michele Fumagalli}
\affiliation{Dipartimento di Fisica G. Occhialini, Università degli Studi di Milano Bicocca, Piazza della Scienza 3, I-20126 Milano, Italy}
\affiliation{INAF - Osservatorio Astronomico di Trieste, via G.B. Tiepolo, 11 I-34143 Trieste, Italy}

\author[0000-0002-2606-5078]{Trystyn A.M. Berg}
\affiliation{Dipartimento di Fisica G. Occhialini, Università degli Studi di Milano Bicocca, Piazza della Scienza 3, I-20126 Milano, Italy}

\author[0000-0003-2344-263X]{George D. Becker}
\affiliation{Department of Physics \& Astronomy, University of California, Riverside, CA 92521, USA}

\author[0000-0002-1768-1899]{Sara Ellison}
\affiliation{Department of Physics \& Astronomy, University of Victoria, Finnerty Road, Victoria, British Columbia V8P 1A1, Canada}

\author[0000-0003-0389-0902]{Sebastian Lopez}
\affiliation{Departamento de Astronom\'ia, Universidad de Chile, Casilla 36-D, Santiago, Chile}

\begin{abstract}
The first stars were born from chemically pristine gas. They were likely massive and thus they rapidly exploded as supernovae, enriching the surrounding gas with the first heavy elements. In the Local Group, the chemical signatures of the first stellar population were identified among low-mass, long-lived, very metal-poor ([Fe/H]$<-2$) stars, characterised by high abundances of carbon over iron ([C/Fe]$>+0.7$): the so-called carbon-enhanced metal-poor stars. Conversely, a similar carbon excess caused by first-star pollution was not found in dense neutral gas traced by absorption systems at different cosmic time. Here we present the detection of 14 very metal-poor, optically thick absorbers at redshift $z\sim3-4$. Among these, 3 are carbon-enhanced and reveal an overabundance with respect to Fe of all the analysed chemical elements (O, Mg, Al, and Si). Their relative abundances show a distribution with respect to [Fe/H] that is in very good agreement to those observed in nearby very metal-poor stars. All the tests we performed support the idea that these C-rich absorbers preserve the chemical yields of the first stars. Our new findings suggest that the first star signatures can survive in optically thick but relatively diffuse absorbers, which are not sufficiently dense to sustain star formation and hence, are not dominated by the chemical products of normal stars.
\end{abstract}



\section{Introduction} 
\label{sec:intro}
Cosmological simulations show that the first (PopIII) stars are likely more massive than present-day ``normal" stars, with a characteristic mass of $\sim 10\,M_\odot$ and a maximum-mass possibly extending up to $\sim1000\,M_\odot$ \citep[e.g.][]{Hosokawa2011,Hirano2014}.
Among such a variety of stellar masses there are many channels to produce supernovae (SNe), and thus to contaminate the surrounding environment with the heavy elements newly produced by Pop~III stars. Very massive first stars, $140M_{\odot}\leq M_{PopIII}\leq260M_{\odot}$, explode as Pair Instability SNe yielding chemical abundance ratios that exhibit a strong odd-even effect \citep{Heger2002,Takahashi2018} and a unique lack of cobalt and zinc over iron \citep{Salvadori2019}. First stars with intermediate masses, $10M_{\odot}\leq M_{PopIII}\leq100M_{\odot}$, also evolve as SNe but they can have a variety of explosion energies, thus yielding very different chemical element ratios depending upon the mass of the progenitor star and the SN explosion energy (e.g. \citealt{Heger2010,Limongi2018} for theory and \citealt{Placco21,Skuladottir21} for observations).
Low-mass stars, born from the first stars-polluted gas \citep{Bromm2003,Schneider2003}, can survive until the present-day and retain in their atmospheres a record of the chemical elements produced by these first stars.

The search for the chemical signatures of Pop III stars has focussed on ancient metal-poor stars in our cosmic neighbourhood. In particular, stars in the Milky Way (MW) halo and Local Group dwarf galaxies are prime targets as we can uniquely study individual stars \citep[e.g.][]{Beers2005,Frebel2015,Simon2019}. Historically a new class of Carbon-Enhanced Metal-Poor (CEMP) stars was first recognized from the observations of \cite{Beers1992}. A few years later two simultaneous studies, by \cite{Norris1997} and \cite{Bonifacio1998}, identify for the first time an object pertaining to the class of CEMP-no stars, i.e. stars that are very metal-poor ([Fe/H]$< -2$), strongly enhanced in carbon with respect to iron ([C/Fe]$> +0.7$), and not enriched in neutron capture elements ([Ba/Fe]$< 0.0$). At the moment many CEMP-no stars have been discovered \citep[e.g.][]{Christlieb2002,Rossi2005,Bonifacio2015,Norris2013,Zepeda2022,Aguado2022} and very recently it has been confirmed by \cite{aguado2023} that these objects are most likely the descendants of massive first stars that exploded as low-energy supernovae \citep[e.g.][]{Iwamoto2005,Marassi2014}.

Indeed, when the explosion energy of a supernova is not high enough to expel Fe-peak elements from the innermost layers, a large fraction of them fall back onto the remnant \citep[e.g.][]{Heger2010}. During these faint SN explosions, therefore, only the outermost layers rich of carbon and other light elements are ejected, yielding large values of [C/Fe].
The idea that CEMP-no stars formed in an environment polluted by low-energy primordial SNe is further supported by the increasing frequency of CEMP-no stars towards lower [Fe/H] \citep[e.g.][]{Marsteller2005,Beers2005,Lucatello2006,Lee2013,Placco2014,deBennassuti2017,Yoon2018,Liu2021}.

Among the very metal-poor stars enriched with carbon another population exists: stars that exhibit an excess in heavy elements formed by slow (or rapid) neutron capture processes, dubbed CEMP-s (or -r) stars. 
The available data suggest that most CEMP-s stars dwell in binary systems \citep{Lucatello2005,Starkenburg2014,Hansen2016_CEMPs,Arentsen2019} and thus that their carbon-excess is not inherited from the natal cloud but acquired via mass transfer. The surplus of carbon likely comes from an evolved star that has passed through the asymptotic giant branch (AGB) phase \citep{Abate2015}, during which s-elements are also produced \citep{Karakas2014}. 
On the other hand, the carbon-excess in CEMP-no stars is expected to be representative of the environment of formation \citep[e.g.][]{Hansen2016_CEMPno,Zepeda2022}, even in the rare case in which CEMP-no stars are found to dwell in binary systems \citep[][]{Aguado2022,aguado2023}.

Based on these results from ancient nearby stars, we expect that at high redshifts it could be possible to find very metal-poor gaseous environments primarily enriched by the first stars \citep{Pallottini2014}, thus showing a carbon-excess.

Quasar absorption lines provide an important gateway to infer observational constraints on galaxy formation and evolution and to look for the signatures of the first stars in gas at high-redshifts. Detecting gas exhibiting similar abundance patterns as CEMP-no stars would open a new window to investigate the properties of the first stars and galaxies in the early Universe. Yet, despite long searches at $z>2-3$, this distinctive chemical signature has not been discovered in dense absorbers (\citealt{Cooke2011b,Dutta2014}), such as damped Lyman-alpha systems (DLAs). These DLAs, which have neutral hydrogen column density $\log (N_{\rm HI}/ {\rm cm}^{-2})>20.3$, trace most of the neutral gas in the Universe together with the galaxies interstellar medium (ISM).

A claim of detection of a CEMP-DLA (QSO J0035-0918) at $z_{abs}=2.340$ with [Fe/H]$=-3.04$ and [C/Fe]$=+1.53$ was published by \cite{Cooke2011a}. However, two subsequent works have disproved the result reporting a much lower carbon-to-iron ratio for the same DLA absorption system, i.e. respectively equal to [C/Fe]$=+0.51\pm0.10$ and [C/Fe]$=+0.45\pm0.19$ \citep{Carswell2012,Dutta2014}.
Another DLA with [C/Fe]$=+0.59$, [Fe/H]$=-2.84$ at $z=3.07$ has been reported by \cite{Cooke2012}, which is however below the limit value of [C/Fe]$=+0.7$. 
Two recent work \cite{Welsh2019,Welsh2022} presenting a collection of all the very metal-poor DLAs in the literature confirm the absence of carbon-enhancement claims.

Aiming at detecting the chemical evidence of gas enriched by the first stars, Lyman limit systems (LLSs) and sub-damped Lyman-$\alpha$ systems (sub-DLAs), with $17.2\leq \log(N_{\rm HI}/{\rm cm}^{-2})\leq20.3$, represent promising gaseous environments to look for the fingerprints of PopIII stars. Indeed, they are less dense than DLAs and therefore 
metal poorer \citep{Fumagalli2016} and likely not strongly contaminated by subsequent generations of normal (Pop II) stars \citep{Salvadori2012}, which are expected to form early on in the ISM of Pop~III enriched galaxies. On the other hand, these systems trace optically thick, relatively diffuse gas that is not sufficiently dense to self-shield the UV radiation; consequently, they are likely characterized by more complex ionization patterns. 

In this work, we exploited the XQ-100 quasar legacy survey \citep{Lopez2016} to collect a sample of 54 absorption systems at redshift $z\sim3-4$ selected by the presence of the Mg~II absorption doublet. Among these systems, we identified a sub-sample of 37 diffuse optically thick LLSs and sub-DLAs absorbers that we have studied in detail. We performed Voigt profile fitting of metal absorption features and hydrogen Lyman lines in the quasar spectra to measure column densities. To derive relative abundances of different elements, we applied photoionization-model corrections to the measured ionic abundances. 

In Sect.\S\ref{sec:analysis} we present our data-set including the line profile fitting of the absorption systems and the determination of the chemical abundances. In Sect.\S\ref{sec:results} and \S\ref{sec:discussion} we present and discuss the results. The conclusions are drawn in Sect.\S\ref{sec:conclusion}.

\section{Data analysis} 
\label{sec:analysis}
The Large Programme ``Quasars and their absorption lines: a legacy survey of the high-redshift Universe with VLT/X-shooter" \citep{Lopez2016} has produced a homogeneous and high-quality sample of echelle spectra of 100 quasars (QSOs) with emission redshift $z\sim3.5-4.5$. The targets were observed with the X-shooter spectrograph \citep{Vernet2011} mounted at the ESO Very Large Telescope (VLT, Cerro Paranal, Chile). X-shooter is characterised by three arms that allow to cover in one observation the full spectral range between the atmospheric cutoff at 300 nm and the near-infrared K-band at 2500 nm, at an intermediate resolving power. The full spectral coverage, along with a well-defined target selection and the high signal-to-noise achieved (median $SNR=30$), clearly makes XQ-100 a unique data-set to study the rest-frame UV/optical spectra of high-z QSOs in a single, homogeneous, and statistically significant sample. The adopted slit widths were 1.0$''$ in the UVB arm and 0.9$''$ in the VIS and NIR arms, to match the requested seeing and to account for its wavelength dependence. These slit widths provide  nominal resolving powers of 5400, 8900 and 5600 for the UVB, VIS and NIR arm respectively\footnote{The nominal resolutions are different from those of \cite{Lopez2016} since at that time, due to a problem in X-shooter's data reduction, the resolutions were calculated incorrectly. \url{https://www.eso.org/sci/facilities/paranal/instruments/xshooter.html}}. The XQ-100 survey was designed to cover many science cases: from the detailed study of the intergalactic medium, to the detection of galaxies in absorption \citep{Sanchez2016,Berg2016,Christensen2017}, from the properties of QSO themselves \citep{Perrotta2016,Perrotta2018} to cosmology \citep{Irsic2017,Irsic2017b}. The spectra, reduced from the collaboration, were delivered to the public\footnote{\url{https://www.eso.org/qi/catalog/show/73}}. 
Two types of reduced data are provided for each target: (i) a joint spectrum of the three arms together;  (ii) individual UVB, VIS, and NIR arm spectra, which includes telluric correction and fitted QSO continuum. When a target is observed more than once at different epochs, there is one spectrum for each epoch and a combined spectrum, putting together all epochs. The telluric corrections are calculated on the spectra corresponding to the single epochs, while the continuum is calculated only on the combined spectrum.

Since for our purposes we needed telluric corrected spectra, we created a coadded spectrum when multiple observations were present and then we re-determined the intrinsic QSO continuum in the framework of the {\it Astrocook}\footnote{\url{https://github.com/DAS-OATs/astrocook}} Python software package \citep{Cupani2020}. In {\it Astrocook}, the emission continuum is estimated by first masking the most prominent absorption features and then interpolating the non-masked regions with a univariate spline of chosen degree.

The final spectrum for each QSO was created by cutting the noisy edges of each arm and "stitching" the three arms together. Also these operations were carried out using {\it Astrocook}. 

Before creating the final spectra, 
we estimated the “effective” resolving power determined by the atmospheric conditions during observations. Indeed, if the seeing during observations is smaller than the width of the slit, the “effective” resolving power of the obtained spectrum will be larger than the nominal one. We recomputed the value of the resolving power (R) for each spectrum (Table \ref{xq100}) based on the average value of the seeing during the observations (reported in the ESO archive as $DIMM$) and assuming a linear relation between resolving power and slit width. For example in the VIS arm, if $\langle DIMM \rangle<0.9$ arcsec, the new resolving power is obtained as $R_{eff}\sim (0.9/\langle DIMM \rangle)\times\,R_{nom}$ \citep[see][]{Dodorico2022}. Note however that these determinations have uncertainties of the order of 10\% (see D'Odorico et al. 2023, in prep.).

\begin{deluxetable*}{lccccc|lccccc}[t!]
\tabletypesize{\footnotesize}
\tablecaption{Redshift range within which the absorption systems were searched. Corrected resolving power for the three arms of each QSO (UVB, VIS, NIR).
\label{xq100}}
\tablehead{\colhead{\textbf{Quasar}} & \colhead{$z_{\rm max}$} & \colhead{$z_{\rm min}$} & \colhead{R (UVB)} & \colhead{R (VIS)} & \colhead{R (NIR)} & 
\colhead{\textbf{Quasar}} & \colhead{$z_{\rm max}$} & \colhead{$z_{\rm min}$} & \colhead{R (UVB)} & \colhead{R (VIS)} & \colhead{R (NIR)}}
\startdata
J0003-2603 & 4.125 & 3.339 & 5400 & 8900 & 5600 &  J1034+1102 & 4.290 & 3.478 & 6200 & 9200 & 5800  \\
J0006-6208 & 4.440 & 3.807 & 5400 & 8900 & 5600 &  J1036-0343 & 4.531 & 3.682 & 7000 & 10400 & 6500 \\
J0030-5129 & 4.173 & 3.379 & 5400 & 8900 & 5600 &  J1037+2135 & 3.634 & 2.923 & 5400 & 8900 & 5600 \\
J0034+1639 & 4.292 & 3.723 & 5400 & 8900 & 5600 &  J1037+0704 & 4.141 & 3.352 & 5400 & 8900 & 5600 \\
J0042-1020 & 3.882 & 3.133 & 7900 & 11800 & 7400 & J1042+1957 & 3.636 & 2.925 & 5400 & 8900 & 5600  \\
J0048-2442 & 4.083 & 3.303 & 7800 & 11600 & 7300 & J1053+0103 & 3.674 & 2.957 & 5400 & 8900 & 5600  \\
J0056-2808 & 3.635 & 3.086 & 5400 & 8900 & 5600 &  J1054+0215 & 3.973 & 3.210 & 7700 & 11400 & 7200 \\
J0057-2643 & 3.661 & 2.946 & 6600 & 9800 & 6100 &  J1057+1910 & 4.137 & 3.349 & 8000 & 12000 & 7500 \\
J0100-2708 & 3.546 & 2.849 & 5400 & 8900 & 5600 &  J1058+1245 & 4.341 & 3.522 & 5400 & 8900 & 5600 \\
J0113-2803 & 4.314 & 3.499 & 7800 & 11600 & 7300 & J1103+1004 & 3.595 & 2.890 & 5400 & 8900 & 5600  \\
J0117+1552 & 4.243 & 3.622 & 5400 & 8900 & 5600 &  J1108+1209 & 3.672 & 3.095 & 6500 & 9700 & 6100 \\
J0121+0347 & 4.125 & 3.339 & 5400 & 8900 & 5600 &  J1110+0244 & 4.158 & 3.367 & 6600 & 9800 & 6100 \\
J0124+0044 & 3.840 & 3.098 & 5400 & 8900 & 5600 &  J1111-0804 & 3.922 & 3.291 & 9200 & 13600 & 8500 \\
J0132+1341 & 4.152 & 3.553 & 6700 & 9900 & 6200 &  J1117+1311 & 3.629 & 2.919 & 5400 & 8900 & 5600 \\
J0133+0400 & 4.185 & 3.563 & 8200 & 12100 & 7600 & J1126-0126 & 3.617 & 2.909 & 5400 & 8900 & 5600 \\
J0137-4224 & 3.971 & 3.208 & 5400 & 8900 & 5600 &  J1126-0124 & 3.737 & 3.072 & 5400 & 8900 & 5600\\
J0153-0011 & 4.206 & 3.407 & 5400 & 8900 & 5600 &  J1135+0842 & 3.847 & 3.100 & 5400 & 8900 & 5600\\
J0211+1107 & 3.973 & 3.210 & 7300 & 10800 & 6800 & J1201+1206 & 3.522 & 3.040 & 9200 & 13600 & 8500 \\
J0214-0518 & 3.985 & 3.221 & 6400 & 9500 & 6000 &  J1202-0054 & 3.593 & 2.889 & 5400 & 8900 & 5600\\
J0234-1806 & 4.305 & 3.689 & 5400 & 8900 & 5600 &  J1248+1304 & 3.709 & 2.986 & 5400 & 8900 & 5600\\
J0244-0134 & 4.055 & 3.439 & 8100 & 12100 & 7600 & J1249-0159 & 3.655 & 3.039 & 5400 & 8900 & 5600 \\
J0247-0556 & 4.234 & 3.600 & 5400 & 8900 & 5600 &  J1304+0239 & 3.618 & 2.910 & 5400 & 8900 & 5600\\
J0248+1802 & 4.439 & 3.604 & 8100 & 12100 & 7600 & J1312+0841 & 3.735 & 3.009 & 7100 & 10500 & 6600 \\
J0255+0048 & 4.003 & 3.387 & 7100 & 10500 & 6600 & J1320-0523 & 3.717 & 2.993 & 6200 & 9200 & 5800 \\
J0307-4945 & 4.716 & 3.896 & 7000 & 10400 & 6500 & J1323+1405 & 4.067 & 3.316 & 6900 & 10300 & 6500 \\
J0311-1722 & 4.034 & 3.262 & 6600 & 9800 & 6100 &  J1330-2522 & 3.949 & 3.190 & 7700 & 11400 & 7200\\
J0401-1711 & 4.227 & 3.669 & 8900 & 13100 & 8300 & J1331+1015 & 3.845 & 3.102 & 7800 & 11600 & 7300 \\
J0415-4357 & 4.073 & 3.510 & 6700 & 10000 & 6300 & J1332+0052 & 3.507 & 2.942 & 5400 & 8900 & 5600 \\
J0424-2209 & 4.329 & 3.762 & 5400 & 8900 & 5600 &  J1336+0243 & 3.810 & 3.072 & 7800 & 11600 & 7300\\
J0523-3345 & 4.385 & 3.559 & 8700 & 12900 & 8100 & J1352+1303 & 3.693 & 2.973 & 5400 & 8900 & 5600 \\
J0529-3526 & 4.418 & 3.787 & 6500 & 9600 & 6100 &  J1401+0244 & 4.418 & 3.718 & 8300 & 12300 & 7800\\
J0529-3552 & 4.172 & 3.538 & 7300 & 10800 & 6800 & J1416+1811 & 3.602 & 2.896 & 5400 & 8900 & 5600 \\
J0714-6455 & 4.465 & 3.872 & 10600 & 15700 & 9900 & J1421-0643 & 3.688 & 2.978 & 5400 & 8900 & 5600\\
J0747+2739 & 4.133 & 3.389 & 5400 & 8900 & 5600 &  J1442+0920 & 3.529 & 2.834 & 6700 & 10000 & 6300\\
J0755+1345 & 3.674 & 2.957 & 5400 & 8900 & 5600 &  J1445+0958 & 3.562 & 2.862 & 7000 & 10400 & 6500\\
J0800+1920 & 3.947 & 3.420 & 5400 & 8900 & 5600 &  J1503+0419 & 3.664 & 2.948 & 5400 & 8900 & 5600\\
J0818+0958 & 3.694 & 3.041 & 7800 & 11600 & 7300 & J1517+0511 & 3.559 & 2.860 & 5400 & 8900 & 5600 \\
J0833+0959 & 3.713 & 2.990 & 6400 & 9400 & 5900 &  J1524+2123 & 3.592 & 2.981 & 5400 & 8900 & 5600\\
J0835+0650 & 3.990 & 3.430 & 7900 & 11800 & 7400 & J1542+0955 & 3.992 & 3.226 & 7100 & 10500 & 6600 \\
J0839+0318 & 4.234 & 3.596 & 6500 & 9600 & 6100 &  J1552+1005 & 3.715 & 3.179 & 7100 & 10500 & 6600\\
J0920+0725 & 3.636 & 2.925 & 6900 & 10300 & 6400 & J1621-0042 & 3.710 & 2.987 & 5400 & 8900 & 5600 \\
J0935+0022 & 3.739 & 3.012 & 5400 & 8900 & 5600 &  J1633+1411 & 4.379 & 3.686 & 7100 & 10500 & 6600\\
J0937+0828 & 3.703 & 2.981 & 7000 & 10300 & 6500 & J1658-0739 & 3.750 & 3.193 & 5400 & 8900 & 5600 \\
J0955-0130 & 4.418 & 3.618 & 6900 & 10300 & 6500 & J1723+2243 & 4.531 & 3.815 & 5400 & 8900 & 5600 \\
J0959+1312 & 4.064 & 3.385 & 5400 & 8900 & 5600 &  J2215-1611 & 3.995 & 3.372 & 10400 & 15400 & 9700\\
J1013+0650 & 3.790 & 3.055 & 9800 & 14600 & 9200 & J2216-6714 & 4.479 & 3.725 & 6600 & 9800 & 6200 \\
J1018+0548 & 3.514 & 2.914 & 7100 & 10500 & 6600 & J2239-0552 & 4.557 & 3.704 & 5400 & 8900 & 5600 \\
J1020+0922 & 3.655 & 2.941 & 7400 & 11000 & 6900 & J2251-1227 & 4.157 & 3.678 & 10600 & 15700 & 9900 \\
J1024+1819 & 3.525 & 2.831 & 6700 & 9900 & 6200 &  J2344+0342 & 4.248 & 3.443 & 7200 & 10700 & 6700\\
J1032+0927 & 4.003 & 3.288 & 7000 & 10400 & 6500 & J2349-3712 & 4.219 & 3.567 & 6700 & 9900 & 6200 \\
\enddata
\end{deluxetable*}

\begin{deluxetable*}{lcccccccccc}[h]
\tabletypesize{\scriptsize}
\tablecaption{Absorption redshift, neutral hydrogen column density, iron abundance, and relative chemical abundances for all absorption systems with a measure/upper limit of Fe~II (30).\\ $*$ Very metal-poor absorption systems with [Fe/H]$<-2$. In bold the carbon enhanced very metal-poor absorbers.\\ $\dagger$ [C/Fe] is computed directly by using the carbon and iron column densities.}
\label{abundances}
\tablehead{\colhead {Quasar} & \colhead{$z_{\rm abs}$} & \colhead{$N_{\rm HI}$} & \colhead{log(Z/Z$_{\odot}$)} & \colhead{[Fe/H]} & \colhead{[C/H]} & \colhead{[O/H]} & \colhead{[Mg/H]} & \colhead{[Al/H]} & \colhead{[Si/H]} & \colhead{[C/Fe]$^\dagger$}}
\startdata
J0042-1020         &3.62953 &18.6$\pm$0.3   &-1.55 &-1.3$\pm$0.3   &$>$-1.1        &-1.6$\pm$0.3   &-1.1$\pm$0.3   &-1.8$\pm$0.3   &-1.2$\pm$0.3   & $>$+0.11\\
J0056-2808         &3.58045 &17.4$\pm$0.2   &-0.97 &+0.70$\pm$0.19  &$>$+2.1         &+1.7$\pm$0.2    &$>$+0.3         &-0.6$\pm$0.3   &+0.1$\pm$0.2    & $>$+1.41\\
J0124+0347         &3.67488 &17.9$\pm$0.2   &-1.98 &$<$-0.96       &-2.46$\pm$0.18 &               &-1.84$\pm$0.17 &-2.8$\pm$0.2   &-2.07$\pm$0.17 & $>$-1.50\\
J0133+0400         &3.99668 &17.4$\pm$0.2   &-1.85 &+0.8$\pm$0.2    &-0.8$\pm$0.2   &+2.2$\pm$0.2    &               &-1.6$\pm$0.2   &-0.5$\pm$0.2   & -1.59$\pm$0.16\\
J0211+1107$*$      &3.50250 &19.9$\pm$0.2   &-1.69 &-2.0$\pm$0.2   &-1.3$\pm$0.2   &               &-1.7$\pm$0.2   &-1.5$\pm$0.2   &-0.9$\pm$0.2   & +0.66$\pm$0.12\\
J0234-1806         &4.22817 &19.2$\pm$0.2   &-1.06 &-0.9$\pm$0.2   &$>$+0.5         &-0.5$\pm$0.2   &$>$-0.7        &-0.9$\pm$0.2   &-0.4$\pm$0.3   & $>$+1.54\\
J0247-0556$*$      &4.13952 &18.9$\pm$0.2   &-2.26 &-2.0$\pm$0.2   &-1.6$\pm$0.2   &-1.1$\pm$0.2   &               &-2.6$\pm$0.2   &-1.7$\pm$0.2   & +0.43$\pm$0.11\\
J0307-4945         &4.21345 &17.2$\pm$0.2   &-1.49 &$<$-0.82       &$>$+1.0         &               &-0.8$\pm$0.2   &-1.6$\pm$0.2   &-0.5$\pm$0.2   & $>$+1.89\\
J0529-3552         &4.06561 &18.6$\pm$0.2   &-2.00 &-1.6$\pm$0.2   &-1.4$\pm$0.2   &-1.2$\pm$0.2   &               &-2.4$\pm$0.2   &-1.4$\pm$0.2   & +0.2$\pm$0.3\\
J0800+1920$*$      &3.42856 &19.9$\pm$0.2   &-2.85 &-3.0$\pm$0.2   &-2.3$\pm$0.2   &               &-2.2$\pm$0.2   &-2.9$\pm$0.2   &-2.3$\pm$0.2   & +0.65$\pm$0.09\\
J0818+0958$*$      &3.45615 &18.8$\pm$0.2   &-2.51 &-2.0$\pm$0.2   &-2.1$\pm$0.2   &$<$-2.3        &-1.9$\pm$0.2   &-2.4$\pm$0.2   &-2.0$\pm$0.2   & -0.04$\pm$0.09\\
J0818+0958         &3.53141 &18.0$\pm$0.2   &-2.30 &$<$+0.07        &-2.09$\pm$0.16 &               &-1.79$\pm$0.16 &-2.08$\pm$0.16 &-2.34$\pm$0.16 & $>$-2.17\\
\bf{J0835+0650$*$} &3.51256 &18.7$\pm$0.2   &-2.49 &$<$-2.59       &-1.9$\pm$0.2   &               &-1.9$\pm$0.2   &-2.5$\pm$0.2   &-2.1$\pm$0.2   & $>$+0.69\\
J1013+0650         &3.23534 &17.3$\pm$0.2   &-1.49 &$<$-0.5        &$>$-1.1        &               &-1.1$\pm$0.2   &-1.8$\pm$0.2   &-1.2$\pm$0.2   & $>$-0.56\\
J1018+0548$*$      &3.38500 &19.3$\pm$0.2   &-2.52 &-2.3$\pm$0.2   &-2.1$\pm$0.2   &-1.5$\pm$0.2   &-1.9$\pm$0.2   &-3.2$\pm$0.2   &-2.1$\pm$0.2   & +0.20$\pm$0.13\\
J1111-0804$*$      &3.48170 &19.9$\pm$0.2   &-2.22 &-1.91$\pm$0.16 &-1.72$\pm$0.16 &-1.51$\pm$0.16 &-1.65$\pm$0.16 &-2.24$\pm$0.15 &-1.75$\pm$0.15 & +0.19$\pm$0.08\\
\bf{J1111-0804$*$} &3.75837 &18.6$\pm$0.2   &-2.42 &$<$-2.91       &-2.1$\pm$0.2   &$<$-2.5        &-2.0$\pm$0.2   &-2.4$\pm$0.2   &-2.1$\pm$0.2   & $>$+0.78\\
J1117+1311         &3.27522 &17.9$\pm$0.2   &-1.93 &$<$-0.92       &-0.64$\pm$0.16 &               &-1.45$\pm$0.18 &-2.01$\pm$0.14 &-1.77$\pm$0.15 & $>$+0.29\\
J1117+1311$*$      &3.43372 &18.7$\pm$0.2   &-2.91 &$<$-2.47       &-2.55$\pm$0.15 &$<$-1.9        &-2.13$\pm$0.18 &-3.08$\pm$0.14 &-2.4$\pm$0.2   & $>$-0.07\\
J1249-0159         &3.10265 &17.70$\pm$0.15 &-1.06 &-0.26$\pm$0.13 &-0.68$\pm$0.13 &               &$>$-0.82       &-1.18$\pm$0.13 &-0.79$\pm$0.13 & -0.42$\pm$0.08\\
J1304+0239         &3.21072 &18.30$\pm$0.15 &-1.41 &-0.11$\pm$0.13 &$>$+1.73        &               &$>$+0.96        &-0.09$\pm$0.18 &0.04$\pm$0.13  & $>$+1.86\\
J1332+0052$*$      &3.42107 &18.5$\pm$0.2   &-2.48 &-2.0$\pm$0.2   &-2.0$\pm$0.2   &               &-2.2$\pm$0.2   &-2.8$\pm$0.2   &-2.0$\pm$0.2   & -0.06$\pm$0.09\\
J1352+1303         &3.00680 &18.80$\pm$0.15 &-0.99 &-1.27$\pm$0.13 &-1.26$\pm$0.13 &               &$>$-0.5        &-1.15$\pm$0.13 &-1.16$\pm$0.13 & 0.00$\pm$0.07\\
J1542+0955         &3.28223 &17.30$\pm$0.15 &-0.25 &-0.08$\pm$0.12 &+0.19$\pm$0.17  &               &$>$-0.42       &-0.26$\pm$0.11 &-0.34$\pm$0.12 & +0.28$\pm$0.15\\
J1552+1005$*$      &3.44250 &19.0$\pm$0.2   &-2.70 &-2.2$\pm$0.2   &-2.0$\pm$0.2   &-1.9$\pm$0.2   &-2.2$\pm$0.2   &-2.9$\pm$0.2   &-2.2$\pm$0.2   & +0.17$\pm$0.16\\
J1621-0042         &3.10570 &19.60$\pm$0.15 &-1.43 &-1.41$\pm$0.16 &-1.26$\pm$0.16 &               &$>$+0.1         &-1.39$\pm$0.16 &-0.99$\pm$0.16 & +0.16$\pm$0.07\\
\bf{J1658-0739$*$} &3.54604 &19.0$\pm$0.2   &-2.88 &$<$-3.30       &-1.9$\pm$0.2   &               &-2.11$\pm$0.14 &               &-2.51$\pm$0.13 & $>$+1.41\\
J1658-0739$*$      &3.69551 &18.5$\pm$0.2   &-2.83 &$<$-2.36       &-2.3$\pm$0.2   &$<$-1.9        &-2.52$\pm$0.16 &               &-2.51$\pm$0.19 & $>$+0.10\\
J1723+2243$*$      &4.24710 &18.8$\pm$0.2   &-2.57 &-1.9$\pm$0.2   &-1.93$\pm$0.17 &-1.63$\pm$0.16 &-1.91$\pm$0.17 &-2.49$\pm$0.16 &-2.07$\pm$0.17 & -0.01$\pm$0.10\\
J2215-1611         &3.70140 &19.20$\pm$0.15 &-1.75 &-1.46$\pm$0.16 &$>$-1.47       &-1.29$\pm$0.16 &$>$-1.43       &-1.97$\pm$0.16 &-1.34$\pm$0.19 & $>$0.00\\
\enddata
\end{deluxetable*}

\subsection{Line Fitting}
\label{method}
The analysis of the spectra has been carried out with the {\it Astrocook} software package. 
In {\it Astrocook}, absorption lines are detected as prominent local minima in the flux density spectrum, then they are identified by cross-matching their wavelengths with a list of ionic transitions commonly observed in QSO spectra and finding coincidences among the obtained redshift values. 

For our study, we need to identify absorption systems with H~I column densities in the range $17.2\leq \log(N_{\rm HI}/{\rm cm}^{-2})\leq20.3$, i.e. LLSs and sub-DLAs, which can trace diffuse gas, such as the one in the outskirts of galaxies and in cosmic filaments \citep{Lofthouse2022}. To this aim, we searched the XQ-100 spectra for singly ionised magnesium doublets, Mg~II, which are good probes of the optically thick, low ionisation gas. Mg~II is one of the best known examples of strong resonance-line doublets: it has rest-frame wavelengths (see Table~\ref{tab:wave_ion}) longer than H~I Lyman-$\alpha$ (Ly$\alpha$), and therefore it appears on the red side of the \Lya\ emission line in the QSO spectrum. For this reason, it is relatively easy to identify as it is not embedded in the thick \Lya\ forest. The search for Mg~II doublets was carried out using an automatic recipe of the {\it Astrocook} software. Subsequently, the detection of the absorption lines was confirmed visually.
In our analysis we have not make any attempt to reach or determine the completeness of the absorber sample since it was not relevant for our scientific purpose.

In each line of sight,  we restricted our search to a redshift range that avoids the proximity region of the quasar ($5000$ \kms\ from the quasar emission redshift) and that allows to cover the H~I \Lya\  and  Lyman-$\beta$ (Ly$\beta$) transitions of a given absorption system, for a more reliable determination of the H~I column density.  Furthermore, we have excluded the interval 13500-14500 \AA\ affected by strong telluric lines. This wavelength range corresponds to the redshift range $z\simeq 3.83-4.18$ for Mg~II $\lambda 2796$ \AA.

Detected systems are then modelled with Voigt profiles in the context of {\it Astrocook}. The Voigt profile fitting provides the central line redshift, $z$, the column density, $N$, and the Doppler broadening parameter, $b$. After the identification of the Mg~II doublets, we proceeded with the search of other low ionization lines (see Table~\ref{tab:wave_ion}) at the same redshift. Assuming that they originate from the same gas and that turbulent motion is dominant over thermal one, we fitted all 
with the same redshift components having the same Doppler parameters.  
On the other hand, C~IV and Si~IV absorption lines, if present, have generally a different velocity structure and they were fitted separately. To better constrain the chemical properties of our absorbers we also estimated column density upper limits, in particular for Fe~II and O~I if they were not detected. We used equations 2 and 3 of \cite{Dodorico2016} adapted for $1\sigma$ limits assuming a Doppler broadening parameter equal to the one of the other low ionisation transitions in the system and the spectral velocity bin (UVB: 20 \kms; VIS: 11 \kms; NIR: 19 \kms) from \cite{Lopez2016}.
The neutral hydrogen column density was determined by considering in the fit all the lines in the Lyman series free from strong blending. 

A well known issue related to absorption line fitting is the identification of saturated lines. Indeed, with the resolving power provided by X-shooter, in many cases the lines are not resolved and therefore it is difficult to understand when they are saturated. We adopted the following empirical procedure for metal lines: when the difference between the normalized flux and the depth of the line was $>$0.7 we assumed that the line was saturated and considered the determined column density as a lower limit. In this hypothesis we are neglecting the possible dependence on the Doppler parameter value.

We selected 51 Mg~II absorption systems plus three previously identified LLS systems whose Mg~II absorption fall in the telluric band (S. Cristiani priv. comm.), for a total of 54 absorption systems. We compared our sample with those of prior studies of the XQ-100 survey investigating  DLAs and Sub-DLAs \citep{Berg2016,Berg2017,Berg2019,Berg2021}, finding matches for 17 DLAs and 16 Sub-DLAs. We excluded the DLAs from our sample and used the HI column density determined in \cite{Berg2019} for the sub-DLAs. Our analysis was then focused on a sample of 16 Sub-DLAs and 21 LLSs.

Three of the QSOs in our sample (J0247-0556, J1111-0804 and J1723+2243) have a reduced UVES spectrum available from the SQUAD database \citep{Murphy2019}. We verified with {\it Astrocook} that for our systems there is a good agreement between column densities measured in the X-shooter and UVES spectra.

\begin{deluxetable}{c|c}[]
\tabletypesize{\normalsize}
\tablecaption{Ionic transitions considered in our study and their rest-frame wavelength. \label{tab:wave_ion}}
\tablehead{\colhead{Ion} & \colhead{$\lambda$ (\AA)}}
\startdata
HI Ly$\alpha$ & $1215.6701$ \\
HI Ly$\beta$ & $1025.7223$ \\
MgII & $2796.35$, $2803.53$\\
FeII & $1608.45$, $2344.21$, $2382.77$, $2586.65$, $2600.17$\\
SiII & $1260.42$, $1304.37$, $1526.71$, $1808.01$ \\
OI & $1302.17$ \\
CII & $1334.53$ \\
AlII & $1670.79$ \\
AlIII & $1854.72$, $1862.79$ \\
ZnII & $2026.137$, $2062.66$ \\
CIV & $1548.20$, $1550.78$ \\
SiIV & $1393.76$, $1402.77$ \\
\enddata
\end{deluxetable}

\subsection{Determination of chemical abundances} 
The gas in LLSs and Sub-DLAs is not fully neutral, therefore it is necessary to apply a ionisation correction to translate the observed ionic column densities into element abundances and thus, into gas metallicity. 

To infer the chemical composition and the physical state of the absorbing gas, we have used ionisation models based on radiative transfer calculations at equilibrium and for a single gas phase. These calculations are the input of a Bayesian formalism that exploits Markov Chain Monte Carlo (MCMC) techniques to derive the posterior probability distribution function for quantities of interest, such as the metallicity, $Z$, and the physical density, $n_{\rm H}$, of the absorbing gas. In particular, the MCMC method we have adopted \citep{Fumagalli2016} uses a grid of Cloudy models. Cloudy (v. 17, \citealt{Ferland2017}) is an open-source photoionization code, that maps each set of input parameters into the corresponding column densities of all metal ions that we considered in this work. The MCMC sampler searches $Z$ and $n_{\rm H}$ for models with the column density pattern that best matches the neutral hydrogen column density, $N_{\rm HI}$, and the metal ion column densities fitted by the software {\it Astrocook}. The output of the MCMC modelling is a probability distribution of the sets of parameters $Z$ and $n_{\rm H}$. In this work, we adopt the minimal model parameters \citep{Fumagalli2016}, which assumes a slab of gas at constant density illuminated on one side by both the UV background \citep{Haardt2012}  and the cosmic microwave background. All metals are assumed to be in the gas phase with a solar abundance pattern \citep{Asplund2009}. 

To obtain the ionization corrections ion by ion, we run again Cloudy system by system optimizing the values of Z and $n_{\rm H}$ determined by the MCMC run based on the observed column densities. We vary the parameters in an interval corresponding to the tenth and ninetieth percentile of their posterior distribution function. We considered a narrow range because the optimization on the full parameter space had already been performed by the MCMC code. Finally, from the photoionization model we obtained the corrections for each ion of each element.

Once the column densities obtained from the Voigt fit have been corrected for ionization, we derived the absolute abundances of the various elements. The corrected column density of an element X has been determined using the formula $N(X)=N(X_{i})/IC(X_{i})$, where $N(X_{i})$ is the column density of the ion fitted from the spectrum and $IC(X_{i})$ the ionization correction obtained from Cloudy. Then, we derived the relative abundances of each element as:
\begin{equation}
[X/{\rm H}]=\log\left(\frac{N_{X}}{N_{\rm H}}\right)-\log\left(\frac{N_{X}}{N_{\rm H}}\right)_{\odot}
\end{equation}
where $N_{\rm H}$ and $N_{X}$ are respectively the column density of hydrogen and of a specific element X and $\log(N_{X}/N_{\rm H})_{\odot}$ is the solar abundance \citep{Asplund2009}.

\begin{figure}[h]
    \centering
    \includegraphics[width=0.48\textwidth]{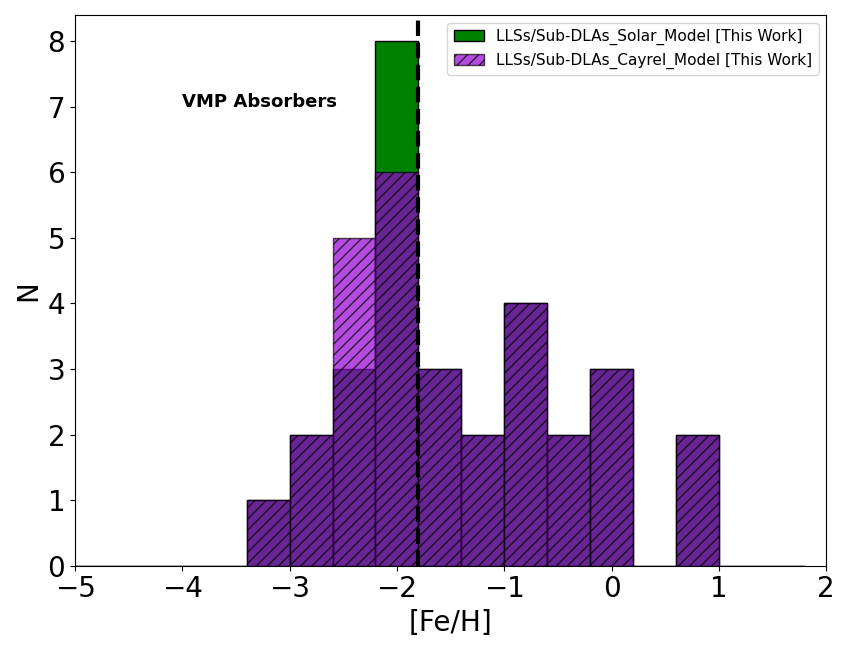}
    \caption{Metallicity distribution function (MDF) of our 30 absorbers assuming solar relative abundances (green) and {\it Cayrel's} relative abundances (purple) in the photoionization model.  A dashed line at [Fe/H]$=-1.8$ separates the very metal-poor (VMP) absorbers from the others.}
    \label{mdf_solar}
\end{figure}

In Fig.~\ref{mdf_solar}, we show the 
Metallicity Distribution Function (MDF) derived for our absorbers, i.e. the number of systems in different bins of [Fe/H], assuming solar relative abundances in the photoionization model (Solar model). 
Note that we restricted our analysis only to the 30 absorbers for which we could determine a measure/upper limit of Fe~II (with respect to the 37 previously selected LLSs and Sub-DLAs).
Furthermore, we are here assuming that all our [Fe/H] values are measurements, although for 10 absorbers (5 if we consider those very metal-poor) we only have upper limits on Fe~II (see Fig.~\ref{ac}).

The MDF is characterized by a pronounced peak at [Fe/H]$\sim-2$ and it extends over a broad range in [Fe/H], both at higher and lower values. In particular, there are 14 very metal-poor absorption systems with [Fe/H]$<-2$, where we include also those systems for which [Fe/H]$- 1\sigma < -2$. 
We have verified that these [Fe/H]$<-2$ systems are also characterised by a low total metallicity with respect to the solar value (see Fig. \ref{metallicity} and Tab. \ref{abundances}) and therefore that they are not just iron-poor because of dust depletion. On the other hand, the shape of the MDF at [Fe/H]$>-2$ is less constrained due to the fact that the iron measurements could be affected by dust depletion: the effect would be to increase the value of [Fe/H].

Note that the chemical abundance pattern of very metal-poor stars, at [Fe/H]$<-2$, is well established to be different from the solar value \citep[e.g.][]{Cayrel2004,Bonifacio2009,Yong2013}. For this reason, we recompute the ionisation corrections for our very metal-poor absorbers by assuming the chemical abundance pattern derived by \cite{Cayrel2004} for a sample of C-normal giant stars with $-4<$[Fe/H]$<-2$ in the Milky Way halo (hereafter we will refer to this pattern as {\it Cayrel}). This average chemical abundance pattern is indeed characterised by a very small star-to-star scatter. Hence, it can be considered as a ``reference" for very metal-poor environments.

\begin{deluxetable}{c|c}[]
\tabletypesize{\large}
\tablecaption{Average abundance ratios, [X/Fe], for the sample of very metal-poor giant stars in the MW halo studied by \cite{Cayrel2004} and corrected to account for internal mixing along with NLTE effects (see text). \label{tab:cayrel_model}}
\tablehead{\colhead{\hspace{1cm}Element}\hspace{1cm} & \colhead{\hspace{1cm}[X/Fe]}\hspace{1cm} }
\startdata
C & +0.45\\
O & +0.67\\
Mg & +0.61\\
Al & -0.10\\
Si & +0.44\\
\enddata
\end{deluxetable}

Table~\ref{tab:cayrel_model} reports the values assumed in the new Cloudy models which are those derived by \cite{Cayrel2004} corrected for 3D and/or Non-Local Thermodynamical Equilibrium (NLTE) effects and for stellar physical processes that can alter the measured abundances.
These corrections are required to use the stellar abundance values for the gaseous component.

\begin{deluxetable}{lccc}[h]
\tabletypesize{\normalsize}
\tablecaption{Modified $\chi^2$ parameter that estimates the agreement between the observed column densities of a given chemical ion and the column densities computed by the model. We report the value for all the very metal-poor absorbers. \label{chisquare}}
\tablehead{\colhead{\hspace{0.4cm}Quasar}\hspace{0.4cm} & \colhead{\hspace{0.4cm}$z_{\rm abs}$}\hspace{0.4cm} & \colhead{\hspace{0.4cm}$\chi^2_{Solar}$}\hspace{0.4cm} & \colhead{\hspace{0.4cm}$\chi^2_{Cayrel}$}\hspace{0.4cm}}
\startdata
J0211+1107 & 3.50250 & 9 & 19\\
J0247-0556 & 4.13952 & 34 & 16\\
J0800+1920 & 3.42856 & 67 & 7\\
J0818+0958 & 3.45615 & 178 & 10\\
J0835+0650 & 3.51256 & 14 & 7\\
J1018+0548 & 3.38500 & 32 & 8\\
J1111-0804 & 3.48170 & 27 & 21\\
J1111-0804 & 3.75837 & 13 & 38\\
J1117+1311 & 3.43372 & 15 & 14\\
J1332+0052 & 3.42107 & 32 & 31\\
J1552+1005 & 3.44250 & 33 & 9\\
J1658-0739 & 3.54604 & 38 & 24\\
J1658-0739 & 3.69551 & 16 & 19\\
J1723+2243 & 4.24710 & 19 & 13\\
\enddata
\end{deluxetable}

In particular we adopted: (i) the [C/Fe] value derived for dwarfs \citep{Bonifacio2009}, since the photospheric carbon abundances in giant stars can be altered by convective mixing. In other words we assumed that the difference between giant and dwarfs is due to the first dredge-up; (ii) the original [O/Fe] value derived by \cite{Cayrel2004}, since \cite{Bonifacio2009} demonstrated that in giant stars 3D effects for oxygen are not important; (iii) the magnesium value derived by \cite{Andrievsky2010} in order to account for NLTE effects; (iv) the [Al/Fe] and [Si/Fe] values reported by \cite{Cayrel2004}, since aluminium was already corrected for NLTE effects and NLTE effects for silicon are expected to be very weak.

We defined a modified chi square to establish which model (solar or Cayrel) reproduces better our observed column densities, $\chi^2=(\sum_{i}((N^{i}_{model} - N^{i}_{obs})/\sigma)^{2})^{1/2}$, where $N_{model}$ and $N_{obs}$ are the modelled and observed ionic column density values and $\sigma$ is the relative error in the observed value. More than 75\% (11/14) of the very metal-poor absorbers are better modeled with Cayrel's abundance pattern than with solar (see Table~\ref{chisquare}). 

The observed column densities are better reproduced for most of the ions. In addition, the relative abundances obtained with the new photoionization model are consistent with the abundances obtained with the solar model for most chemical elements.

The results of the Voigt profile fitting for the 14 very metal-poor absorption systems are available online as supplemental material. The tables reporting the fit parameters are available for {\it all} the 37 absorption systems.\footnote{Note that for the 16 absorbers with [Fe/H]$>-2$ and for the 7 without iron detection (or upper limits), we fit each ionic transition (or multiplet) independently. We only refined the analysis for the 14 very metal-poor absorption systems, [Fe/H]$<-2$, being the main focus of the work.}

An example figure of two metal-poor absorption systems is
shown in the Appendix together with a portion of the fit
parameters table for guidance.

\section{Results} \label{sec:results}
\subsection{Metallicity Distribution Function}

Figure~\ref{mdf_solar} shows the new MDF obtained for our absorbers, 
adopting the Cayrel abundances for very metal-poor systems and the solar values for those with [Fe/H]$>-2$. We can first note that the shape of the MDF remains essentially unvaried with respect to the one which was obtained by assuming solar values for all absorbers. This is because most of the changes are small, with a [Fe/H] difference $<$\,0.4\,dex, i.e. smaller than the width of the MDF bin.
By further inspecting Fig.~\ref{mdf_solar} we see that the MDF of our gaseous absorbers is roughly bi-modal: the first broad peak is around [Fe/H]$\sim-0.8$ while the second one, which is more pronounced, appears at [Fe/H]$\sim-2$. This result implies that these absorbers are likely a variegated population \citep[e.g.][]{Fumagalli2016}.

\begin{figure}[h]
    \centering
    \includegraphics[width=0.48\textwidth]{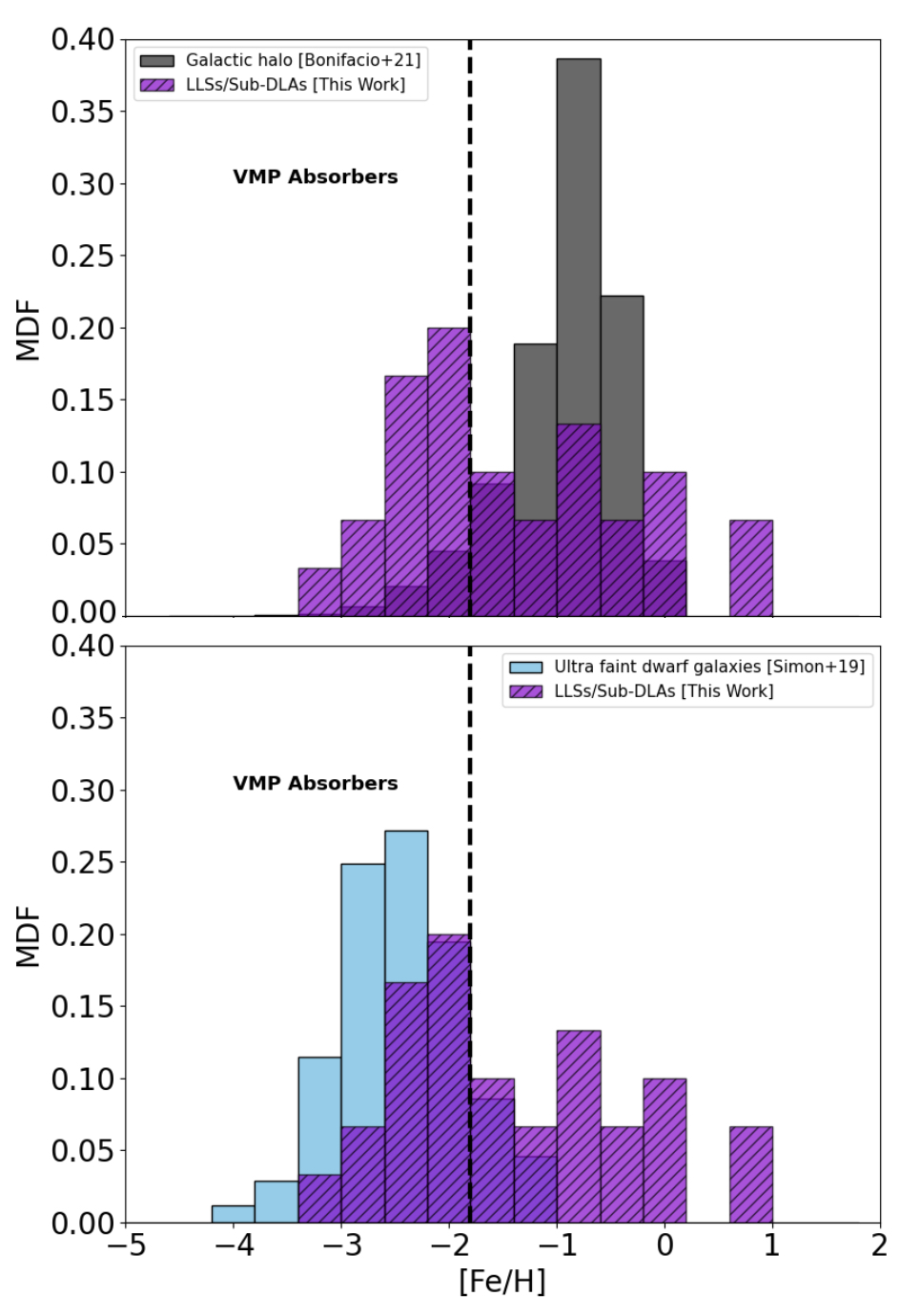}
    \caption{Comparison between the Metallicity distribution functions (MDFs) of our 30 absorbers, assuming {\it Cayrel's} relative abundances in the photoionization model, and the MDF of $\sim140000$ stars in the Galactic halo \citep{Bonifacio2021} (top panel) and of 350 stars in 26 UFDs \citep{Simon2019} (bottom panel). A dashed line at [Fe/H]$=-1.8$ separates the very metal-poor (VMP) absorbers from the others.}
    \label{mdf}
\end{figure}

We compare the normalized MDF of our high redshift absorption systems (Cayrel model) with those of ancient very metal-poor stars observed in the Galactic halo \citep{Bonifacio2021} and in Local Group ultra-faint dwarf galaxies (UFDs) \citep{Simon2019}, respectively in the upper and lower panel of Fig. \ref{mdf}. 
We are aware that we are comparing the chemical abundances of diffuse high redshift gas, which can be located in the outskirts of galaxies, with the ones of ancient local stars. Still, the main point behind our comparison is that stars are born from gas.

The MDF of our sample spans a wide range of [Fe/H], covering the values measured in stars of both the MW stellar halo and nearby UFDs. The distribution of the halo stars covers a similar range of [Fe/H] with respect to our absorbers, but has a different shape. Indeed the stars of the Galactic halo have a unique peak, which is more pronounced, and appears around [Fe/H]$\sim-1$. Note that, although very rare and hence almost invisible in the normalised MDF, stars with [Fe/H]$<-2.5$ in the Galactic halo are those showing the chemical imprint of the first stars (see Fig. \ref{allratio}). The low-Fe peak of our absorbers, therefore, suggests similarities between these high-z absorbers and the gas that may have hosted the imprint of the first stars, like the birth environment of very metal-poor halo stars.

The MDF of UFDs covers a narrower range of iron abundances, $-4<$[Fe/H]$<-1$, than the MDF of our absorption systems and shows a peak at [Fe/H]$= -2.5$, almost overlapping with the low-Fe peak of the MDF of our absorbers (Fig.~\ref{mdf}, lower panel). These findings suggest that our absorption systems might represent the gas-rich counterpart of UFDs at high redshift \citep{Salvadori2012,Skuladottir2018}.
The absence of ultra metal-poor absorption systems ([Fe/H]$<-4$) which are observed among Galactic halo stars, could be due to observational biases. On the one hand, we select our systems by the presence of the singly ionised magnesium doublet, on the other, the resolution and signal-to-noise ratio of the XQ-100 spectra do not allow us to put very stringent upper limits on the iron column density, when the ionic lines are not detected.

\subsection{Carbon-enhanced systems:\\stellar relics vs high-z absorbers}
\label{Carbon-enhanced systems}
\begin{figure}[h]
    \centering
    \includegraphics[width=0.48\textwidth]{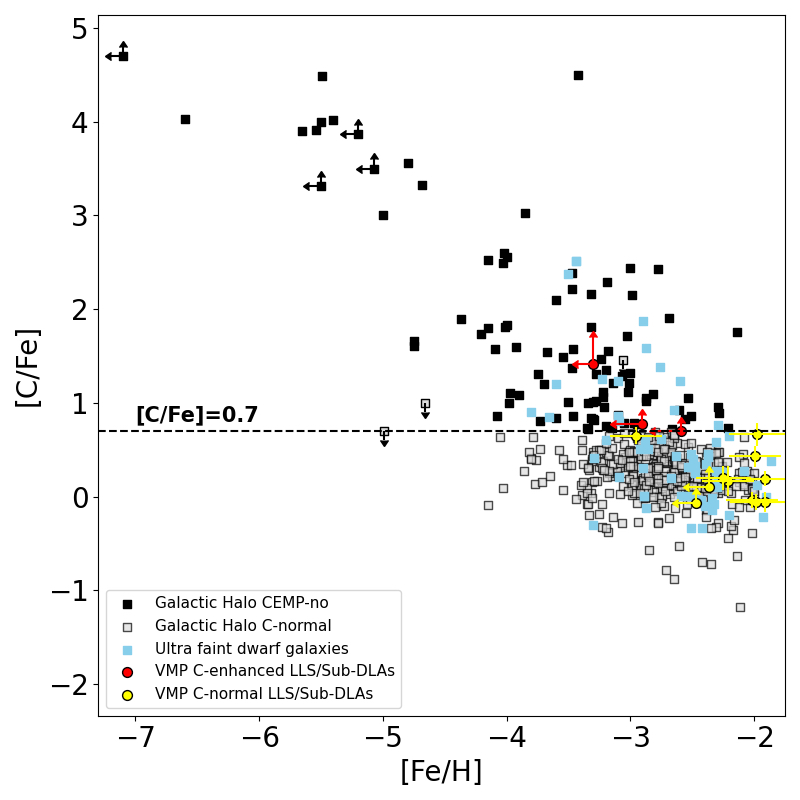}
    \caption{Comparison between the carbon-to-iron ratio of the gaseous absorption systems at different [Fe/H] and the very metal-poor compilation of stars in UFDs (light blue squares) and the Galactic halo.
    CEMP-no halo stars are shown as black squares, C-normal halo stars as grey squares. Stellar measurements have typical 1$\sigma$ errors of 0.2\,dex. Red and yellow circles are respectively C-enhanced and C-normal very metal-poor absorbers, which include Lyman Limit systems (LLSs) and sub-damped Lyman-$\alpha$ systems (Sub-DLAs). }
    \label{ratio}
\end{figure}

In Fig.~\ref{ratio} we compare the carbon-to-iron ratio, [C/Fe] measured in our very metal-poor gaseous systems and in Local Group stars (halo and UFDs) as a function of [Fe/H]. 
The stellar data we used were taken from \cite{Salvadori2015} and updated with new measurements for stars in UFDs \citep{Ji2016,Spite2018} and newly discovered extremely metal-poor stars in the Milky Way \citep{Stankenburg2018,Francois2018,Bonifacio2018,Aguado2019,Gonzalez2020}. All [C/Fe] values are corrected to account for internal mixing processes \citep{Placco2014}.
We see that our very metal-poor absorbers exhibit the same trend in [C/Fe] vs [Fe/H] that is observed in Local Group stars, i.e. an increasing [C/Fe] value for decreasing [Fe/H]. In particular, we see a nice overlapping in the [C/Fe] and [Fe/H] values of our absorbers and of stars in UFDs. This  suggests an additional link between the ISM of UFDs at the time of formation of their stellar populations and our gas-rich absorbers.

We also notice that among the 14 very metal-poor absorbers, 3 are carbon-enhanced, with [C/Fe]$>+0.7$. These C-enhanced absorbers have an upper limit on Fe~II, implying that their true [C/Fe] could be even larger. 
Note that there are other two absorption systems with [C/Fe]$\approx +0.7$. Still, since the definition of C-enhanced very metal-poor absorbers varies in the literature, e.g. [C/Fe]$\geq +0.7$ \citep[e.g.][]{Beers2005} and [C/Fe]$\geq +1.0$ \citep{Bonifacio2015} we decided to include them in the C-normal sub-sample. For the 3 carbon-enhanced systems we checked the reliability of the column density measurements against our hypothesis of turbulent broadening with respect to thermal broadening. To this end, we fit the detected lines (Si~II, C~II and Al~II) assuming they arise in the same gas and they are thermally broadened and we find negligible variation of the column density ( $\leq 0.02$) for all the transitions and $\leq 0.1$ for the C~II in J1658-0739 at z=3.54604. 
 
In Figure \ref{allratio} (upper left panel) we show the [C/Fe] abundance as a function of [Fe/H] for {\it all} our absorption systems, i.e. including those at [Fe/H]$>-2$, some of which are also C-enhanced. To make a full and detailed comparison with the stellar chemical abundances, we show in the same Figure the [C/Fe] vs [Fe/H] values for {\it all} the observed stars in the Milky Way and nearby dwarf galaxies (both UFDs and dwarf spheroidal), including also CEMP-s stars \citep{Salvadori2015,Placco2019}.

When considering this complete stellar sample, the excellent agreement between the chemical abundances measured in stars and gaseous absorbers at [Fe/H]$<-2$ is even more evident, for both C-enhanced and C-normal systems. In particular, we see that very metal-poor C-enhanced absorbers never overlap with CEMP-s stars, which are shifted towards higher [Fe/H] at any given carbon-to-iron ratio. Conversely, we notice that systems with [Fe/H]$>-2$ cover a wider range of [C/Fe] abundances than what is observed both in very metal-poor absorbers and in stars at the same [Fe/H].
Among these [Fe/H]$>-2$ absorbers, we observe four C-enhanced systems, three of which nicely overlap with CEMP-s stars. We also see many (9) C-normal absorbers that reside in the same region of C-normal stars. At [Fe/H]$>0$, where a few star measurements are available, we notice one C-enhanced absorber and two systems that are strongly C-deficient with respect to stars ([C/Fe]$<-1$). The large [C/Fe] scatter of the absorbers in this [Fe/H]$\geq -1$ regime, along with the presence of C-deficient systems, suggest that a non-negligible amount of carbon (and iron) is likely depleted onto dust grains.
Ultimately our results, which are in line with previous studies \citep[e.g.][]{Vladilo1998,Quiret2016,Decia2018,Vladilo2018}, confirm that the dust contribution makes the comparison between the chemical abundances of gas and stars challenging at [Fe/H]$\geq -1$, while it can be neglected in the very metal-poor regime. Armed with these new findings we can make further comparison between the chemical abundances of gas and stars in the very metal-poor regime.

\subsection{Other chemical elements} 
\label{otherchemialelemets}
\begin{figure*}[t!]
\vspace{1cm}
\centering
 \makebox[\linewidth]{\includegraphics[scale=0.25]{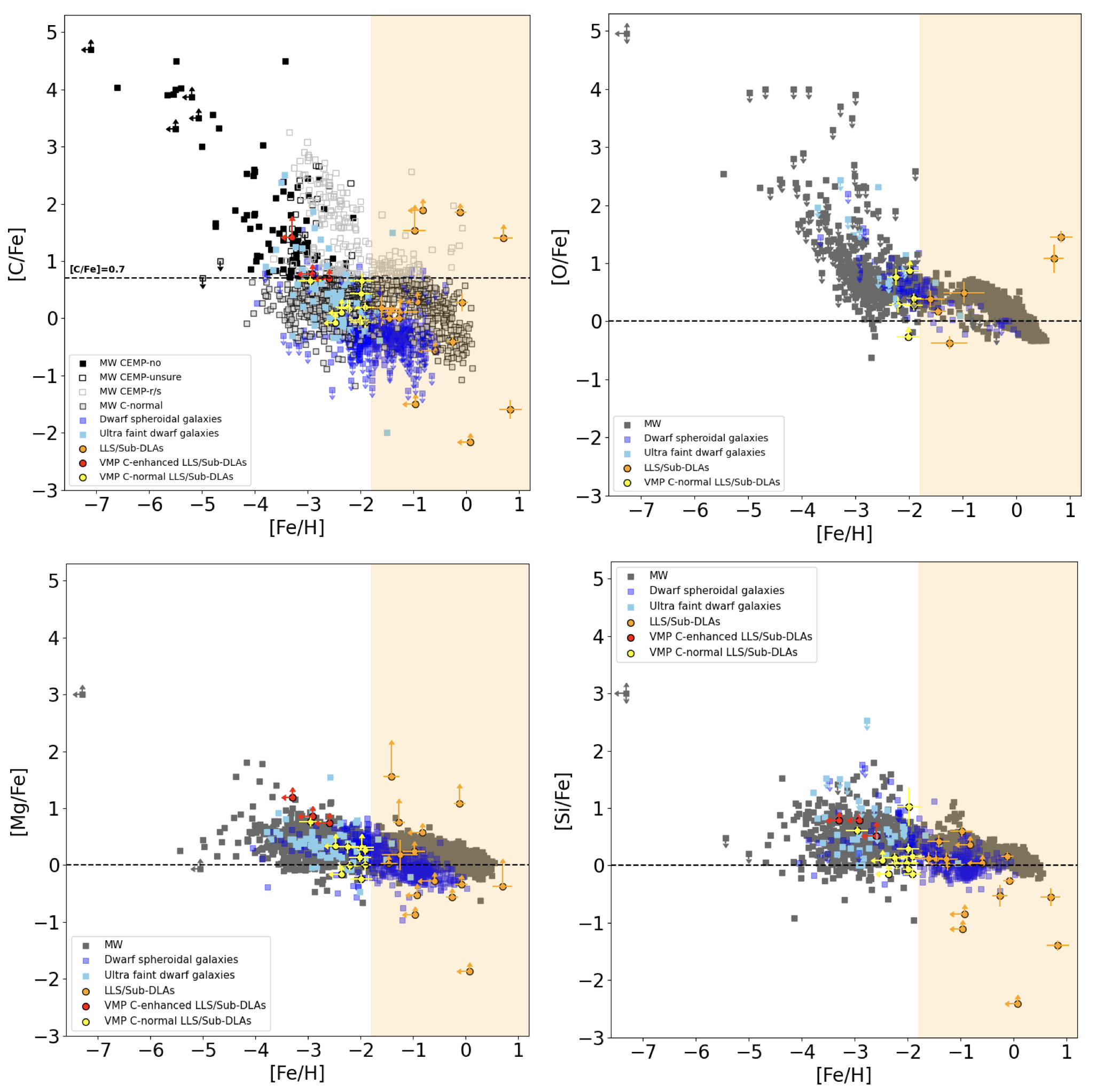}}
\caption{Upper left panel: carbon-to-iron ratio as a function of [Fe/H] of all our gaseous absorption systems (filled circles) and of stars in ultra-faint dwarf galaxies (light blue squares), dwarf spheroidal galaxies (blue squares) and the Milky Way (grey/black squares). CEMP-no halo stars are shown as filled black symbols, C-normal halo stars as filled bordered light grey squares and open symbols are CEMP-s/r stars. Stellar measurements have typical 1$\sigma$ errors of 0.2\,dex. We distinguish among C-enhanced (red) and C-normal (yellow) very metal-poor absorbers. Orange circles are iron-rich absorbers.
Other panels: Comparison between the chemical abundances of the very metal-poor gaseous absorption systems (circles) and of the stars (squares) in the Milky Way (grey), ultra-faint dwarf galaxies (light blue) and dwarf spheroidal galaxies (blue). Red and yellow circles are respectively C-enhanced and C-normal, very metal-poor absorbers. Orange circles are iron-rich absorbers. Stellar data are taken from the SAGA database \citep{Suda2008}. In all the panels the orange shaded area identifies the region where chemical abundances could be affected by dust depletion. \vspace{1cm}}
\label{allratio}
\end{figure*}

To investigate more deeply the chemical enrichment history of our C-enhanced very metal-poor absorbers, in Fig. \ref{allratio} we compare the relative abundances of oxygen, magnesium, and silicon with respect to iron, with those of ancient very metal-poor stars.
The stellar data we used for these elements were taken from the SAGA database \citep{Suda2008}. The sample of stars is different from the one used for carbon-to-iron (Fig. \ref{ratio} and upper left panel of Fig. \ref{allratio}).
For this reason we do not distinguish between CEMP-no and CEMP-s/r stars.
\begin{figure*}[t!]
\centering
\includegraphics[width=0.77\textwidth]{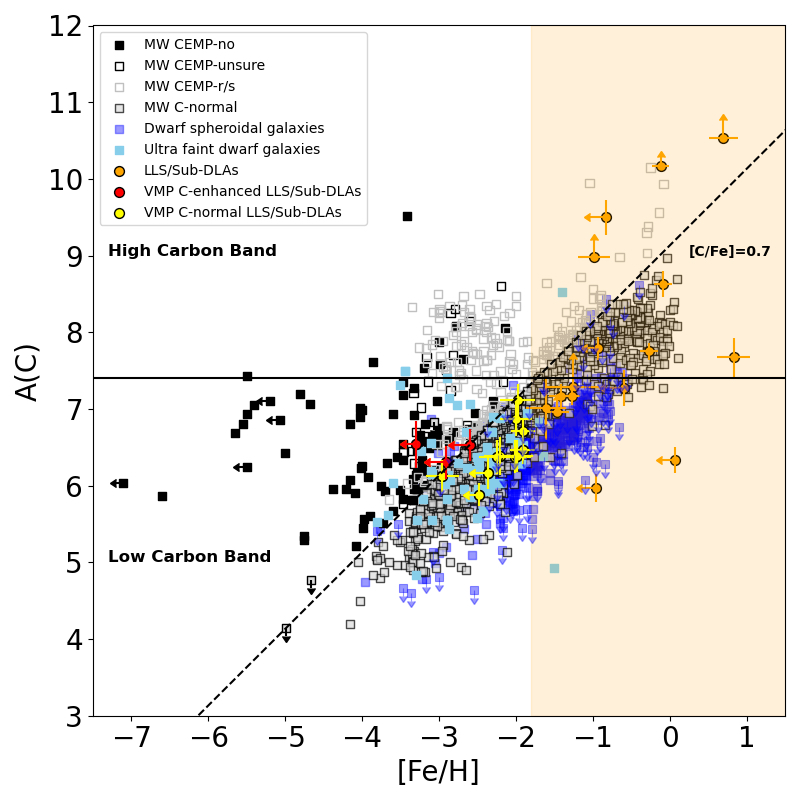}
\caption{Absolute carbon abundance (A(C)) as a function of [Fe/H] of all our gaseous absorption systems (filled circles) and of stars in ultra-faint dwarf galaxies (light blue squares), dwarf spheroidal galaxies (blue squares) and the Milky Way (grey/black squares). CEMP-no halo stars are shown as filled black symbols, C-normal halo stars as filled bordered light grey squares and open symbols are CEMP-s/r stars. We distinguish among C-enhanced (red) and C-normal (yellow) very metal-poor absorbers. Orange circles are iron-rich absorbers. Stellar measurements have typical 1$\sigma$ errors of 0.2\,dex. We used the same stellar sample of Fig.~\ref{allratio} (upper left panel).
The horizontal line separates the low-C from the high-C band, while the dashed one indicates the value [C/Fe]=+0.7. The orange shaded area identifies the region where chemical abundances could be affected by dust depletion.}
    \label{ac}
\end{figure*}

In Fig.~\ref{allratio} we see that there is a general agreement between the chemical abundances of our absorption systems and the abundances of the stars in the MW and dwarf galaxies. Furthermore, at [Fe/H]$< -2$, we note the same trend for all the chemical elements in both stars and gas: the relative abundances increase as the [Fe/H] decreases. In particular, we see that C-enhanced very metal-poor absorbers are always overabundant also in the other chemical elements. The same is observed in CEMP-no stars (Vanni et al.
in prep).

Oxygen (Fig. \ref{allratio}, upper right panel) shows a similar trend to that of carbon even if we have fewer measurements. 
Indeed, we could measure the oxygen abundance only for 15 of the 30 systems for which we have the C abundance. In Fig. \ref{allratio}, however, we report only 12 measurements since for 3 absorbers we have both upper limits for iron and oxygen. For the remaining systems, the O~I absorption lines fall in the Ly$\alpha$ forest and are heavily blended with other absorption lines. 

Magnesium (Fig. \ref{allratio}, lower left panel), which is produced by massive stars and released in the gas during supernova explosions, shows a similar trend with iron to that seen for [C/Fe], although the values are less extreme. In particular, we see that all our C-enhanced very metal-poor systems are rich in magnesium, [Mg/Fe]$>+0.7$, as observed in CEMP-no stars \citep[e.g.][and Vanni et al. in prep. for a global view]{Frebel2005, Keller2014}. The correlation between C and Mg tells us that these elements are probably produced by the same sources, i.e. most likely primordial low-energy supernovae. At higher iron abundances, [Fe/H]$>-2$, we see a large scatter. 

Silicon (Fig. \ref{allratio}, lower right panel) also shows increasing relative abundances with respect to iron as [Fe/H] decreases. Once again, the C-enhanced very metal-poor systems are also enhanced in silicon, showing super solar abundances. Also for silicon, we see that the system-to-system scatter increases at [Fe/H]$>-2$, with abundance ratios spanning a wide range of values, many also in the sub-solar regime. This result suggests that silicon is likely one of the chemical elements that is most affected by dust depletion. 

We recall that the chemical abundances measured in stars might suffer NLTE effects, whose precise estimate, at the moment, is only available at [Fe/H]$<-2$ for a small stellar sub-sample and for a few elements. Such effects can lower the [C/Fe] and [O/Fe] values measured in metal-poor stars by more than $0.3$\,dex \citep{Amarsi2019} while they can increase the [Mg/Fe] value by $+0.4$\,dex \citep{Andrievsky2010}, thus resulting in a better agreement between our C-enhanced very metal-poor absorbers and CEMP-no stars (see Fig.~4). For [Al/Fe] the correction is extremely high, $+0.65$\,dex \citep{Cayrel2004}, and for this reason we decided not to make the comparison with the values measured in our absorbers.

Ultimately, Fig.~\ref{allratio} demonstrates that in the very metal-poor regime, [Fe/H]$<-2$, the chemical abundance ratios measured in our gaseous absorbers are in very good agreement with those of present-day stars and that our C-enhanced very metal-poor absorption systems reside in the same regions of CEMP-no stars and have the same chemical properties.

\subsection{First star signatures}
\label{subsect:ac}
The results described in the previous Sections emphasize the idea that our C-enhanced very metal-poor absorbers are the gaseous analogues of CEMP-no stars, which have been likely imprinted by primordial low-energy SNe. Indeed, in our absorption systems we see an overabundance of Mg and Si, which is also observed in CEMP-no stars. Magnesium and silicon are key elements since they are produced by primordial low-energy supernovae but not by AGB stars, which are the pollutant of CEMP-s stars. AGB stars yield high C, N and O, and also produce Ba via the slow-neutron capture process. 

In CEMP-s stars the C and Ba excess is expected to be acquired via mass transfer from a binary AGB companion, while in CEMP-no stars to be representative of the environment of formation. For this reason barium is used to discriminate between CEMP-s stars ([C/Fe]$>+0.7$, [Ba/Fe]$>+1.0$), and CEMP-no stars ([C/Fe]$>+0.7$, [Ba/Fe]$<0.0$, \cite[e.g.][]{Beers2005}).
Unfortunately, no barium measurements are (nor will be) available for our absorption systems. Thus, to further validate the link between our C-enhanced very metal-poor absorbers and CEMP-no stars we need to carry out additional tests. 

In stellar archaeology, the absolute abundance of carbon, $A(C)$, displayed as a function of [Fe/H], is used to discriminate among CEMP-no and CEMP-s stars when barium abundances are not available.
Indeed, it has been shown \citep[e.g.][]{Spite2013,Bonifacio2015,Yoon2016} that these two different populations dwell in two well separated regions: the high carbon band, $A(C)>7.4$, and the low carbon band, $A(C)<7.4$. The stars belonging to the low carbon band are mostly CEMP-no stars\footnote{Stars belonging to the low carbon band can be subsequently divided into {\it Group II} and {\it Group III}, which might have different astrophysical origin \citep[e.g.][]{Yoon2016,Yoon2019,Zepeda2022} possibly linked with various dust composition in Pop III supernovae ejecta \citep[e.g.][]{Chiaki2017}. However, given the upper limit on [Fe/H] of our CEMP-no absorbers, we are not able to distinguish between the two groups.}, while those of the high carbon band are CEMP-s stars. We can thus use this diagnostic to unveil the nature of our C-enhanced absorption systems.

In Figure \ref{ac} we display the absolute carbon abundance of Local Group stars and of our absorbers as a function of [Fe/H]. For the absorption systems we computed $A(C)=\log(N_{C}/N_{H})+12$, where $N_C$ and $N_H$ are the measured column densities corrected for ionisation.
We see that our three carbon-enhanced very metal-poor absorbers are found in the low C-band, which is consistent with what is found for CEMP-no stars. These systems are clearly separated from the C-enhanced absorbers at [Fe/H]$>-2$, which populate the high carbon band where CEMP-s stars reside. In other words, the division between the low- and the high-C band corresponds to a division in [Fe/H] of our C-enhanced absorbers. Hence, despite the lack of barium measurements, we can conclude that our three C-enhanced absorbers at [Fe/H]$<-2$ are CEMP-no absorption systems, while the four at [Fe/H]$>-2$ are CEMP-s. Note that the C-enhanced absorber recently discovered by \cite{Zou2020} reside in the CEMP-s region ([Fe/H]$=-1.6$, A(C)$=9.0$). Thus, like for our four CEMP-s absorbers, the C-excess of the absorption system studied by \cite{Zou2020} most likely arise from the contribution of AGB stars, and not from the chemical elements yielded by the first stars.
\newpage
\section{Discussion} \label{sec:discussion}
All gathered evidence supports the idea that the newly discovered C-enhanced very metal-poor absorption systems are the gaseous $z\sim3-4$ analogues of present-day CEMP-no stars. Hence, we propose to define {\it CEMP-no absorbers}, LLSs/sub-DLAs with [Fe/H]$<-2$ and [C/Fe]$>+0.7$. An increasing number of theoretical studies is supporting the idea that CEMP-no stars observed in different environments have been most likely enriched by the first stellar generations \citep[e.g.][]{Iwamoto2005, Salvadori2015, Liu2021}. Thus, we are providing the first clues of Pop~III stars-enriched gas in high-z absorbers. Still, two questions naturally arise: what is the nature of these high-z CEMP-no absorption systems? And why CEMP-no absorbers have so far escaped detection?

\subsection{Origin of CEMP-no absorbers}
\label{subsect:origin}
Different semi-analytical models and cosmological simulations have investigated the origin of CEMP-no stars in the Galactic halo \citep[e.g.][]{deBennassuti2017,Hartwig2018,Liu2021} and in ultra-faint dwarf galaxies \citep[e.g.][Rossi et al. in prep]{Salvadori2015,Jeon2021}. We can thus exploit these findings to interpret the properties of our CEMP-no absorption systems.
The moderately high [C/Fe] and relatively high [Fe/H] values of CEMP-no absorbers can be explained as the result of two different enrichment mechanisms (Vanni et al. in prep). These mechanisms are: (i) a pollution solely driven by Pop~III stars, which explode as low-energy supernovae with different masses \citep{Hartwig2018,Welsh2021,Jeon2021}; (ii) an enrichment still dominated by the products of low-energy Pop~III supernovae, but which is transiting towards a C-normal pattern due to the contribution of subsequent generations of normal (Pop II) stars exploding as core-collapse supernovae \citep[][]{deBennassuti2017,Salvadori2015,Jeon2021}. 

In both cases, the chemical enrichment should be dominated by the chemical products of primordial low-energy supernovae, which produce the carbon-over-iron excess (Salvadori et al. 2023). However, CEMP-no relic stars likely form during the first Gyr of cosmic evolution ($z>6$), while our CEMP-no absorbers are observed at lower redshifts, $z\approx 3-4$, when the Universe was $>2$~Gyr old. What are these CEMP-no absorbers and how can they preserve the chemical signature of Pop~III stars? 

Two main possibilities exist. The first, is that our CEMP-no absorption systems are associated to the diffuse circum-galactic medium (CGM) of recently born Pop~III galaxies at $z\approx 3-4$. Still, these objects should be extremely rare. Indeed, direct detection of Pop~III galaxies is still lacking and at $z\approx 3-4$ pristine galaxies are expected to be extremely rare \citep[e.g.][]{Pallottini2014, Jaacks2019}. 
The second, is that our CEMP-no absorption systems are associated to low-mass ``sterile" mini-halos \citep{Salvadori2012}. Radiative feedback processes can indeed increase the gas temperature of metal-enriched mini-halos, which are then too diffuse to form stars but too metal-enhanced to be photo-evaporated. The chemical signature of the first stellar generations can then be preserved in these gas-rich systems until they evolve in isolation. This physical mechanism, which has been proposed by \cite{Salvadori2012}, should be quite common in the early Universe and can turn low-mass UFDs at the beginning of their evolution into ``failed" UFDs: the gas-rich and high-z dark counterpart of ultra-faint dwarf galaxies.

\begin{figure}[h]
\centering
\includegraphics[scale=0.18]{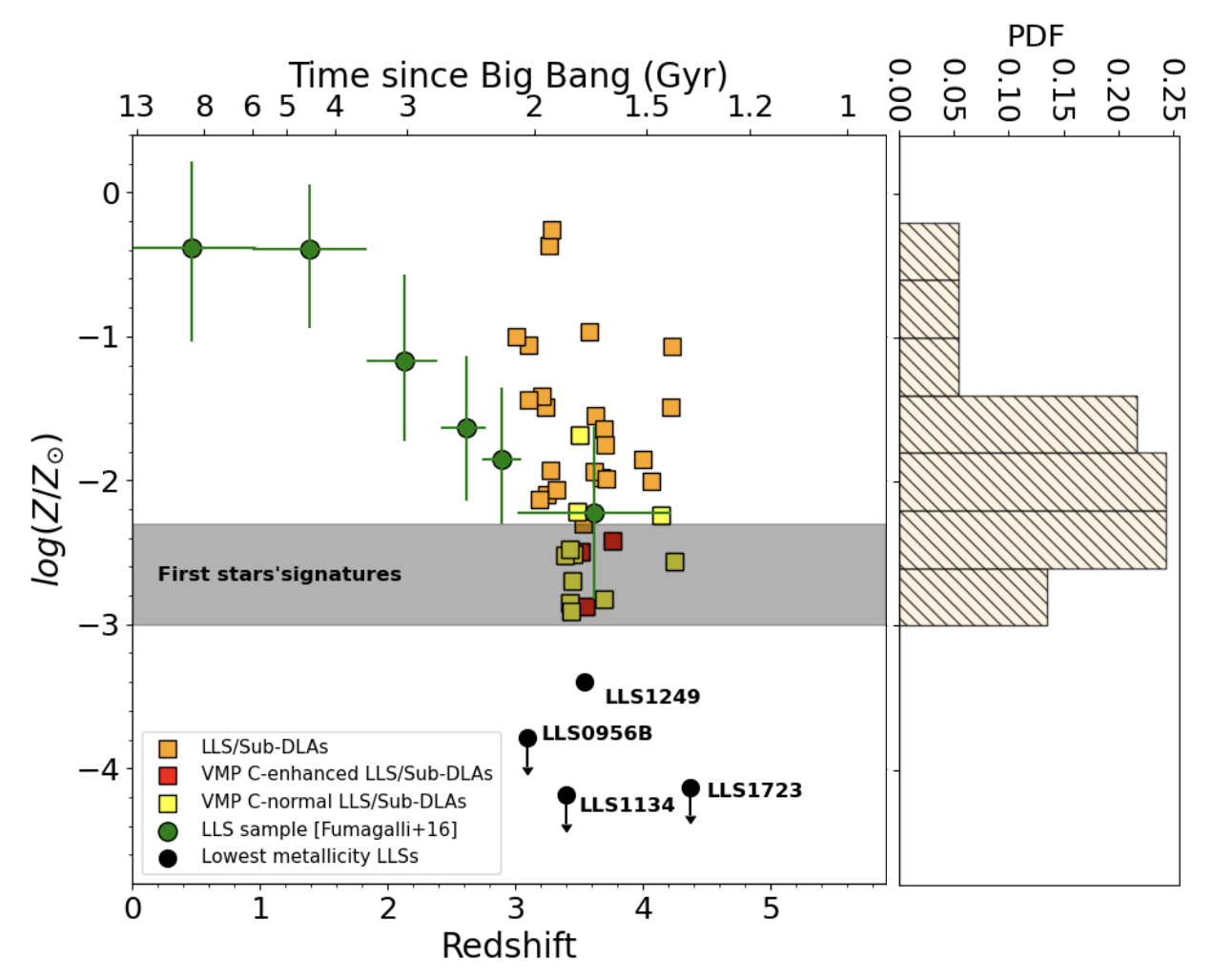}
\caption{Summary of the metallicity distribution of LLSs (green and black circles) in the literature. Upper limits from \cite{Fumagalli2011} for two apparently metal-free LLSs, LLS0958B65 and LLS113465, and from \cite{Robert2019}, LLS172366, are shown with black arrows. The lowest metallicity measurement for a LLS from \cite{Crighton2016}, LLS124967, is shown in black. The green circles and error bars represent the LLS sample of \cite{Fumagalli2016}. Red and yellow squares are respectively C-enhanced and C-normal, very metal-poor absorbers. Orange squares are iron-rich absorbers.
The shaded red region shows the metallicity range, according to our results, for gas enriched by first stars. On the right we show the metallicity posterior distribution function of all our absorbers.}
\label{metallicity}
\end{figure}

\subsection{Comparison with Literature Data}
To examine our results in a global context and understand why CEMP-no absorption systems have not been identified yet, we can compare the {\it total metallicity} of all our 37 diffuse absorbers\footnote{Note that the number of absorption systems with iron measurements is smaller (30) than those with metallicity measurements.} with literature data for LLSs \citep{Fumagalli2011,Fumagalli2016,Crighton2016,Robert2019}. The comparison is shown in Fig. \ref{metallicity}, where the total metallicity of the absorbers is displayed as a function of redshift (or cosmic time) and our systems are coloured to distinguish among [Fe/H]$>-2$ absorbers, CEMP-no systems, and very metal-poor C-normal absorbers (see caption and Fig.~\ref{allratio}). Literature data include: (i) the four serendipitously discovered LLSs that have $Z<10^{-3}Z_{\odot}$ and cover the same redshift range of our absorbers, $3<z<4.5$; and (ii) the sample of LLSs at $0<z<4.5$ collected by \cite{Fumagalli2016}, for which we report the median metallicity computed by the authors using redshift bins containing at least 25 LLSs.

The right panel of Fig. \ref{metallicity} shows the total metallicity distribution function of our absorbers.
Firstly, we see that our absorption systems cover a wide metallicity range, $10^{-3} Z_{\odot}< Z < Z_{\odot}$, and that $>90\%$ of our very metal-poor absorbers, [Fe/H]$<-2$, have a total metallicity $Z< 10^{-2} Z_{\odot}$ (the only exception being one C-normal absorber). This result is a further confirmation that the effect of dust is negligible at [Fe/H]$<-2$, as suggested by previous works (e.g. \citealt{Vladilo1998,Vladilo2018}).
Secondly, we note that the peak of the distribution of our absorption systems, $Z\approx 10^{-2.5} Z_{\odot}$, is in perfect agreement with the average metallicity value of LLSs derived by \cite{Fumagalli2016} in the corresponding redshift bin, $3<z<4.5$. 
Then, we note that among our absorbers there is a lack of extremely metal-poor absorption systems, $Z<10^{-3}Z_{\odot}$, although our systems cover exactly the same redshift range of the most pristine LLSs. This is 
certainly due to our initial selection which was based on the presence of metal absorption lines (see Sec.~\ref{method}).
Finally, we note that CEMP-no absorbers showing the chemical signature of the first stars are {\it not} the most pristine objects. This result, which is in line with what is found in the stellar halo, is a consequence of the key chemical signature left by primordial low-energy supernovae: an excess of C (O, Mg, and Si) over iron. To identify the footprint of primordial low energy supernovae in the gas at high-z we should then look for {\it the most iron-poor absorbers, which are not the most metal-poor} but rather have $10^{-3}Z_{\odot}< Z < 10^{-2.3}Z_{\odot}$. Among the 12 absorption systems with available carbon and iron measurements at $Z<10^{-2.5\pm 0.1}Z_{\odot}$, we have that three are CEMP-no, implying that the fraction of CEMP-no systems is $F_{CEMP-no}(Z<10^{-2.5}Z_{\odot})\geq 25\%$.
We note that this is just an indicative fraction since we have not determined the completeness of our selection of absorption systems based on the automatic detection of  Mg~II absorption doublets.

Ultimately, we suggest that metal-poor LLSs are the most promising candidates as CEMP-no absorption systems. Indeed, from the one hand, DLAs are likely too dense to stop forming stars and thus are naturally dominated by the chemical products of normal Pop~II stars. From the other hand, the most pristine LLSs are too metal-poor to be enriched by the chemical products of primordial low-energy supernovae.

\section{Conclusions} \label{sec:conclusion}
In this paper, we have studied the chemical abundances of 37 optically thick Lyman-$\alpha$ absorbers (LLSs and Sub-DLAs) at $z\sim3-4.5$ identified in the spectra of the XQ-100 quasar legacy survey \citep{Lopez2016}. 
Column densities of the absorption features have been derived through fitting with Voigt profiles and ionization corrections have been applied in order to derive the chemical abundances. 
The main results of our study can be summarised as follows: 
\begin{itemize}
\item The MDF of our absorption systems covers a broad range in iron abundance and is bi-modal: it has a broad peak at [Fe/H]$\sim-0.8$ and a more pronounced one at [Fe/H]$\sim-2$. 

\item The low-Fe tail of the MDF of our absorption systems almost overlaps with the stellar MDF observed in UFDs. 

\item Among the 30 analysed absorbers with available iron measurements, we identified 14 very metal-poor systems, [Fe/H]$<-2$, three of which are carbon-enhanced, [C/Fe]$>+0.7$ (J0835+0650 at z=3.51256, J1111-0804 at z=3.75837 and J1658-0739 at z=3.54604).

\item The chemical abundance ratios (C, O, Mg, and Si over Fe) measured in very metal-poor absorbers are in good agreement with those of very metal-poor stars in the Galactic halo and dwarf galaxies.

\item Conversely, at [Fe/H]$>-2$, the absorbers' abundance ratios exhibit a larger scatter than present-day stars, suggesting that dust contribution is no longer negligible at these [Fe/H] values.

\item The three C-enhanced very metal-poor absorbers also show an overabundance of Mg and Si, which are produced by first stars exploding as low-energy supernovae and not by AGB stars. These overabundances are also observed in CEMP-no stars.

\item All C-enhanced very metal-poor absorption systems have $A(C)<7.4$, i.e. they reside in the so-called low carbon band, as it is observed for CEMP-no stars.

\item Conversely, all C-enhanced systems at [Fe/H]$>-2$ dwell in the high carbon band, $A(C)>7.4$, consistent with CEMP-s stars.

\item CEMP-no absorbers are the most iron-poor among our diffuse systems, but they are not the most metal-poor ones.

\item Among the 12 absorption systems with available C and Fe measurements at $Z<10^{-2.5\pm 0.1}Z_{\odot}$, we have that three are CEMP-no, implying that $F_{CEMP-no}(Z<10^{-2.5}Z_{\odot})\geq 25\%$.

\end{itemize}

Ultimately, our CEMP-no absorption systems seem to be the gaseous high-redshift analogues of locally observed CEMP-no stars, which have been more likely imprinted by primordial low-energy supernovae. Our new discovery of CEMP-no absorbers suggests that optically thick, relatively diffuse absorption systems are the best environments to identify the missing chemical signature of the first stellar generations in the gaseous component. These absorbers are likely too diffuse to be star-forming, a key requirement to prevent further chemical pollution and thus to preserve the first-star signature. Theoretical investigations and further observational studies are required to fully understand their nature. 

In the very near future, extremely large samples of high-redshift quasars will be available from several  surveys (DESI \citealt{Abareshi2022}; WEAVE-QSO \citealt{Pieri2016}; 4-MOST \citealt{deJong2019}). They will allow us to select more of these CEMP-no absorbers to statistically characterize their properties and number density.  A significant step forward in the analysis of these systems will be represented by ANDES, the high-resolution ($R\sim100,000$) spectrograph for the Extremely Large Telescope (ELT), foreseen for the beginning of the 30s \citep{Marconi2022}. The ANDES spectral coverage extending to the NIR and the collecting power of ELT will allow us to carry out detailed studies of these systems resolving the metal absorptions and determining significant constraints on key elements like Zn.  

In the end, our work, which fully complements stellar archaeology, represents a fresh start for the searches of first-star pollution in high-z environments that can provide unique insight on both the early phases of reionization and on the physical processes shaping the evolution of the first galaxies \citep{Pallottini2014,Salvadori2012}.

\begin{acknowledgments}
\textbf{Acknowledgements.} This project has received funding from the European Research Council (ERC) under the European Union’s Horizon 2020 research and innovation programme (grant agreement No 804240) (PI S. Salvadori). A.S. acknowledges support from ED127 and DIM-ACAV+, S.S. and V.D. from the PRIN-MIUR17, The quest for the first stars, prot. n. 2017T4ARJ5, M.F. from the ERC Starting Grant FEEDGALAXIES H2020/757535, SL was funded by FONDECYT grant number 1231187, and G.B. was supported by the National Science Foundation through grant AST-1751404. The authors are grateful to Stefano Cristiani, Paolo Molaro, Asa Skuladottir, Monique Spite, Irene Vanni, Susanna D. Vergani and Louise Welsh for insightful discussions and critical reading of the first draft of the manuscript. We thank the referee, Timothy Beers, for an enthusiastic report. 
\end{acknowledgments}

\bibliography{bibliografia}{}

\begin{thebibliography}{}
\expandafter\ifx\csname natexlab\endcsname\relax\def\natexlab#1{#1}\fi
\providecommand{\url}[1]{\href{#1}{#1}}
\providecommand{\dodoi}[1]{doi:~\href{http://doi.org/#1}{\nolinkurl{#1}}}
\providecommand{\doeprint}[1]{\href{http://ascl.net/#1}{\nolinkurl{http://ascl.net/#1}}}
\providecommand{\doarXiv}[1]{\href{https://arxiv.org/abs/#1}{\nolinkurl{https://arxiv.org/abs/#1}}}

\bibitem[{{Abareshi} {et~al.}(2022){Abareshi}, {Aguilar}, {Ahlen}, {Alam},
  {Alexander}, {Alfarsy}, {Allen}, {Allende Prieto}, {Alves}, {Ameel},
  {Armengaud}, {Asorey}, {Aviles}, {Bailey}, {Balaguera-Antol{\'\i}nez},
  {Ballester}, {Baltay}, {Bault}, {Beltran}, {Benavides}, {BenZvi}, {Berti},
  {Besuner}, {Beutler}, {Bianchi}, {Blake}, {Blanc}, {Blum}, {Bolton}, {Bose},
  {Bramall}, {Brieden}, {Brodzeller}, {Brooks}, {Brownewell}, {Buckley-Geer},
  {Cahn}, {Cai}, {Canning}, {Carnero Rosell}, {Carton}, {Casas}, {Castander},
  {Cervantes-Cota}, {Chabanier}, {Chaussidon}, {Chuang}, {Circosta}, {Cole},
  {Cooper}, {da Costa}, {Cousinou}, {Cuceu}, {Davis}, {Dawson}, {de la
  Cruz-Noriega}, {de la Macorra}, {de Mattia}, {Della Costa}, {Demmer},
  {Derwent}, {Dey}, {Dey}, {Dhungana}, {Ding}, {Dobson}, {Doel},
  {Donald-McCann}, {Donaldson}, {Douglass}, {Duan}, {Dunlop}, {Edelstein},
  {Eftekharzadeh}, {Eisenstein}, {Enriquez-Vargas}, {Escoffier}, {Evatt},
  {Fagrelius}, {Fan}, {Fanning}, {Fawcett}, {Ferraro}, {Ereza}, {Flaugher},
  {Font-Ribera}, {Forero-Romero}, {Frenk}, {Fromenteau}, {G{\"a}nsicke},
  {Garcia-Quintero}, {Garrison}, {Gazta{\~n}aga}, {Gerardi}, {Gil-Mar{\'\i}n},
  {Gontcho}, {Gonzalez-Morales}, {Gonzalez-de-Rivera}, {Gonzalez-Perez},
  {Gordon}, {Graur}, {Green}, {Grove}, {Gruen}, {Gutierrez}, {Guy}, {Hahn},
  {Harris}, {Herrera}, {Herrera-Alcantar}, {Honscheid}, {Howlett}, {Huterer},
  {Ir{\v{s}}i{\v{c}}}, {Ishak}, {Jelinsky}, {Jiang}, {Jimenez}, {Jing},
  {Joyce}, {Jullo}, {Juneau}, {Kara{\c{c}}ayl{\i}}, {Karamanis}, {Karcher},
  {Karim}, {Kehoe}, {Kent}, {Kirkby}, {Kisner}, {Kitaura}, {Koposov},
  {Kov{\'a}cs}, {Kremin}, {Krolewski}, {L'Huillier}, {Lahav}, {Lambert},
  {Lamman}, {Lan}, {Landriau}, {Lane}, {Lang}, {Lange}, {Lasker}, {Le Guillou},
  {Leauthaud}, {Le Van Suu}, {Levi}, {Li}, {Magneville}, {Manera}, {Manser},
  {Marshall}, {McCollam}, {McDonald}, {Meisner}, {Mezcua}, {Miller}, {Miquel},
  {Montero-Camacho}, {Moon}, {Martini}, {Meneses-Rizo}, {Moustakas}, {Mueller},
  {Mu{\~n}oz-Guti{\'e}rrez}, {Myers}, {Nadathur}, {Najita}, {Napolitano},
  {Neilsen}, {Newman}, {Nie}, {Ning}, {Niz}, {Norberg}, {Noriega}, {O'Brien},
  {Obuljen}, {Palanque-Delabrouille}, {Palmese}, {Zhiwei}, {Pappalardo},
  {Peng}, {Percival}, {Perruchot}, {Pogge}, {Poppett}, {Porredon}, {Prada},
  {Prochaska}, {Pucha}, {P{\'e}rez-Fern{\'a}ndez}, {P{\'e}rez-R{\'a}fols},
  {Rabinowitz}, {Raichoor}, {Ramirez-Solano}, {Ram{\'\i}rez-P{\'e}rez},
  {Ravoux}, {Reil}, {Rezaie}, {Rocher}, {Rockosi}, {Roe}, {Roodman}, {Ross},
  {Rossi}, {Ruggeri}, {Ruhlmann-Kleider}, {Sabiu}, {Safonova}, {Said},
  {Saintonge}, {Salas Catonga}, {Samushia}, {Sanchez}, {Saulder}, {Schaan},
  {Schlafly}, {Schlegel}, {Schmoll}, {Scholte}, {Schubnell}, {Secroun}, {Seo},
  {Serrano}, {Sharples}, {Sholl}, {Silber}, {Silva}, {Sirk}, {Siudek}, {Smith},
  {Sprayberry}, {Staten}, {Stupak}, {Tan}, {Tarl{\'e}}, {Sien Tie}, {Tojeiro},
  {Ure{\~n}a-L{\'o}pez}, {Valdes}, {Valenzuela}, {Valluri},
  {Vargas-Maga{\~n}a}, {Verde}, {Walther}, {Wang}, {Wang}, {Weaver},
  {Weaverdyck}, {Wechsler}, {Wilson}, {Yang}, {Yu}, {Yuan}, {Y{\`e}che},
  {Zhang}, {Zhang}, {Zhao}, {Zhou}, {Zhou}, {Zou}, {Zou}, {Zou}, \&
  {Zu}}]{Abareshi2022}
{Abareshi}, B., {Aguilar}, J., {Ahlen}, S., {et~al.} 2022, arXiv e-prints,
  arXiv:2205.10939.
\newblock \doarXiv{2205.10939}

\bibitem[{{Abate} {et~al.}(2015){Abate}, {Pols}, {Izzard}, \&
  {Karakas}}]{Abate2015}
{Abate}, C., {Pols}, O.~R., {Izzard}, R.~G., \& {Karakas}, A.~I. 2015, \aap,
  581, A22, \dodoi{10.1051/0004-6361/201525876}

\bibitem[{{Aguado} {et~al.}(2022){Aguado}, {Molaro}, {Caffau}, {Gonz{\'a}lez
  Hern{\'a}ndez}, {Zapatero Osorio}, {Bonifacio}, {Allende Prieto}, {Rebolo},
  {Damasso}, {Su{\'a}rez Mascare{\~n}o}, {Howell}, {Furlan}, {Cristiani},
  {Cupani}, {Di Marcantonio}, {D'Odorico}, {Lovis}, {Martins}, {Milakovic},
  {Murphy}, {Nunes}, {Pepe}, {Santos}, {Schmidt}, \& {Sozzetti}}]{Aguado2022}
{Aguado}, D., {Molaro}, P., {Caffau}, E., {et~al.} 2022, MNRAS accepted,
  arXiv:2210.04910.
\newblock \doarXiv{2210.04910}

\bibitem[{{Aguado} {et~al.}(2019){Aguado}, {Gonz{\'a}lez Hern{\'a}ndez},
  {Allende Prieto}, \& {Rebolo}}]{Aguado2019}
{Aguado}, D.~S., {Gonz{\'a}lez Hern{\'a}ndez}, J.~I., {Allende Prieto}, C., \&
  {Rebolo}, R. 2019, \apjl, 874, L21, \dodoi{10.3847/2041-8213/ab1076}

\bibitem[{{Aguado} {et~al.}(2023){Aguado}, {Caffau}, {Molaro}, {Allende
  Prieto}, {Bonifacio}, {Gonz{\'a}lez Hern{\'a}ndez}, {Rebolo}, {Salvadori},
  {Zapatero Osorio}, {Cristiani}, {Pepe}, {Santos}, {Cupani}, {Di Marcantonio},
  {D'Odorico}, {Lovis}, {Nunes}, {Martins}, {Milakovi}, {Rodrigues}, {Schmidt},
  {Sozzetti}, \& {Su{\'a}rez Mascare{\~n}o}}]{aguado2023}
{Aguado}, D.~S., {Caffau}, E., {Molaro}, P., {et~al.} 2023, \aap, 669, L4,
  \dodoi{10.1051/0004-6361/202245392}

\bibitem[{{Amarsi} {et~al.}(2019){Amarsi}, {Nissen}, \&
  {Sk{\'u}lad{\'o}ttir}}]{Amarsi2019}
{Amarsi}, A.~M., {Nissen}, P.~E., \& {Sk{\'u}lad{\'o}ttir}, {\'A}. 2019, \aap,
  630, A104, \dodoi{10.1051/0004-6361/201936265}

\bibitem[{{Andrievsky} {et~al.}(2010){Andrievsky}, {Spite}, {Korotin}, {Spite},
  {Bonifacio}, {Cayrel}, {Fran{\c{c}}ois}, \& {Hill}}]{Andrievsky2010}
{Andrievsky}, S.~M., {Spite}, M., {Korotin}, S.~A., {et~al.} 2010, \aap, 509,
  A88, \dodoi{10.1051/0004-6361/200913223}

\bibitem[{{Arentsen} {et~al.}(2019){Arentsen}, {Starkenburg}, {Shetrone},
  {Venn}, {Depagne}, \& {McConnachie}}]{Arentsen2019}
{Arentsen}, A., {Starkenburg}, E., {Shetrone}, M.~D., {et~al.} 2019, \aap, 621,
  A108, \dodoi{10.1051/0004-6361/201834146}

\bibitem[{{Asplund} {et~al.}(2009){Asplund}, {Grevesse}, {Sauval}, \&
  {Scott}}]{Asplund2009}
{Asplund}, M., {Grevesse}, N., {Sauval}, A.~J., \& {Scott}, P. 2009, \araa, 47,
  481, \dodoi{10.1146/annurev.astro.46.060407.145222}

\bibitem[{{Beers} \& {Christlieb}(2005)}]{Beers2005}
{Beers}, T.~C., \& {Christlieb}, N. 2005, \araa, 43, 531,
  \dodoi{10.1146/annurev.astro.42.053102.134057}

\bibitem[{{Beers} {et~al.}(1992){Beers}, {Preston}, \& {Shectman}}]{Beers1992}
{Beers}, T.~C., {Preston}, G.~W., \& {Shectman}, S.~A. 1992, \aj, 103, 1987,
  \dodoi{10.1086/116207}

\bibitem[{{Berg} {et~al.}(2016){Berg}, {Ellison}, {S{\'a}nchez-Ram{\'\i}rez},
  {Prochaska}, {Lopez}, {D'Odorico}, {Becker}, {Christensen}, {Cupani},
  {Denney}, \& {Worseck}}]{Berg2016}
{Berg}, T.~A.~M., {Ellison}, S.~L., {S{\'a}nchez-Ram{\'\i}rez}, R., {et~al.}
  2016, \mnras, 463, 3021, \dodoi{10.1093/mnras/stw2232}

\bibitem[{{Berg} {et~al.}(2017){Berg}, {Ellison}, {Prochaska},
  {S{\'a}nchez-Ram{\'\i}rez}, {Lopez}, {D'Odorico}, {Becker}, {Christensen},
  {Cupani}, {Denney}, \& {Worseck}}]{Berg2017}
{Berg}, T.~A.~M., {Ellison}, S.~L., {Prochaska}, J.~X., {et~al.} 2017, \mnras,
  464, L56, \dodoi{10.1093/mnrasl/slw185}

\bibitem[{{Berg} {et~al.}(2019){Berg}, {Ellison}, {S{\'a}nchez-Ram{\'\i}rez},
  {L{\'o}pez}, {D'Odorico}, {Becker}, {Christensen}, {Cupani}, {Denney}, \&
  {Worseck}}]{Berg2019}
{Berg}, T. A.~M., {Ellison}, S.~L., {S{\'a}nchez-Ram{\'\i}rez}, R., {et~al.}
  2019, \mnras, 488, 4356, \dodoi{10.1093/mnras/stz2012}

\bibitem[{{Berg} {et~al.}(2021){Berg}, {Fumagalli}, {D'Odorico}, {Ellison},
  {L{\'o}pez}, {Becker}, {Christensen}, {Cupani}, {Denney},
  {S{\'a}nchez-Ram{\'\i}rez}, \& {Worseck}}]{Berg2021}
{Berg}, T. A.~M., {Fumagalli}, M., {D'Odorico}, V., {et~al.} 2021, \mnras, 502,
  4009, \dodoi{10.1093/mnras/stab184}

\bibitem[{{Bonifacio} {et~al.}(1998){Bonifacio}, {Molaro}, {Beers}, \&
  {Vladilo}}]{Bonifacio1998}
{Bonifacio}, P., {Molaro}, P., {Beers}, T.~C., \& {Vladilo}, G. 1998, \aap,
  332, 672.
\newblock \doarXiv{astro-ph/9712227}

\bibitem[{{Bonifacio} {et~al.}(2009){Bonifacio}, {Spite}, {Cayrel}, {Hill},
  {Spite}, {Fran{\c{c}}ois}, {Plez}, {Ludwig}, {Caffau}, {Molaro}, {Depagne},
  {Andersen}, {Barbuy}, {Beers}, {Nordstr{\"o}m}, \& {Primas}}]{Bonifacio2009}
{Bonifacio}, P., {Spite}, M., {Cayrel}, R., {et~al.} 2009, \aap, 501, 519,
  \dodoi{10.1051/0004-6361/200810610}

\bibitem[{{Bonifacio} {et~al.}(2015){Bonifacio}, {Caffau}, {Spite}, {Limongi},
  {Chieffi}, {Klessen}, {Fran{\c{c}}ois}, {Molaro}, {Ludwig}, {Zaggia},
  {Spite}, {Plez}, {Cayrel}, {Christlieb}, {Clark}, {Glover}, {Hammer}, {Koch},
  {Monaco}, {Sbordone}, \& {Steffen}}]{Bonifacio2015}
{Bonifacio}, P., {Caffau}, E., {Spite}, M., {et~al.} 2015, \aap, 579, A28,
  \dodoi{10.1051/0004-6361/201425266}

\bibitem[{{Bonifacio} {et~al.}(2018){Bonifacio}, {Caffau}, {Spite}, {Spite},
  {Sbordone}, {Monaco}, {Fran{\c{c}}ois}, {Plez}, {Molaro}, {Gallagher},
  {Cayrel}, {Christlieb}, {Klessen}, {Koch}, {Ludwig}, {Steffen}, {Zaggia}, \&
  {Abate}}]{Bonifacio2018}
---. 2018, \aap, 612, A65, \dodoi{10.1051/0004-6361/201732320}

\bibitem[{{Bonifacio} {et~al.}(2021){Bonifacio}, {Monaco}, {Salvadori},
  {Caffau}, {Spite}, {Sbordone}, {Spite}, {Ludwig}, {Di Matteo}, {Haywood},
  {Fran{\c{c}}ois}, {Koch-Hansen}, {Christlieb}, \& {Zaggia}}]{Bonifacio2021}
{Bonifacio}, P., {Monaco}, L., {Salvadori}, S., {et~al.} 2021, \aap, 651, A79,
  \dodoi{10.1051/0004-6361/202140816}

\bibitem[{{Bromm} \& {Loeb}(2003)}]{Bromm2003}
{Bromm}, V., \& {Loeb}, A. 2003, \nat, 425, 812, \dodoi{10.1038/nature02071}

\bibitem[{{Carswell} {et~al.}(2012){Carswell}, {Becker}, {Jorgenson}, {Murphy},
  \& {Wolfe}}]{Carswell2012}
{Carswell}, R.~F., {Becker}, G.~D., {Jorgenson}, R.~A., {Murphy}, M.~T., \&
  {Wolfe}, A.~M. 2012, \mnras, 422, 1700,
  \dodoi{10.1111/j.1365-2966.2012.20746.x}

\bibitem[{{Cayrel} {et~al.}(2004){Cayrel}, {Depagne}, {Spite}, {Hill}, {Spite},
  {Fran{\c{c}}ois}, {Plez}, {Beers}, {Primas}, {Andersen}, {Barbuy},
  {Bonifacio}, {Molaro}, \& {Nordstr{\"o}m}}]{Cayrel2004}
{Cayrel}, R., {Depagne}, E., {Spite}, M., {et~al.} 2004, \aap, 416, 1117,
  \dodoi{10.1051/0004-6361:20034074}

\bibitem[{{Chiaki} {et~al.}(2017){Chiaki}, {Tominaga}, \&
  {Nozawa}}]{Chiaki2017}
{Chiaki}, G., {Tominaga}, N., \& {Nozawa}, T. 2017, \mnras, 472, L115,
  \dodoi{10.1093/mnrasl/slx163}

\bibitem[{{Christensen} {et~al.}(2017){Christensen}, {Vergani}, {Schulze},
  {Annau}, {Selsing}, {Fynbo}, {de Ugarte Postigo}, {Ca{\~n}ameras}, {Lopez},
  {Passi}, {Cort{\'e}s-Zuleta}, {Ellison}, {D'Odorico}, {Becker}, {Berg},
  {Cano}, {Covino}, {Cupani}, {D'Elia}, {Goldoni}, {Gomboc}, {Hammer},
  {Heintz}, {Jakobsson}, {Japelj}, {Kaper}, {Malesani}, {M{\o}ller},
  {Petitjean}, {Pugliese}, {S{\'a}nchez-Ram{\'\i}rez}, {Tanvir}, {Th{\"o}ne},
  {Vestergaard}, {Wiersema}, \& {Worseck}}]{Christensen2017}
{Christensen}, L., {Vergani}, S.~D., {Schulze}, S., {et~al.} 2017, \aap, 608,
  A84, \dodoi{10.1051/0004-6361/201731382}

\bibitem[{{Christlieb} {et~al.}(2002){Christlieb}, {Bessell}, {Beers},
  {Gustafsson}, {Korn}, {Barklem}, {Karlsson}, {Mizuno-Wiedner}, \&
  {Rossi}}]{Christlieb2002}
{Christlieb}, N., {Bessell}, M.~S., {Beers}, T.~C., {et~al.} 2002, \nat, 419,
  904, \dodoi{10.1038/nature01142}

\bibitem[{{Cooke} {et~al.}(2012){Cooke}, {Pettini}, \& {Murphy}}]{Cooke2012}
{Cooke}, R., {Pettini}, M., \& {Murphy}, M.~T. 2012, \mnras, 425, 347,
  \dodoi{10.1111/j.1365-2966.2012.21470.x}

\bibitem[{{Cooke} {et~al.}(2011{\natexlab{a}}){Cooke}, {Pettini}, {Steidel},
  {Rudie}, \& {Jorgenson}}]{Cooke2011a}
{Cooke}, R., {Pettini}, M., {Steidel}, C.~C., {Rudie}, G.~C., \& {Jorgenson},
  R.~A. 2011{\natexlab{a}}, \mnras, 412, 1047,
  \dodoi{10.1111/j.1365-2966.2010.17966.x}

\bibitem[{{Cooke} {et~al.}(2011{\natexlab{b}}){Cooke}, {Pettini}, {Steidel},
  {Rudie}, \& {Nissen}}]{Cooke2011b}
{Cooke}, R., {Pettini}, M., {Steidel}, C.~C., {Rudie}, G.~C., \& {Nissen},
  P.~E. 2011{\natexlab{b}}, \mnras, 417, 1534,
  \dodoi{10.1111/j.1365-2966.2011.19365.x}

\bibitem[{{Crighton} {et~al.}(2016){Crighton}, {O'Meara}, \&
  {Murphy}}]{Crighton2016}
{Crighton}, N. H.~M., {O'Meara}, J.~M., \& {Murphy}, M.~T. 2016, \mnras, 457,
  L44, \dodoi{10.1093/mnrasl/slv191}

\bibitem[{Cupani {et~al.}(2020)Cupani, D'Odorico, Cristiani, Russo, Calderone,
  \& Taffoni}]{Cupani2020}
Cupani, G., D'Odorico, V., Cristiani, S., {et~al.} 2020, in Software and
  Cyberinfrastructure for Astronomy VI, ed. J.~C. Guzman \& J.~Ibsen, Vol.
  11452, International Society for Optics and Photonics (SPIE), 372 -- 388,
  \dodoi{10.1117/12.2561343}

\bibitem[{{de Bennassuti} {et~al.}(2017){de Bennassuti}, {Salvadori},
  {Schneider}, {Valiante}, \& {Omukai}}]{deBennassuti2017}
{de Bennassuti}, M., {Salvadori}, S., {Schneider}, R., {Valiante}, R., \&
  {Omukai}, K. 2017, \mnras, 465, 926, \dodoi{10.1093/mnras/stw2687}

\bibitem[{{De Cia} {et~al.}(2018){De Cia}, {Ledoux}, {Petitjean}, \&
  {Savaglio}}]{Decia2018}
{De Cia}, A., {Ledoux}, C., {Petitjean}, P., \& {Savaglio}, S. 2018, \aap, 611,
  A76, \dodoi{10.1051/0004-6361/201731970}

\bibitem[{{de Jong} {et~al.}(2019){de Jong}, {Agertz}, {Berbel}, {Aird},
  {Alexander}, {Amarsi}, {Anders}, {Andrae}, {Ansarinejad}, {Ansorge},
  {Antilogus}, {Anwand-Heerwart}, {Arentsen}, {Arnadottir}, {Asplund}, {Auger},
  {Azais}, {Baade}, {Baker}, {Baker}, {Balbinot}, {Baldry}, {Banerji},
  {Barden}, {Barklem}, {Barth{\'e}l{\'e}my-Mazot}, {Battistini}, {Bauer},
  {Bell}, {Bellido-Tirado}, {Bellstedt}, {Belokurov}, {Bensby}, {Bergemann},
  {Bestenlehner}, {Bielby}, {Bilicki}, {Blake}, {Bland-Hawthorn}, {Boeche},
  {Boland}, {Boller}, {Bongard}, {Bongiorno}, {Bonifacio}, {Boudon}, {Brooks},
  {Brown}, {Brown}, {Br{\"u}ggen}, {Brynnel}, {Brzeski}, {Buchert},
  {Buschkamp}, {Caffau}, {Caillier}, {Carrick}, {Casagrande}, {Case}, {Casey},
  {Cesarini}, {Cescutti}, {Chapuis}, {Chiappini}, {Childress}, {Christlieb},
  {Church}, {Cioni}, {Cluver}, {Colless}, {Collett}, {Comparat}, {Cooper},
  {Couch}, {Courbin}, {Croom}, {Croton}, {Daguis{\'e}}, {Dalton}, {Davies},
  {Davis}, {de Laverny}, {Deason}, {Dionies}, {Disseau}, {Doel}, {D{\"o}scher},
  {Driver}, {Dwelly}, {Eckert}, {Edge}, {Edvardsson}, {Youssoufi}, {Elhaddad},
  {Enke}, {Erfanianfar}, {Farrell}, {Fechner}, {Feiz}, {Feltzing}, {Ferreras},
  {Feuerstein}, {Feuillet}, {Finoguenov}, {Ford}, {Fotopoulou}, {Fouesneau},
  {Frenk}, {Frey}, {Gaessler}, {Geier}, {Gentile Fusillo}, {Gerhard},
  {Giannantonio}, {Giannone}, {Gibson}, {Gillingham},
  {Gonz{\'a}lez-Fern{\'a}ndez}, {Gonzalez-Solares}, {Gottloeber}, {Gould},
  {Grebel}, {Gueguen}, {Guiglion}, {Haehnelt}, {Hahn}, {Hansen}, {Hartman},
  {Hauptner}, {Hawkins}, {Haynes}, {Haynes}, {Heiter}, {Helmi}, {Aguayo},
  {Hewett}, {Hinton}, {Hobbs}, {Hoenig}, {Hofman}, {Hook}, {Hopgood},
  {Hopkins}, {Hourihane}, {Howes}, {Howlett}, {Huet}, {Irwin}, {Iwert},
  {Jablonka}, {Jahn}, {Jahnke}, {Jarno}, {Jin}, {Jofre}, {Johl}, {Jones},
  {J{\"o}nsson}, {Jordan}, {Karovicova}, {Khalatyan}, {Kelz}, {Kennicutt},
  {King}, {Kitaura}, {Klar}, {Klauser}, {Kneib}, {Koch}, {Koposov},
  {Kordopatis}, {Korn}, {Kosmalski}, {Kotak}, {Kovalev}, {Kreckel}, {Kripak},
  {Krumpe}, {Kuijken}, {Kunder}, {Kushniruk}, {Lam}, {Lamer}, {Laurent},
  {Lawrence}, {Lehmitz}, {Lemasle}, {Lewis}, {Li}, {Lidman}, {Lind}, {Liske},
  {Lizon}, {Loveday}, {Ludwig}, {McDermid}, {Maguire}, {Mainieri}, {Mali},
  {Mandel}, {Mandel}, {Mannering}, {Martell}, {Martinez Delgado}, {Matijevic},
  {McGregor}, {McMahon}, {McMillan}, {Mena}, {Merloni}, {Meyer}, {Michel},
  {Micheva}, {Migniau}, {Minchev}, {Monari}, {Muller}, {Murphy},
  {Muthukrishna}, {Nandra}, {Navarro}, {Ness}, {Nichani}, {Nichol}, {Nicklas},
  {Niederhofer}, {Norberg}, {Obreschkow}, {Oliver}, {Owers}, {Pai},
  {Pankratow}, {Parkinson}, {Paschke}, {Paterson}, {Pecontal}, {Parry},
  {Phillips}, {Pillepich}, {Pinard}, {Pirard}, {Piskunov}, {Plank},
  {Pl{\"u}schke}, {Pons}, {Popesso}, {Power}, {Pragt}, {Pramskiy}, {Pryer},
  {Quattri}, {Queiroz}, {Quirrenbach}, {Rahurkar}, {Raichoor}, {Ramstedt},
  {Rau}, {Recio-Blanco}, {Reiss}, {Renaud}, {Revaz}, {Rhode}, {Richard},
  {Richter}, {Rix}, {Robotham}, {Roelfsema}, {Romaniello}, {Rosario},
  {Rothmaier}, {Roukema}, {Ruchti}, {Rupprecht}, {Rybizki}, {Ryde}, {Saar},
  {Sadler}, {Sahl{\'e}n}, {Salvato}, {Sassolas}, {Saunders}, {Saviauk},
  {Sbordone}, {Schmidt}, {Schnurr}, {Scholz}, {Schwope}, {Seifert}, {Shanks},
  {Sheinis}, {Sivov}, {Sk{\'u}lad{\'o}ttir}, {Smartt}, {Smedley}, {Smith},
  {Smith}, {Sorce}, {Spitler}, {Starkenburg}, {Steinmetz}, {Stilz}, {Storm},
  {Sullivan}, {Sutherland}, {Swann}, {Tamone}, {Taylor}, {Teillon}, {Tempel},
  {ter Horst}, {Thi}, {Tolstoy}, {Trager}, {Traven}, {Tremblay}, {Tresse},
  {Valentini}, {van de Weygaert}, {van den Ancker}, {Veljanoski}, {Venkatesan},
  {Wagner}, {Wagner}, {Walcher}, {Waller}, {Walton}, {Wang}, {Winkler},
  {Wisotzki}, {Worley}, {Worseck}, {Xiang}, {Xu}, {Yong}, {Zhao}, {Zheng},
  {Zscheyge}, \& {Zucker}}]{deJong2019}
{de Jong}, R.~S., {Agertz}, O., {Berbel}, A.~A., {et~al.} 2019, The Messenger,
  175, 3, \dodoi{10.18727/0722-6691/5117}

\bibitem[{{D'Odorico} {et~al.}(2016){D'Odorico}, {Cristiani}, {Pomante},
  {Carswell}, {Viel}, {Barai}, {Becker}, {Calura}, {Cupani}, {Fontanot},
  {Haehnelt}, {Kim}, {Miralda-Escud{\'e}}, {Rorai}, {Tescari}, \&
  {Vanzella}}]{Dodorico2016}
{D'Odorico}, V., {Cristiani}, S., {Pomante}, E., {et~al.} 2016, \mnras, 463,
  2690, \dodoi{10.1093/mnras/stw2161}

\bibitem[{{D'Odorico} {et~al.}(2022){D'Odorico}, {Finlator}, {Cristiani},
  {Cupani}, {Perrotta}, {Calura}, {C{\`e}nturion}, {Becker}, {Berg}, {Lopez},
  {Ellison}, \& {Pomante}}]{Dodorico2022}
{D'Odorico}, V., {Finlator}, K., {Cristiani}, S., {et~al.} 2022, \mnras, 512,
  2389, \dodoi{10.1093/mnras/stac545}

\bibitem[{{Dutta} {et~al.}(2014){Dutta}, {Srianand}, {Rahmani}, {Petitjean},
  {Noterdaeme}, \& {Ledoux}}]{Dutta2014}
{Dutta}, R., {Srianand}, R., {Rahmani}, H., {et~al.} 2014, \mnras, 440, 307,
  \dodoi{10.1093/mnras/stu260}

\bibitem[{{Ferland} {et~al.}(2017){Ferland}, {Chatzikos}, {Guzm{\'a}n},
  {Lykins}, {van Hoof}, {Williams}, {Abel}, {Badnell}, {Keenan}, {Porter}, \&
  {Stancil}}]{Ferland2017}
{Ferland}, G.~J., {Chatzikos}, M., {Guzm{\'a}n}, F., {et~al.} 2017, \rmxaa, 53,
  385.
\newblock \doarXiv{1705.10877}

\bibitem[{{Fran{\c{c}}ois} {et~al.}(2018){Fran{\c{c}}ois}, {Caffau}, {Wanajo},
  {Aguado}, {Spite}, {Aoki}, {Aoki}, {Bonifacio}, {Gallagher}, {Salvadori}, \&
  {Spite}}]{Francois2018}
{Fran{\c{c}}ois}, P., {Caffau}, E., {Wanajo}, S., {et~al.} 2018, \aap, 619,
  A10, \dodoi{10.1051/0004-6361/201833824}

\bibitem[{{Frebel} \& {Norris}(2015)}]{Frebel2015}
{Frebel}, A., \& {Norris}, J.~E. 2015, \araa, 53, 631,
  \dodoi{10.1146/annurev-astro-082214-122423}

\bibitem[{{Frebel} {et~al.}(2005){Frebel}, {Aoki}, {Christlieb}, {Ando},
  {Asplund}, {Barklem}, {Beers}, {Eriksson}, {Fechner}, {Fujimoto}, {Honda},
  {Kajino}, {Minezaki}, {Nomoto}, {Norris}, {Ryan}, {Takada-Hidai},
  {Tsangarides}, \& {Yoshii}}]{Frebel2005}
{Frebel}, A., {Aoki}, W., {Christlieb}, N., {et~al.} 2005, Nature, 434, 871,
  \dodoi{10.1038/nature03455}

\bibitem[{{Fumagalli} {et~al.}(2011){Fumagalli}, {O'Meara}, \&
  {Prochaska}}]{Fumagalli2011}
{Fumagalli}, M., {O'Meara}, J.~M., \& {Prochaska}, J.~X. 2011, Science, 334,
  1245, \dodoi{10.1126/science.1213581}

\bibitem[{{Fumagalli} {et~al.}(2016){Fumagalli}, {O'Meara}, \&
  {Prochaska}}]{Fumagalli2016}
---. 2016, \mnras, 455, 4100, \dodoi{10.1093/mnras/stv2616}

\bibitem[{{Gonz{\'a}lez Hern{\'a}ndez} {et~al.}(2020){Gonz{\'a}lez
  Hern{\'a}ndez}, {Aguado}, {Allende Prieto}, {Burgasser}, \&
  {Rebolo}}]{Gonzalez2020}
{Gonz{\'a}lez Hern{\'a}ndez}, J.~I., {Aguado}, D.~S., {Allende Prieto}, C.,
  {Burgasser}, A.~J., \& {Rebolo}, R. 2020, \apjl, 889, L13,
  \dodoi{10.3847/2041-8213/ab62ae}

\bibitem[{{Haardt} \& {Madau}(2012)}]{Haardt2012}
{Haardt}, F., \& {Madau}, P. 2012, \apj, 746, 125,
  \dodoi{10.1088/0004-637X/746/2/125}

\bibitem[{{Hansen} {et~al.}(2016{\natexlab{a}}){Hansen}, {Andersen},
  {Nordstr{\"o}m}, {Beers}, {Placco}, {Yoon}, \& {Buchhave}}]{Hansen2016_CEMPs}
{Hansen}, T.~T., {Andersen}, J., {Nordstr{\"o}m}, B., {et~al.}
  2016{\natexlab{a}}, \aap, 588, A3, \dodoi{10.1051/0004-6361/201527409}

\bibitem[{{Hansen} {et~al.}(2016{\natexlab{b}}){Hansen}, {Andersen},
  {Nordstr{\"o}m}, {Beers}, {Placco}, {Yoon}, \&
  {Buchhave}}]{Hansen2016_CEMPno}
---. 2016{\natexlab{b}}, \aap, 586, A160, \dodoi{10.1051/0004-6361/201527235}

\bibitem[{{Hartwig} {et~al.}(2018){Hartwig}, {Yoshida}, {Magg}, {Frebel},
  {Glover}, {G{\'o}mez}, {Griffen}, {Ishigaki}, {Ji}, {Klessen}, {O'Shea}, \&
  {Tominaga}}]{Hartwig2018}
{Hartwig}, T., {Yoshida}, N., {Magg}, M., {et~al.} 2018, \mnras, 478, 1795,
  \dodoi{10.1093/mnras/sty1176}

\bibitem[{Heger \& Woosley(2002)}]{Heger2002}
Heger, A., \& Woosley, S.~E. 2002, The Astrophysical Journal, 567, 532

\bibitem[{{Heger} \& {Woosley}(2010)}]{Heger2010}
{Heger}, A., \& {Woosley}, S.~E. 2010, \apj, 724, 341,
  \dodoi{10.1088/0004-637X/724/1/341}

\bibitem[{{Hirano} {et~al.}(2014){Hirano}, {Hosokawa}, {Yoshida}, {Umeda},
  {Omukai}, {Chiaki}, \& {Yorke}}]{Hirano2014}
{Hirano}, S., {Hosokawa}, T., {Yoshida}, N., {et~al.} 2014, \apj, 781, 60,
  \dodoi{10.1088/0004-637X/781/2/60}

\bibitem[{{Hosokawa} {et~al.}(2011){Hosokawa}, {Omukai}, {Yoshida}, \&
  {Yorke}}]{Hosokawa2011}
{Hosokawa}, T., {Omukai}, K., {Yoshida}, N., \& {Yorke}, H.~W. 2011, Science,
  334, 1250, \dodoi{10.1126/science.1207433}

\bibitem[{{Ir{\v{s}}i{\v{c}}} {et~al.}(2017{\natexlab{a}}){Ir{\v{s}}i{\v{c}}},
  {Viel}, {Haehnelt}, {Bolton}, {Cristiani}, {Becker}, {D'Odorico}, {Cupani},
  {Kim}, {Berg}, {L{\'o}pez}, {Ellison}, {Christensen}, {Denney}, \&
  {Worseck}}]{Irsic2017}
{Ir{\v{s}}i{\v{c}}}, V., {Viel}, M., {Haehnelt}, M.~G., {et~al.}
  2017{\natexlab{a}}, \prd, 96, 023522, \dodoi{10.1103/PhysRevD.96.023522}

\bibitem[{{Ir{\v{s}}i{\v{c}}} {et~al.}(2017{\natexlab{b}}){Ir{\v{s}}i{\v{c}}},
  {Viel}, {Berg}, {D'Odorico}, {Haehnelt}, {Cristiani}, {Cupani}, {Kim},
  {L{\'o}pez}, {Ellison}, {Becker}, {Christensen}, {Denney}, {Worseck}, \&
  {Bolton}}]{Irsic2017b}
{Ir{\v{s}}i{\v{c}}}, V., {Viel}, M., {Berg}, T. A.~M., {et~al.}
  2017{\natexlab{b}}, \mnras, 466, 4332, \dodoi{10.1093/mnras/stw3372}

\bibitem[{{Iwamoto} {et~al.}(2005){Iwamoto}, {Umeda}, {Tominaga}, {Nomoto}, \&
  {Maeda}}]{Iwamoto2005}
{Iwamoto}, N., {Umeda}, H., {Tominaga}, N., {Nomoto}, K., \& {Maeda}, K. 2005,
  Science, 309, 451, \dodoi{10.1126/science.1112997}

\bibitem[{{Jaacks} {et~al.}(2019){Jaacks}, {Finkelstein}, \&
  {Bromm}}]{Jaacks2019}
{Jaacks}, J., {Finkelstein}, S.~L., \& {Bromm}, V. 2019, mnras, 488, 2202,
  \dodoi{10.1093/mnras/stz1529}

\bibitem[{{Jeon} {et~al.}(2021){Jeon}, {Bromm}, {Besla}, {Yoon}, \&
  {Choi}}]{Jeon2021}
{Jeon}, M., {Bromm}, V., {Besla}, G., {Yoon}, J., \& {Choi}, Y. 2021, mnras,
  502, 1, \dodoi{10.1093/mnras/staa4017}

\bibitem[{{Ji} {et~al.}(2016){Ji}, {Frebel}, {Ezzeddine}, \& {Casey}}]{Ji2016}
{Ji}, A.~P., {Frebel}, A., {Ezzeddine}, R., \& {Casey}, A.~R. 2016, \apjl, 832,
  L3, \dodoi{10.3847/2041-8205/832/1/L3}

\bibitem[{{Karakas} \& {Lattanzio}(2014)}]{Karakas2014}
{Karakas}, A.~I., \& {Lattanzio}, J.~C. 2014, pasa, 31, e030,
  \dodoi{10.1017/pasa.2014.21}

\bibitem[{{Keller} {et~al.}(2014){Keller}, {Bessell}, {Frebel}, {Casey},
  {Asplund}, {Jacobson}, {Lind}, {Norris}, {Yong}, {Heger}, {Magic}, {da
  Costa}, {Schmidt}, \& {Tisserand}}]{Keller2014}
{Keller}, S.~C., {Bessell}, M.~S., {Frebel}, A., {et~al.} 2014, \nat, 506, 463,
  \dodoi{10.1038/nature12990}

\bibitem[{{Lee} {et~al.}(2013){Lee}, {Beers}, {Masseron}, {Plez}, {Rockosi},
  {Sobeck}, {Yanny}, {Lucatello}, {Sivarani}, {Placco}, \& {Carollo}}]{Lee2013}
{Lee}, Y.~S., {Beers}, T.~C., {Masseron}, T., {et~al.} 2013, \aj, 146, 132,
  \dodoi{10.1088/0004-6256/146/5/132}

\bibitem[{{Limongi} \& {Chieffi}(2018)}]{Limongi2018}
{Limongi}, M., \& {Chieffi}, A. 2018, \apjs, 237, 13,
  \dodoi{10.3847/1538-4365/aacb24}

\bibitem[{{Liu} {et~al.}(2021){Liu}, {Sibony}, {Meynet}, \& {Bromm}}]{Liu2021}
{Liu}, B., {Sibony}, Y., {Meynet}, G., \& {Bromm}, V. 2021, mnras, 506, 5247,
  \dodoi{10.1093/mnras/stab2057}

\bibitem[{{Lofthouse} {et~al.}(2022){Lofthouse}, {Fumagalli}, {Fossati},
  {Dutta}, {Galbiati}, {Arrigoni Battaia}, {Cantalupo}, {Christensen}, {Cooke},
  {Longobardi}, {Murphy}, \& {Xavier. Prochaska}}]{Lofthouse2022}
{Lofthouse}, E.~K., {Fumagalli}, M., {Fossati}, M., {et~al.} 2022, arXiv
  e-prints, arXiv:2209.15021.
\newblock \doarXiv{2209.15021}

\bibitem[{{L{\'o}pez} {et~al.}(2016){L{\'o}pez}, {D'Odorico}, {Ellison},
  {Becker}, {Christensen}, {Cupani}, {Denney}, {P{\^a}ris}, {Worseck}, {Berg},
  {Cristiani}, {Dessauges-Zavadsky}, {Haehnelt}, {Hamann}, {Hennawi},
  {Ir{\v{s}}i{\v{c}}}, {Kim}, {L{\'o}pez}, {Lund Saust}, {M{\'e}nard},
  {Perrotta}, {Prochaska}, {S{\'a}nchez-Ram{\'\i}rez}, {Vestergaard}, {Viel},
  \& {Wisotzki}}]{Lopez2016}
{L{\'o}pez}, S., {D'Odorico}, V., {Ellison}, S.~L., {et~al.} 2016, \aap, 594,
  A91, \dodoi{10.1051/0004-6361/201628161}

\bibitem[{{Lucatello} {et~al.}(2006){Lucatello}, {Beers}, {Christlieb},
  {Barklem}, {Rossi}, {Marsteller}, {Sivarani}, \& {Lee}}]{Lucatello2006}
{Lucatello}, S., {Beers}, T.~C., {Christlieb}, N., {et~al.} 2006, \apjl, 652,
  L37, \dodoi{10.1086/509780}

\bibitem[{{Lucatello} {et~al.}(2005){Lucatello}, {Tsangarides}, {Beers},
  {Carretta}, {Gratton}, \& {Ryan}}]{Lucatello2005}
{Lucatello}, S., {Tsangarides}, S., {Beers}, T.~C., {et~al.} 2005, \apj, 625,
  825, \dodoi{10.1086/428104}

\bibitem[{{Marassi} {et~al.}(2014){Marassi}, {Chiaki}, {Schneider}, {Limongi},
  {Omukai}, {Nozawa}, {Chieffi}, \& {Yoshida}}]{Marassi2014}
{Marassi}, S., {Chiaki}, G., {Schneider}, R., {et~al.} 2014, apj, 794, 100,
  \dodoi{10.1088/0004-637X/794/2/100}

\bibitem[{{Marconi} {et~al.}(2022){Marconi}, {Abreu}, {Adibekyan}, {Alberti},
  {Albrecht}, {Alcaniz}, {Aliverti}, {Allende Prieto}, {Alvarado G{\'o}mez},
  {Amado}, {Amate}, {Andersen}, {Artigau}, {Baker}, {Baldini}, {Balestra},
  {Barnes}, {Baron}, {Barros}, {Bauer}, {Beaulieu}, {Bellido-Tirado},
  {Benneke}, {Bensby}, {Bergin}, {Biazzo}, {Bik}, {Birkby}, {Blind}, {Boisse},
  {Bolmont}, {Bonaglia}, {Bonfils}, {Borsa}, {Brandeker}, {Brandner}, {Broeg},
  {Brogi}, {Brousseau}, {Brucalassi}, {Brynnel}, {Buchhave}, {Buscher},
  {Cabral}, {Calderone}, {Calvo-Ortega}, {Canto Martins}, {Cantalloube},
  {Carbonaro}, {Chauvin}, {Chazelas}, {Cheffot}, {Cheng}, {Chiavassa},
  {Christensen}, {Cirami}, {Cook}, {Cooke}, {Coretti}, {Covino}, {Cowan},
  {Cresci}, {Cristiani}, {Cunha Parro}, {Cupani}, {D'Odorico}, {de Castro
  Le{\~a}o}, {De Cia}, {De Medeiros}, {Debras}, {Debus}, {Demangeon},
  {Dessauges-Zavadsky}, {Di Marcantonio}, {Dionies}, {Doyon}, {Dunn},
  {Ehrenreich}, {Faria}, {Feruglio}, {Fisher}, {Fontana}, {Fumagalli}, {Fusco},
  {Fynbo}, {Gabella}, {Gaessler}, {Gallo}, {Gao}, {Genolet}, {Genoni},
  {Giacobbe}, {Giro}, {Gon{\c{c}}alves}, {Gonzalez}, {Gonz{\'a}lez
  Hern{\'a}ndez}, {Gracia T{\'e}mich}, {Haehnelt}, {Haniff}, {Hatzes},
  {Helled}, {Hoeijmakers}, {Huke}, {J{\"a}rvinen}, {J{\"a}rvinen}, {Kaminski},
  {Korn}, {Kouach}, {Kowzan}, {Kreidberg}, {Landoni}, {Lanotte}, {Lavail},
  {Li}, {Liske}, {Lovis}, {Lucatello}, {Lunney}, {MacIntosh}, {Madhusudhan},
  {Magrini}, {Maiolino}, {Malo}, {Man}, {Marquart}, {Marques}, {Martins},
  {Martins}, {Maslowski}, {Mason}, {Mason}, {McCracken}, {Mergo}, {Micela},
  {Mitchell}, {Molli{\`e}re}, {Monteiro}, {Montgomery}, {Mordasini}, {Morin},
  {Mucciarelli}, {Murphy}, {N'Diaye}, {Neichel}, {Niedzielski}, {Niemczura},
  {Nortmann}, {Noterdaeme}, {Nunes}, {Oggioni}, {Oliva}, {{\"O}nel}, {Origlia},
  {{\"O}stlin}, {Palle}, {Papaderos}, {Pariani}, {Pe{\~n}ate Castro}, {Pepe},
  {Perreault Levasseur}, {Petit}, {Pino}, {Piqueras}, {Pollo}, {Poppenhaeger},
  {Quirrenbach}, {Rauscher}, {Rebolo}, {Redaelli}, {Reffert}, {Reid},
  {Reiners}, {Richter}, {Riva}, {Rivoire}, {Rodr{\'\i}guez-L{\'o}pez},
  {Roederer}, {Romano}, {Rousseau}, {Rowe}, {Salvadori}, {Santos}, {Santos
  Diaz}, {Sanz-Forcada}, {Sarajlic}, {Sauvage}, {Sch{\"a}fer}, {Schiavon},
  {Schmidt}, {Selmi}, {Sivanandam}, {Sordet}, {Sordo}, {Sortino}, {Sosnowska},
  {Sousa}, {Stempels}, {Strassmeier}, {Su{\'a}rez Mascare{\~n}o}, {Sulich},
  {Sun}, {Tanvir}, {Tenegi-Sangin{\'e}s}, {Thibault}, {Thompson}, {Tozzi},
  {Turbet}, {Vall{\'e}e}, {Varas}, {Venn}, {V{\'e}ran}, {Verma}, {Viel},
  {Wade}, {Waring}, {Weber}, {Weder}, {Wehbe}, {Weingrill}, {Woche}, {Xompero},
  {Zackrisson}, {Zanutta}, {Zapatero Osorio}, {Zechmeister}, \&
  {Zimara}}]{Marconi2022}
{Marconi}, A., {Abreu}, M., {Adibekyan}, V., {et~al.} 2022, in Society of
  Photo-Optical Instrumentation Engineers (SPIE) Conference Series, Vol. 12184,
  Ground-based and Airborne Instrumentation for Astronomy IX, ed. C.~J.
  {Evans}, J.~J. {Bryant}, \& K.~{Motohara}, 1218424,
  \dodoi{10.1117/12.2628689}

\bibitem[{{Marsteller} {et~al.}(2005){Marsteller}, {Beers}, {Rossi},
  {Christlieb}, {Bessell}, \& {Rhee}}]{Marsteller2005}
{Marsteller}, B., {Beers}, T.~C., {Rossi}, S., {et~al.} 2005, \nphysa, 758,
  312, \dodoi{10.1016/j.nuclphysa.2005.05.056}

\bibitem[{{Murphy} {et~al.}(2019){Murphy}, {Kacprzak}, {Savorgnan}, \&
  {Carswell}}]{Murphy2019}
{Murphy}, M.~T., {Kacprzak}, G.~G., {Savorgnan}, G. A.~D., \& {Carswell}, R.~F.
  2019, \mnras, 482, 3458, \dodoi{10.1093/mnras/sty2834}

\bibitem[{{Norris} {et~al.}(1997){Norris}, {Ryan}, \& {Beers}}]{Norris1997}
{Norris}, J.~E., {Ryan}, S.~G., \& {Beers}, T.~C. 1997, \apjl, 489, L169,
  \dodoi{10.1086/316787}

\bibitem[{{Norris} {et~al.}(2013){Norris}, {Yong}, {Bessell}, {Christlieb},
  {Asplund}, {Gilmore}, {Wyse}, {Beers}, {Barklem}, {Frebel}, \&
  {Ryan}}]{Norris2013}
{Norris}, J.~E., {Yong}, D., {Bessell}, M.~S., {et~al.} 2013, \apj, 762, 28,
  \dodoi{10.1088/0004-637X/762/1/28}

\bibitem[{{Pallottini} {et~al.}(2014){Pallottini}, {Ferrara}, {Gallerani},
  {Salvadori}, \& {D'Odorico}}]{Pallottini2014}
{Pallottini}, A., {Ferrara}, A., {Gallerani}, S., {Salvadori}, S., \&
  {D'Odorico}, V. 2014, \mnras, 440, 2498, \dodoi{10.1093/mnras/stu451}

\bibitem[{{Perrotta} {et~al.}(2016){Perrotta}, {D'Odorico}, {Prochaska},
  {Cristiani}, {Cupani}, {Ellison}, {L{\'o}pez}, {Becker}, {Berg},
  {Christensen}, {Denney}, {Hamann}, {P{\^a}ris}, {Vestergaard}, \&
  {Worseck}}]{Perrotta2016}
{Perrotta}, S., {D'Odorico}, V., {Prochaska}, J.~X., {et~al.} 2016, \mnras,
  462, 3285, \dodoi{10.1093/mnras/stw1703}

\bibitem[{{Perrotta} {et~al.}(2018){Perrotta}, {D'Odorico}, {Hamann},
  {Cristiani}, {Prochaska}, {Ellison}, {Lopez}, {Cupani}, {Becker}, {Berg},
  {Christensen}, {Denney}, \& {Worseck}}]{Perrotta2018}
{Perrotta}, S., {D'Odorico}, V., {Hamann}, F., {et~al.} 2018, \mnras, 481, 105,
  \dodoi{10.1093/mnras/sty2205}

\bibitem[{{Pieri} {et~al.}(2016){Pieri}, {Bonoli}, {Chaves-Montero},
  {P{\^a}ris}, {Fumagalli}, {Bolton}, {Viel}, {Noterdaeme},
  {Miralda-Escud{\'e}}, {Busca}, {Rahmani}, {Peroux}, {Font-Ribera}, \&
  {Trager}}]{Pieri2016}
{Pieri}, M.~M., {Bonoli}, S., {Chaves-Montero}, J., {et~al.} 2016, in
  SF2A-2016: Proceedings of the Annual meeting of the French Society of
  Astronomy and Astrophysics, ed. C.~{Reyl{\'e}}, J.~{Richard},
  L.~{Cambr{\'e}sy}, M.~{Deleuil}, E.~{P{\'e}contal}, L.~{Tresse}, \&
  I.~{Vauglin}, 259--266.
\newblock \doarXiv{1611.09388}

\bibitem[{{Placco} {et~al.}(2014){Placco}, {Frebel}, {Beers}, \&
  {Stancliffe}}]{Placco2014}
{Placco}, V.~M., {Frebel}, A., {Beers}, T.~C., \& {Stancliffe}, R.~J. 2014,
  \apj, 797, 21, \dodoi{10.1088/0004-637X/797/1/21}

\bibitem[{{Placco} {et~al.}(2019){Placco}, {Santucci}, {Beers}, {Chanam{\'e}},
  {Sep{\'u}lveda}, {Coronado}, {Rossi}, {Lee}, {Starkenburg}, {Youakim},
  {Barrientos}, {Ezzeddine}, {Frebel}, {Hansen}, {Holmbeck}, {Ji}, {Rasmussen},
  {Roederer}, {Sakari}, \& {Whitten}}]{Placco2019}
{Placco}, V.~M., {Santucci}, R.~M., {Beers}, T.~C., {et~al.} 2019, \apj, 870,
  122, \dodoi{10.3847/1538-4357/aaf3b9}

\bibitem[{{Placco} {et~al.}(2021){Placco}, {Roederer}, {Lee},
  {Almeida-Fernandes}, {Herpich}, {Perottoni}, {Schoenell}, {Ribeiro}, \&
  {Kanaan}}]{Placco21}
{Placco}, V.~M., {Roederer}, I.~U., {Lee}, Y.~S., {et~al.} 2021, \apjl, 912,
  L32, \dodoi{10.3847/2041-8213/abf93d}

\bibitem[{{Quiret} {et~al.}(2016){Quiret}, {P{\'e}roux}, {Zafar}, {Kulkarni},
  {Jenkins}, {Milliard}, {Rahmani}, {Popping}, {Rao}, {Turnshek}, \&
  {Monier}}]{Quiret2016}
{Quiret}, S., {P{\'e}roux}, C., {Zafar}, T., {et~al.} 2016, \mnras, 458, 4074,
  \dodoi{10.1093/mnras/stw524}

\bibitem[{{Robert} {et~al.}(2019){Robert}, {Murphy}, {O'Meara}, {Crighton}, \&
  {Fumagalli}}]{Robert2019}
{Robert}, P.~F., {Murphy}, M.~T., {O'Meara}, J.~M., {Crighton}, N. H.~M., \&
  {Fumagalli}, M. 2019, \mnras, 483, 2736, \dodoi{10.1093/mnras/sty3287}

\bibitem[{{Rossi} {et~al.}(2005){Rossi}, {Beers}, {Sneden}, {Sevastyanenko},
  {Rhee}, \& {Marsteller}}]{Rossi2005}
{Rossi}, S., {Beers}, T.~C., {Sneden}, C., {et~al.} 2005, \aj, 130, 2804,
  \dodoi{10.1086/497164}

\bibitem[{{Salvadori} {et~al.}(2019){Salvadori}, {Bonifacio}, {Caffau},
  {Korotin}, {Andreevsky}, {Spite}, \& {Sk{\'u}lad{\'o}ttir}}]{Salvadori2019}
{Salvadori}, S., {Bonifacio}, P., {Caffau}, E., {et~al.} 2019, \mnras, 487,
  4261, \dodoi{10.1093/mnras/stz1464}

\bibitem[{{Salvadori} \& {Ferrara}(2012)}]{Salvadori2012}
{Salvadori}, S., \& {Ferrara}, A. 2012, \mnras, 421, L29,
  \dodoi{10.1111/j.1745-3933.2011.01200.x}

\bibitem[{{Salvadori} {et~al.}(2015){Salvadori}, {Sk{\'u}lad{\'o}ttir}, \&
  {Tolstoy}}]{Salvadori2015}
{Salvadori}, S., {Sk{\'u}lad{\'o}ttir}, {\'A}., \& {Tolstoy}, E. 2015, \mnras,
  454, 1320, \dodoi{10.1093/mnras/stv1969}

\bibitem[{{S{\'a}nchez-Ram{\'\i}rez} {et~al.}(2016){S{\'a}nchez-Ram{\'\i}rez},
  {Ellison}, {Prochaska}, {Berg}, {L{\'o}pez}, {D'Odorico}, {Becker},
  {Christensen}, {Cupani}, {Denney}, {P{\^a}ris}, {Worseck}, \&
  {Gorosabel}}]{Sanchez2016}
{S{\'a}nchez-Ram{\'\i}rez}, R., {Ellison}, S.~L., {Prochaska}, J.~X., {et~al.}
  2016, \mnras, 456, 4488, \dodoi{10.1093/mnras/stv2732}

\bibitem[{{Schneider} {et~al.}(2003){Schneider}, {Ferrara}, {Salvaterra},
  {Omukai}, \& {Bromm}}]{Schneider2003}
{Schneider}, R., {Ferrara}, A., {Salvaterra}, R., {Omukai}, K., \& {Bromm}, V.
  2003, \nat, 422, 869, \dodoi{10.1038/nature01579}

\bibitem[{{Simon}(2019)}]{Simon2019}
{Simon}, J.~D. 2019, \araa, 57, 375,
  \dodoi{10.1146/annurev-astro-091918-104453}

\bibitem[{{Sk{\'u}lad{\'o}ttir} {et~al.}(2018){Sk{\'u}lad{\'o}ttir},
  {Salvadori}, {Pettini}, {Tolstoy}, \& {Hill}}]{Skuladottir2018}
{Sk{\'u}lad{\'o}ttir}, {\'A}., {Salvadori}, S., {Pettini}, M., {Tolstoy}, E.,
  \& {Hill}, V. 2018, \aap, 615, A137, \dodoi{10.1051/0004-6361/201732359}

\bibitem[{{Sk{\'u}lad{\'o}ttir} {et~al.}(2021){Sk{\'u}lad{\'o}ttir},
  {Salvadori}, {Amarsi}, {Tolstoy}, {Irwin}, {Hill}, {Jablonka}, {Battaglia},
  {Starkenburg}, {Massari}, {Helmi}, \& {Posti}}]{Skuladottir21}
{Sk{\'u}lad{\'o}ttir}, {\'A}., {Salvadori}, S., {Amarsi}, A.~M., {et~al.} 2021,
  \apjl, 915, L30, \dodoi{10.3847/2041-8213/ac0dc2}

\bibitem[{{Spite} {et~al.}(2013){Spite}, {Caffau}, {Bonifacio}, {Spite},
  {Ludwig}, {Plez}, \& {Christlieb}}]{Spite2013}
{Spite}, M., {Caffau}, E., {Bonifacio}, P., {et~al.} 2013, \aap, 552, A107,
  \dodoi{10.1051/0004-6361/201220989}

\bibitem[{{Spite} {et~al.}(2018){Spite}, {Spite}, {Fran{\c{c}}ois},
  {Bonifacio}, {Caffau}, \& {Salvadori}}]{Spite2018}
{Spite}, M., {Spite}, F., {Fran{\c{c}}ois}, P., {et~al.} 2018, \aap, 617, A56,
  \dodoi{10.1051/0004-6361/201833548}

\bibitem[{{Starkenburg} {et~al.}(2014){Starkenburg}, {Shetrone}, {McConnachie},
  \& {Venn}}]{Starkenburg2014}
{Starkenburg}, E., {Shetrone}, M.~D., {McConnachie}, A.~W., \& {Venn}, K.~A.
  2014, \mnras, 441, 1217, \dodoi{10.1093/mnras/stu623}

\bibitem[{{Starkenburg} {et~al.}(2018){Starkenburg}, {Aguado}, {Bonifacio},
  {Caffau}, {Jablonka}, {Lardo}, {Martin}, {S{\'a}nchez-Janssen}, {Sestito},
  {Venn}, {Youakim}, {Allende Prieto}, {Arentsen}, {Gentile}, {Gonz{\'a}lez
  Hern{\'a}ndez}, {Kielty}, {Koppelman}, {Longeard}, {Tolstoy}, {Carlberg},
  {C{\^o}t{\'e}}, {Fouesneau}, {Hill}, {McConnachie}, \&
  {Navarro}}]{Stankenburg2018}
{Starkenburg}, E., {Aguado}, D.~S., {Bonifacio}, P., {et~al.} 2018, \mnras,
  481, 3838, \dodoi{10.1093/mnras/sty2276}

\bibitem[{{Suda} {et~al.}(2008){Suda}, {Katsuta}, {Yamada}, {Suwa}, {Ishizuka},
  {Komiya}, {Sorai}, {Aikawa}, \& {Fujimoto}}]{Suda2008}
{Suda}, T., {Katsuta}, Y., {Yamada}, S., {et~al.} 2008, \pasj, 60, 1159,
  \dodoi{10.1093/pasj/60.5.1159}

\bibitem[{{Takahashi} {et~al.}(2018){Takahashi}, {Yoshida}, \&
  {Umeda}}]{Takahashi2018}
{Takahashi}, K., {Yoshida}, T., \& {Umeda}, H. 2018, apj, 857, 111,
  \dodoi{10.3847/1538-4357/aab95f}

\bibitem[{{Vernet} {et~al.}(2011){Vernet}, {Dekker}, {D'Odorico}, {Kaper},
  {Kjaergaard}, {Hammer}, {Randich}, {Zerbi}, {Groot}, {Hjorth}, {Guinouard},
  {Navarro}, {Adolfse}, {Albers}, {Amans}, {Andersen}, {Andersen}, {Binetruy},
  {Bristow}, {Castillo}, {Chemla}, {Christensen}, {Conconi}, {Conzelmann},
  {Dam}, {de Caprio}, {de Ugarte Postigo}, {Delabre}, {di Marcantonio},
  {Downing}, {Elswijk}, {Finger}, {Fischer}, {Flores}, {Fran{\c{c}}ois},
  {Goldoni}, {Guglielmi}, {Haigron}, {Hanenburg}, {Hendriks}, {Horrobin},
  {Horville}, {Jessen}, {Kerber}, {Kern}, {Kiekebusch}, {Kleszcz}, {Klougart},
  {Kragt}, {Larsen}, {Lizon}, {Lucuix}, {Mainieri}, {Manuputy}, {Martayan},
  {Mason}, {Mazzoleni}, {Michaelsen}, {Modigliani}, {Moehler}, {M{\o}ller},
  {Norup S{\o}rensen}, {N{\o}rregaard}, {P{\'e}roux}, {Patat}, {Pena}, {Pragt},
  {Reinero}, {Rigal}, {Riva}, {Roelfsema}, {Royer}, {Sacco}, {Santin},
  {Schoenmaker}, {Spano}, {Sweers}, {Ter Horst}, {Tintori}, {Tromp}, {van
  Dael}, {van der Vliet}, {Venema}, {Vidali}, {Vinther}, {Vola}, {Winters},
  {Wistisen}, {Wulterkens}, \& {Zacchei}}]{Vernet2011}
{Vernet}, J., {Dekker}, H., {D'Odorico}, S., {et~al.} 2011, \aap, 536, A105,
  \dodoi{10.1051/0004-6361/201117752}

\bibitem[{{Vladilo}(1998)}]{Vladilo1998}
{Vladilo}, G. 1998, \apj, 493, 583, \dodoi{10.1086/305148}

\bibitem[{{Vladilo} {et~al.}(2018){Vladilo}, {Gioannini}, {Matteucci}, \&
  {Palla}}]{Vladilo2018}
{Vladilo}, G., {Gioannini}, L., {Matteucci}, F., \& {Palla}, M. 2018, \apj,
  868, 127, \dodoi{10.3847/1538-4357/aae8dc}

\bibitem[{{Welsh} {et~al.}(2019){Welsh}, {Cooke}, \& {Fumagalli}}]{Welsh2019}
{Welsh}, L., {Cooke}, R., \& {Fumagalli}, M. 2019, \mnras, 487, 3363,
  \dodoi{10.1093/mnras/stz1526}

\bibitem[{{Welsh} {et~al.}(2021){Welsh}, {Cooke}, \& {Fumagalli}}]{Welsh2021}
---. 2021, \mnras, 500, 5214, \dodoi{10.1093/mnras/staa3342}

\bibitem[{{Welsh} {et~al.}(2022){Welsh}, {Cooke}, {Fumagalli}, \&
  {Pettini}}]{Welsh2022}
{Welsh}, L., {Cooke}, R., {Fumagalli}, M., \& {Pettini}, M. 2022, \apj, 929,
  158, \dodoi{10.3847/1538-4357/ac4503}

\bibitem[{{Yong} {et~al.}(2013){Yong}, {Norris}, {Bessell}, {Christlieb},
  {Asplund}, {Beers}, {Barklem}, {Frebel}, \& {Ryan}}]{Yong2013}
{Yong}, D., {Norris}, J.~E., {Bessell}, M.~S., {et~al.} 2013, \apj, 762, 27,
  \dodoi{10.1088/0004-637X/762/1/27}

\bibitem[{{Yoon} {et~al.}(2019){Yoon}, {Beers}, {Tian}, \&
  {Whitten}}]{Yoon2019}
{Yoon}, J., {Beers}, T.~C., {Tian}, D., \& {Whitten}, D.~D. 2019, \apj, 878,
  97, \dodoi{10.3847/1538-4357/ab1ead}

\bibitem[{{Yoon} {et~al.}(2016){Yoon}, {Beers}, {Placco}, {Rasmussen},
  {Carollo}, {He}, {Hansen}, {Roederer}, \& {Zeanah}}]{Yoon2016}
{Yoon}, J., {Beers}, T.~C., {Placco}, V.~M., {et~al.} 2016, \apj, 833, 20,
  \dodoi{10.3847/0004-637X/833/1/20}

\bibitem[{{Yoon} {et~al.}(2018){Yoon}, {Beers}, {Dietz}, {Lee}, {Placco}, {Da
  Costa}, {Keller}, {Owen}, \& {Sharma}}]{Yoon2018}
{Yoon}, J., {Beers}, T.~C., {Dietz}, S., {et~al.} 2018, \apj, 861, 146,
  \dodoi{10.3847/1538-4357/aaccea}

\bibitem[{{Zepeda} {et~al.}(2022){Zepeda}, {Beers}, {Placco}, {Shank}, {Gudin},
  {Hirai}, {Mardini}, {Pifer}, {Catapano}, \& {Calagna}}]{Zepeda2022}
{Zepeda}, J., {Beers}, T.~C., {Placco}, V.~M., {et~al.} 2022, arXiv e-prints,
  arXiv:2209.12224.
\newblock \doarXiv{2209.12224}

\bibitem[{{Zou} {et~al.}(2020){Zou}, {Petitjean}, {Noterdaeme}, {Ledoux},
  {Srianand}, {Jiang}, \& {Krogager}}]{Zou2020}
{Zou}, S., {Petitjean}, P., {Noterdaeme}, P., {et~al.} 2020, \apj, 901, 105,
  \dodoi{10.3847/1538-4357/abb092}

\end{thebibliography}
\bibliographystyle{aasjournal}

\section*{Appendix\\Supplemental Material}

\begin{figure}
\figurenum{1.1}
\epsscale{0.75}
\plotone{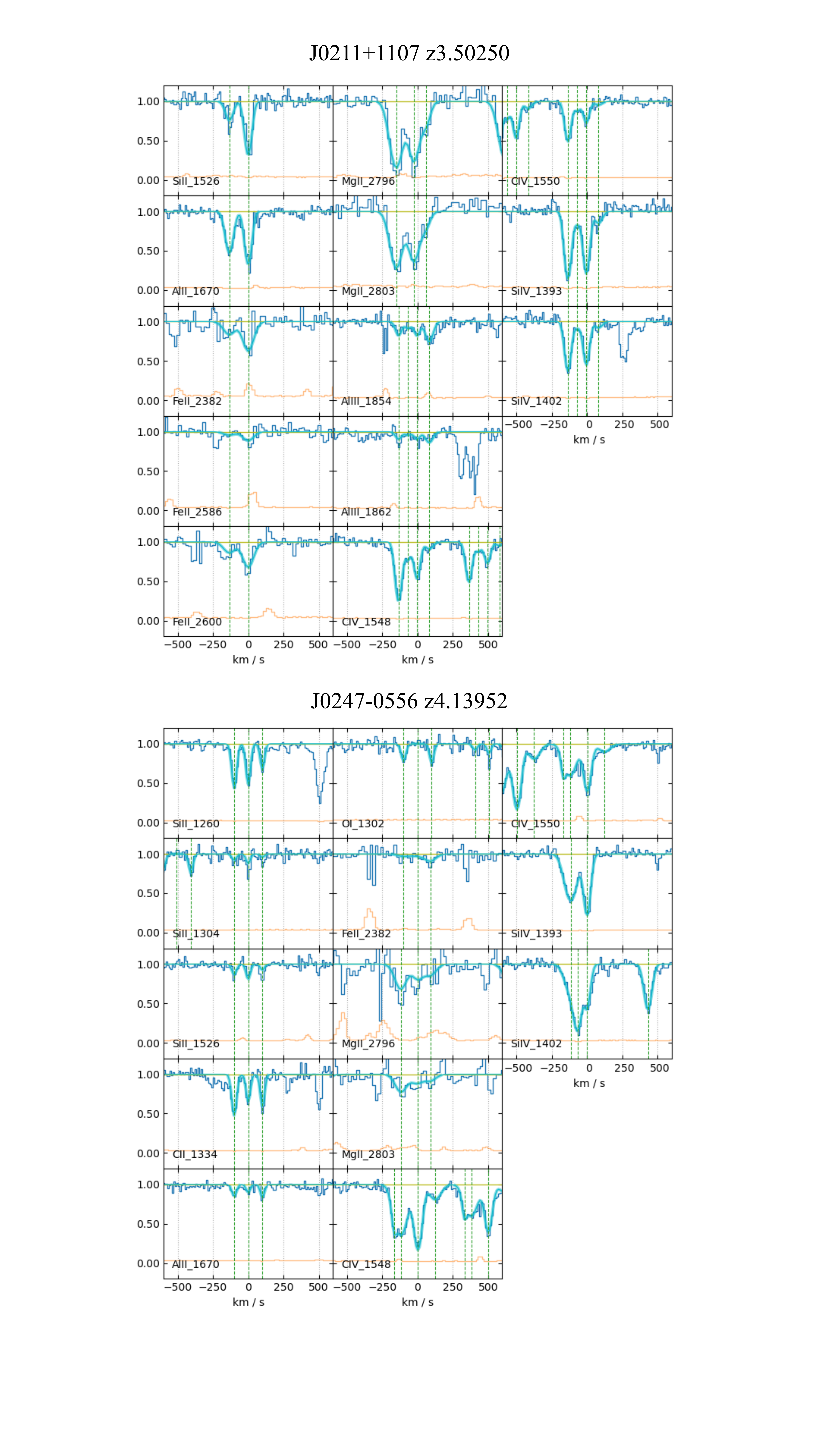}
\caption{Voigt fit profile of J0211+1107 at z=3.50250 and J0247-0556 at z=4.13952.\\Data are in blue, the fit is in cyan, the error spectrum is in orange, the continuum level is in yellow, and the vertical green dashed lines indicate the center of the components.}
\end{figure}

\figsetstart
\figsetnum{}
\figsettitle{Voigt fit profile of the 14 very metal-poor absorption systems.}\\
\figsettitle{``The complete figure set is available in the online journal.''}

\figsetgrpstart
\figsetgrpnum{1.1}
\figsetgrptitle{Voigt fit profile of J0211+1107 at z=3.50250 and J0247-0556 at z=4.13952}
\figsetplot{f1_1.pdf}
\figsetgrpnote{}
\figsetgrpend

\figsetgrpstart
\figsetgrpnum{1.2}
\figsetgrptitle{Voigt fit profile of J0800+1920 at z=3.42856 and J0818+0958 at z=3.45615}
\figsetplot{f1_2.pdf}
\figsetgrpnote{}
\figsetgrpend

\figsetgrpstart
\figsetgrpnum{1.3}
\figsetgrptitle{Voigt fit profile of J0835+0650 at z=3.51256 and J1018+0548 at z=3.38500}
\figsetplot{f1_3.pdf}
\figsetgrpnote{}
\figsetgrpend

\figsetgrpstart
\figsetgrpnum{1.4}
\figsetgrptitle{Voigt fit profile of J1111-0804 at z=3.48170 and J1111-0804 at z=3.75837}
\figsetplot{f1_4.pdf}
\figsetgrpnote{}
\figsetgrpend

\figsetgrpstart
\figsetgrpnum{1.5}
\figsetgrptitle{Voigt fit profile of J1117+1311 at z=3.43372 and J1332+0052 at z=3.42107}
\figsetplot{f1_5.pdf}
\figsetgrpnote{}
\figsetgrpend

\figsetgrpstart
\figsetgrpnum{1.6}
\figsetgrptitle{Voigt fit profile of J1552+1005 at z=3.44250 and J1658-0739 at z=3.54604}
\figsetplot{f1_6.pdf}
\figsetgrpnote{}
\figsetgrpend

\figsetgrpstart
\figsetgrpnum{1.7}
\figsetgrptitle{Voigt fit profile of J1658-0739 at z=3.69551 and J1723+2243 at z=4.24710}
\figsetplot{f1_7.pdf}
\figsetgrpnote{}
\figsetgrpend

\figsetend


\newpage

\begin{deluxetable}{c|c|c|c|c|c|c}
\tablenum{1.1}
\tabletypesize{\scriptsize}
\tablecaption{Voigt Fit Parameters}
\tablehead{\colhead{System} & \colhead{Transition} & \colhead{Redshift} & \colhead{$logN$} & \colhead{$\Delta logN$} & \colhead{$b$} & \colhead{$\Delta b$}}
\tablecomments{Table 1.1 is published in its entirety in the machine-readable format.
      A portion is shown here for guidance regarding its form and content.}
\startdata
J0211+1107 z3.50250\\                                                                                               
&MgII-2796.MgII-2803                             &   3.5014070 & 13.719  & 0.046 & 37.773 & 1.987\\           
&AlII-1670                                       &   3.5016674 & 12.825  & 0.025 & 37.773 & 1.987\\ 
&SiII-1526                                       &   3.5016674 & 13.548  & 0.090 & 37.773 & 1.987\\ 
&FeII-2344.FeII-2382.FeII-2586.FeII-2600         &   3.5016674 & 12.992  & 0.078 & 37.773 & 1.987\\ 
&SiIV-1393.SiIV-1402                             &   3.5016728 & 13.759  & 0.018 & 28.752 & 0.983\\ 
&CIV-1548.CIV-1550                               &   3.5016728 & 14.021  & 0.017 & 28.752 & 0.983\\ 
&AlIII-1854.AlIII-1862                           &   3.5016728 & 12.623  & 0.066 & 28.752 & 0.983\\ 
&SiIV-1393.SiIV-1402                             &   3.5035909 & 13.661  & 0.020 & 29.551 & 1.503\\ 
&CIV-1548.CIV-1550                               &   3.5035909 & 13.682  & 0.026 & 29.551 & 1.503\\ 
&AlIII-1854.AlIII-1862                           &   3.5035909 & 12.641  & 0.071 & 29.551 & 1.503\\ 
&CIV-1548.CIV-1550                               &   3.5026589 & 13.216  & 0.082 & 23.373 & 7.591\\     
&SiIV-1393.SiIV-1402                             &   3.5026589 & 12.647  & 0.129 & 23.373 & 7.591\\
&AlIII-1854.AlIII-1862                           &   3.5026589 & 12.221  & 0.141 & 23.373 & 7.591\\
&MgII-2796.MgII-2803                             &   3.5033051 & 13.671  & 0.061 & 28.884 & 1.607\\ 
&AlII-1670                                       &   3.5036733 & 12.922  & 0.028 & 28.884 & 1.607\\ 
&SiII-1526                                       &   3.5036733 & 13.952  & 0.038 & 28.884 & 1.607\\ 
&FeII-2344.FeII-2382.FeII-2586.FeII-2600         &   3.5036733 & 13.397  & 0.047 & 28.884 & 1.607\\ 
&MgII-2796.MgII-2803                             &   3.5045587 & 12.976  & 0.166 & 9.951  & 4.374\\ 
&SiIV-1393.SiIV-1402                             &   3.5049457 & 12.718  & 0.097 & 30.884 & 5.803\\ 
&CIV-1548.CIV-1550                               &   3.5049457 & 12.857  & 0.131 & 30.884 & 5.803\\ 
&AlIII-1854.AlIII-1862                           &   3.5049457 & 12.838  & 0.067 & 30.884 & 5.803\\ 
J0247-0556 z4.13952\\               
&CIV-1548.CIV-1550                               &   4.1355034 & 13.644  & 0.167 & 6.000  & 2.703\\
&CIV-1548.CIV-1550                               &   4.1363172 & 14.153  & 0.019 & 54.048 & 3.693\\  
&SiIV-1393.SiIV-1402                             &   4.1363915 & 13.685  & 0.022 & 54.048 & 3.693\\  
&SiII-1260.SiII-1304.SiII-1526                   &   4.1366451 & 13.058  & 0.043 & 12.487 & 1.721\\ 
&CII-1334                                        &   4.1366451 & 13.938  & 0.050 & 12.487 & 1.721\\ 
&OI-1302                                         &   4.1366451 & 13.819  & 0.082 & 12.487 & 1.721\\ 
&AlII-1670                                       &   4.1366451 & 11.953  & 0.090 & 12.487 & 1.721\\ 
&FeII-2382                                       &   4.1366451 & 12.170  & 0.150 & 12.487 & 1.721\\ 
&MgII-2796.MgII-2803                             &   4.1363254 & 12.937  & 0.179 & 12.487 & 1.721\\ 
&SiII-1260.SiII-1304.SiII-1526                   &   4.1383433 & 13.243  & 0.091 &  7.412 & 0.601\\ 
&CII-1334                                        &   4.1383433 & 13.720  & 0.087 &  7.412 & 0.601\\ 
&AlII-1670                                       &   4.1383433 & 11.749  & 0.160 &  7.412 & 0.601\\ 
&MgII-2796.MgII-2803                             &   4.1383433 & 12.985  & 0.179 &  7.412 & 0.601\\ 
&FeII-2382                                       &   4.1383433 & 12.326  & 0.150 &  7.412 & 0.601\\ 
&OI-1302                                         &   4.1383433 & 11.819  & 0.082 &  7.412 & 0.601\\ 
&SiIV-1393.SiIV-1402                             &   4.1383660 & 13.687  & 0.027 & 23.871 & 2.021\\ 
&CIV-1548.CIV-1550                               &   4.1384004 & 14.182  & 0.014 & 23.871 & 2.021\\ 
&MgII-2796.MgII-2803                             &   4.1399028 & 12.634  & 0.136 & 10.000 & 0.000\\ 
&FeII-2382                                       &   4.1399028 & 12.741  & 0.192 & 10.000 & 0.000\\ 
&SiII-1260.SiII-1304.SiII-1526                   &   4.1400627 & 12.750  & 0.091 &  7.000 & 0.000\\ 
&CII-1334                                        &   4.1400627 & 13.879  & 0.132 &  7.000 & 0.000\\ 
&OI-1302                                         &   4.1400627 & 13.908  & 0.150 &  7.000 & 0.000\\ 
&AlII-1670                                       &   4.1400627 & 11.974  & 0.150 &  7.000 & 0.000\\ 
&CIV-1548.CIV-1550                               &   4.1404764 & 13.444  & 0.047 & 43.605 & 8.025\\ 
\enddata
\end{deluxetable}

Title: Voigt Fit Parameters
Authors: Saccardi et al. 
Table: 1.1
================================================================================
Byte-by-byte Description of file: datafile1.1.txt
--------------------------------------------------------------------------------
   Bytes Format Units Label       Explanations
--------------------------------------------------------------------------------
   1- 21 A21    ---     System      Name of the absorption system
  23- 71 A49    ---     Transition  Absorption line transition
  73- 81 F9.7   ---     Redshift    Redshift of the absorbers
  83- 91 A9     ---     logN        Column density
  93- 98 F6.3   ---     DeltalogN   Error column density
 100-105 F6.3   ---     b           Doppler parameter
 107-115 F9.6   ---     Deltab      Error Doppler parameter
--------------------------------------------------------------------------------
Note (1):   ***ADD LENGTHY NOTES HERE***
Note (2):   ***OR REMOVE THIS SECTION***
--------------------------------------------------------------------------------

J0211+1107 z3.50250\\                                                                                               
                      MgII-2796.MgII-2803                               3.5014070 13.719     0.046 37.773  1.987           
                      AlII-1670                                         3.5016674 12.825     0.025 37.773  1.987 
                      SiII-1526                                         3.5016674 13.548     0.090 37.773  1.987 
                      FeII-2344.FeII-2382.FeII-2586.FeII-2600           3.5016674 12.992     0.078 37.773  1.987 
                      SiIV-1393.SiIV-1402                               3.5016728 13.759     0.018 28.752  0.983 
                      CIV-1548.CIV-1550                                 3.5016728 14.021     0.017 28.752  0.983 
                      AlIII-1854.AlIII-1862                             3.5016728 12.623     0.066 28.752  0.983 
                      SiIV-1393.SiIV-1402                               3.5035909 13.661     0.020 29.551  1.503 
                      CIV-1548.CIV-1550                                 3.5035909 13.682     0.026 29.551  1.503 
                      AlIII-1854.AlIII-1862                             3.5035909 12.641     0.071 29.551  1.503 
                      CIV-1548.CIV-1550                                 3.5026589 13.216     0.082 23.373  7.591     
                      SiIV-1393.SiIV-1402                               3.5026589 12.647     0.129 23.373  7.591
                      AlIII-1854.AlIII-1862                             3.5026589 12.221     0.141 23.373  7.591
                      MgII-2796.MgII-2803                               3.5033051 13.671     0.061 28.884  1.607 
                      AlII-1670                                         3.5036733 12.922     0.028 28.884  1.607 
                      SiII-1526                                         3.5036733 13.952     0.038 28.884  1.607 
                      FeII-2344.FeII-2382.FeII-2586.FeII-2600           3.5036733 13.397     0.047 28.884  1.607 
                      MgII-2796.MgII-2803                               3.5045587 12.976     0.166 9.951   4.374 
                      SiIV-1393.SiIV-1402                               3.5049457 12.718     0.097 30.884  5.803 
                      CIV-1548.CIV-1550                                 3.5049457 12.857     0.131 30.884  5.803 
                      AlIII-1854.AlIII-1862                             3.5049457 12.838     0.067 30.884  5.803

J0247-0556 z4.13952\\               
                      CIV_1548,CIV_1550                                 4.1355034 13.644     0.167 6.000   0.000
                      CIV_1548,CIV_1550                                 4.1363172 14.153     0.019 54.048  3.693  
                      SiIV-1393.SiIV-1402                               4.1363915 13.685     0.022 54.048  3.693  
                      SiII-1260.SiII-1304.SiII-1526                     4.1366451 13.058     0.043 12.487  1.721 
                      CII-1334                                          4.1366451 13.938     0.050 12.487  1.721 
                      OI-1302                                           4.1366451 13.819     0.082 12.487  1.721 
                      AlII-1670                                         4.1366451 11.953     0.090 12.487  1.721 
                      FeII-2382                                         4.1366451 12.170     0.150 12.487  1.721 
                      MgII-2796.MgII-2803                               4.1363254 12.937     0.179 12.487  1.721 
                      SiII-1260.SiII-1304.SiII-1526                     4.1383433 13.243     0.091  7.412  0.601 
                      CII-1334                                          4.1383433 13.720     0.087  7.412  0.601 
                      AlII-1670                                         4.1383433 11.749     0.160  7.412  0.601 
                      MgII-2796.MgII-2803                               4.1383433 12.985     0.179  7.412  0.601 
                      FeII-2382                                         4.1383433 12.326     0.150  7.412  0.601 
                      OI-1302                                           4.1383433 11.819     0.082  7.412  0.601 
                      SiIV-1393.SiIV-1402                               4.1383660 13.687     0.027 23.871  2.021 
                      CIV-1548.CIV-1550                                 4.1384004 14.182     0.014 23.871  2.021 
                      MgII-2796.MgII-2803                               4.1399028 12.634     0.136 10.000  0.000 
                      FeII-2382                                         4.1399028 12.741     0.192 10.000  0.000 
                      SiII-1260.SiII-1304.SiII-1526                     4.1400627 12.750     0.091  7.000  0.000 
                      CII-1334                                          4.1400627 13.879     0.132  7.000  0.000 
                      OI-1302                                           4.1400627 13.908     0.150  7.000  0.000 
                      AlII-1670                                         4.1400627 11.974     0.150  7.000  0.000 
                      CIV-1548.CIV-1550                                 4.1404764 13.444     0.047 43.605  8.025

J0800+1920 z3.42856\\                                                                                               
                      CIV-1548.CIV-1550                                 3.4283095 13.476     0.170 26.723  3.279
                      MgII-2796.MgII-2803                               3.4285568 13.344     0.013 28.508  1.716
                      FeII-2382                                         3.4285568 12.557     0.126 28.508  1.716
                      AlII-1670                                         3.4288175 12.095     0.031 19.728  2.419
                      SiII-1526                                         3.4288175 13.351     0.022 19.728  2.419
                      SiIV-1393.SiIV-1402                               3.4291269 14.049     0.016 26.027  2.141
                      CIV-1548.CIV-1550                                 3.4292404 14.714     0.065 26.027  2.141

J0818+0958 z3.45615\\                                                                                               
                      SiIV-1393.SiIV-1402                               3.4527287 12.444     0.124  17.806  8.875
                      CIV-1548.CIV-1550                                 3.4527695 13.584     0.074  40.950  7.847
                      SiIV-1393.SiIV-1402                               3.4533877 12.649     0.120  6.946   6.271
                      CIV-1548.CIV-1550                                 3.4534656 13.265     0.117  9.829   6.206
                      SiIV-1393.SiIV-1402                               3.4541606 12.651     0.132  6.000   0.000
                      CIV-1548.CIV-1550                                 3.4543490 13.455     0.033  16.547  2.969
                      SiIV-1393.SiIV-1402                               3.4544210 13.212     0.362  6.000   0.000
                      CIV-1548.CIV-1550                                 3.4552341 13.632     0.048  17.167  2.735
                      SiIV-1393.SiIV-1402                               3.4552854 13.357     0.020  24.049  2.134
                      AlII-1670                                         3.4552881 11.743     0.172  23.905  6.818
                      CII-1334                                          3.4552881 13.428     0.099  23.905  6.818
                      MgII-2796.MgII-2803                               3.4552881 12.813     0.123  23.905  6.818
                      FeII-2382                                         3.4552881 11.600     0.121  23.905  6.818
                      SiII-1526                                         3.4552881 12.600     0.114  23.905  6.818
                      SiIV-1393.SiIV-1402                               3.4562259 13.246     0.060  15.920  4.021
                      MgII-2796.MgII-2803                               3.4563966 13.380     0.042  25.418  2.246
                      AlII-1670                                         3.4563966 12.423     0.048  25.418  2.246
                      CII-1334                                          3.4563966 14.155     0.037  25.418  2.246
                      FeII-2382                                         3.4563966 12.608     0.221  25.418  2.246
                      SiII-1526                                         3.4563966 13.211     0.114  25.418  2.246
                      CIV-1548.CIV-1550                                 3.4564296 13.999     0.028  42.079  5.330
                      SiIV-1393.SiIV-1402                               3.4567991 13.253     0.057  17.950  3.309 
                      OI-1302                                           3.4561500 $<$12.95      \\

J0835+0650 z3.51256\\                                                                                               
                      CIV-1548.CIV-1550                                 3.5101223 13.149     0.126 30.318  8.750
                      CIV-1548.CIV-1550                                 3.5121919 13.206     0.081 10.649  6.200
                      MgII-2796.MgII-2803                               3.5122260 13.302     0.268  7.668  2.567
                      AlII-1670                                         3.5124854 12.452     0.043  7.668  2.567
                      SiII-1526                                         3.5124854 12.893     0.385  7.668  2.567
                      CIV-1548.CIV-1550                                 3.5131432 14.198     0.027 24.619  1.500
                      MgII-2796.MgII-2803                               3.5133001 12.786     0.039  6.899  2.010
                      AlII-1670                                         3.5135754 11.152     0.143  6.899  2.010
                      SiII-1526                                         3.5135754 12.167     0.185  6.899  2.010
                      SiIV-1393.SiIV-1402                               3.5133006 13.759     0.019 46.439  2.315
                      CIV-1548.CIV-1550                                 3.5140363 13.449     0.107  5.000  0.000
                      CIV-1548.CIV-1550                                 3.5147242 13.021     0.118  5.000  0.000
                      SiIV-1393.SiIV-1402                               3.5153161 12.945     0.073 28.793  7.145
                      FeII-2344                                         3.5124854 $<$11.93      \\

J1018+0548 z3.38500\\                                                                                               
                      FeII-2382.FeII-2600                               3.3844277 12.830     0.109 22.132  2.812 
                      MgII-2796.MgII-2803                               3.3844277 13.289     0.039 22.132  2.812 
                      OI-1302                                           3.3844277 14.488     0.088 22.132  2.812 
                      SiII-1260.SiII-1304.SiII-1526                     3.3844277 13.344     0.046 22.132  2.812 
                      CII-1334                                          3.3844277 13.944     0.091 22.132  2.812 
                      AlII-1670                                         3.3844277 11.766     0.170 22.132  2.812 
                      CIV-1548.CIV-1550                                 3.3845251 13.257     0.060 22.579  5.729 
                      CIV-1548.CIV-1550                                 3.3852553 13.494     0.035 10.005  3.160 
                      SiIV-1393.SiIV-1402                               3.3853316 13.225     0.072 10.005  3.160 
                      CIV-1548.CIV-1550                                 3.3867113 13.560     0.025 30.706  4.699 
                      SiIV-1393.SiIV-1402                               3.3868889 13.002     0.045 30.706  4.699 
                      MgII-2796.MgII-2803                               3.3869555 12.647     0.131  9.000  2.113
                      SiII-1260.SiII-1304.SiII-1526                     3.3869555 12.396     0.083  9.000  2.113
                      CII-1334                                          3.3869555 13.261     0.130  9.000  2.113
                      AlII-1670                                         3.3869555 11.709     0.070  9.000  2.113
                      FeII-2382.FeII-2600                               3.3869555 12.464     0.090  9.000  2.113
                      OI-1302                                           3.3869555 11.164     0.190  9.000  2.113

J1111-0804 z3.48170\\                                                                                               
                      MgII-2796.MgII-2803                               3.4817281 13.806     0.051 18.377  0.794 
                      FeII-1608.FeII-2344.FeII-2382.FeII-2586.FeII-2600 3.4817281 13.554     0.018 18.377  1.219 
                      CII-1334                                          3.4819429 14.848     0.055 18.377  0.622 
                      AlII-1670                                         3.4819429 12.543     0.034 18.377  0.622 
                      SiII-1260.SiII-1526.SiII-1808                     3.4819429 13.895     0.028 18.377  0.622 
                      OI-1302                                           3.4819429 15.097     0.075 18.377  0.622 
                      SiIV-1393.SiIV-1402                               3.4821471 13.229     0.019 20.502  0.876 
                      CIV-1548.CIV-1550                                 3.4821735 13.632     0.017 20.502  0.876

J1111-0804 z3.75837\\                                                                                               
                      SiIV-1393.SiIV-1402                               3.7565141 13.214     0.034 33.050  3.226
                      CIV-1548.CIV-1550                                 3.7567639 13.821     0.022 40.704  3.365
                      SiIV-1393.SiIV-1402                               3.7578669 12.883     0.042 15.360  3.189
                      MgII-2796.MgII-2803                               3.7578344 12.502     0.098  8.427  1.062
                      SiII-1260.SiII-1304.SiII-1526                     3.7578344 12.367     0.022  8.427  1.062
                      CII-1334                                          3.7578344 13.369     0.043  8.427  1.062
                      AlII-1670                                         3.7578344 11.908     0.049  8.427  1.062
                      CIV-1548.CIV-1550                                 3.7579520 12.892     0.145 17.919  7.578
                      SiIV-1393.SiIV-1402                               3.7588129 12.563     0.151 17.919  7.578
                      MgII-2796.MgII-2803                               3.7594764 12.492     0.060 12.643  6.762
                      AlII-1670                                         3.7594764 12.130     0.107 12.643  6.762 
                      CII-1334                                          3.7594764 13.539     0.090 12.643  6.762
                      SiII-1260.SiII-1304.SiII-1526                     3.7594764 12.638     0.030 12.643  6.762  
                      CIV-1548.CIV-1550                                 3.7593019 13.326     0.034 18.744  3.226
                      SiIV-1393.SiIV-1402                               3.7594728 12.945     0.063 18.744  3.226
                      MgII-2796.MgII-2803                               3.7604750 12.874     0.099  7.327  2.666
                      SiII-1260.SiII-1304.SiII-1526                     3.7604750 13.024     0.128  7.327  2.666
                      AlII-1670                                         3.7604750 11.961     0.103  7.327  2.666
                      CII-1334                                          3.7604750 13.820     0.238  7.327  2.666
                      CIV-1548.CIV-1550                                 3.7604789 13.304     0.029 11.067  1.395
                      SiIV-1393.SiIV-1402                               3.7605808 13.025     0.031 11.067  1.395
                      FeII-2600                                         3.7583700 $<$11.50      \\                  
                      OI-1302                                           3.7583700 $<$12.50      \\

J1117+1311 z3.43372\\                                                                                               
                      CIV-1548.CIV-1550                                 3.4318018 13.175     0.065 31.394  7.801 
                      SiIV-1393.SiIV-1402                               3.4318362 12.621     0.050  7.000  0.000 
                      CIV-1548.CIV-1550                                 3.4335084 14.023     0.305 13.884  6.306 
                      SiIV-1393.SiIV-1402                               3.4336445 13.883     0.110 13.884  6.306 
                      AlII-1670                                         3.4336592 11.867     0.061 20.987  3.660 
                      SiII-1526                                         3.4336592 12.740     0.135 20.987  3.660 
                      MgII-2796.MgII-2803                               3.4336592 13.071     0.124 20.987  3.660 
                      CII-1334                                          3.4336592 13.515     0.044 20.987  3.660 
                      CIV-1548.CIV-1550                                 3.4339745 14.332     0.025 39.464  4.002 
                      SiIV-1393.SiIV-1402                               3.4341055 13.815     0.104 39.464  4.002 
                      FeII-2382                                         3.4337200 $<$11.70      \\                  
                      OI-1302                                           3.4337200 $<$12.77      \\

J1332+0052 z3.42107\\                                                                                               
                      MgII-2796.MgII-2803                               3.4210895 12.881     0.020 13.236  3.555 
                      FeII-2382                                         3.4210895 12.468     0.195 13.236  3.555 
                      AlII-1670                                         3.4210895 11.822     0.101 13.236  3.555 
                      CII-1334                                          3.4210895 13.837     0.059 13.236  3.555 
                      SiII-1260,SiII-1526                               3.4210895 12.936     0.069 13.236  3.555
                      SiIV-1393.SiIV-1402                               3.4210841 13.101     0.020 29.358  1.633 
                      CIV-1548.CIV-1550                                 3.4211560 13.398     0.014 29.358  1.633

J1552+1005 z3.44250\\                                                                                               
                      SiIV-1393.SiIV-1402                               3.4416225 13.431     0.050 16.629  2.362
                      CIV-1548.CIV-1550                                 3.4416382 14.107     0.053 16.629  2.362
                      CIV-1548.CIV-1550                                 3.4422315 13.290     0.265 21.946  5.013
                      SiIV-1393.SiIV-1402                               3.4422852 12.670     0.229 21.946  5.013
                      MgII-2796.MgII-2803                               3.4422404 13.050     0.037  6.000  0.000
                      FeII-2382.FeII-2600                               3.4421932 12.644     0.069  6.000  0.000
                      OI-1302                                           3.4422469 13.430     0.139  6.000  0.000                     
                      CII-1334                                          3.4422979 14.237     0.075  6.000  0.000
                      SiII-1526                                         3.4423142 13.059     0.105  6.000  0.000
                      AlII-1670                                         3.4423737 11.857     0.025  6.000  0.000
                      SiII-1526                                         3.4429701 12.696     0.156 13.530  4.282
                      CII-1334                                          3.4431013 13.316     0.051 13.530  4.282
                      OI-1302                                           3.4431574 13.468     0.083 13.530  4.282
                      AlII-1670                                         3.4433597 11.461     0.146 13.530  4.282
                      SiIV-1393.SiIV-1402                               3.4432938 12.775     0.086 14.588  8.659
                      CIV-1548.CIV-1550                                 3.4432938 12.650     0.135 14.588  8.659
                      CIV-1548.CIV-1550                                 3.4443012 13.659     0.033 19.988  2.862
                      SiIV-1393.SiIV-1402                               3.4443078 12.665     0.097 19.988  2.862

J1658-0739 z3.54604\\                                                                                           
                      CIV-1548.CIV-1550                                 3.5424132 13.241     0.092  7.000  0.000
                      SiIV-1393.SiIV-1402                               3.5441887 13.631     0.056 14.102  1.417
                      CIV-1548.CIV-1550                                 3.5441887 15.268     0.298 14.102  1.417
                      SiIV-1393.SiIV-1402                               3.5460522 13.255     0.046 16.355  1.349
                      CIV-1548.CIV-1550                                 3.5460522 14.360     0.067 16.355  1.349
                      SiIV-1393.SiIV-1402                               3.5470664 13.521     0.037 18.284  1.991
                      CIV-1548.CIV-1550                                 3.5470664 13.926     0.031 18.284  1.991
                      MgII-2796.MgII-2803                               3.5471694 13.175     0.082 11.049  2.258
                      SiII-1260.SiII-1304.SiII-1526                     3.5471694 13.087     0.079 11.049  2.258
                      CII-1334                                          3.5471694 14.009     0.098 11.049  2.258
                      CIV-1548.CIV-1550                                 3.5482491 13.526     0.043 15.441  5.499
                      SiIV-1393.SiIV-1402                               3.5482491 12.553     0.156 15.441  5.499
                      CIV-1548.CIV-1550                                 3.5492112 13.460     0.058 22.276  6.708
                      SiIV-1393.SiIV-1402                               3.5492112 12.750     0.110 22.276  6.708
                      FeII-2344                                         3.5471694 $<$11.56      \\

J1658-0739 z3.69551\\                                                                                                
                      SiIV-1393.SiIV-1402                               3.6953133 13.279     0.029 45.949  4.069
                      CIV-1548.CIV-1550                                 3.6953680 13.689     0.027 45.949  4.069
                      MgII-2796.MgII-2803                               3.6954172 12.626     0.037 14.219  5.920
                      CII-1334                                          3.6954172 13.178     0.131 14.219  5.920
                      SiII-1260.SiII-1304.SiII-1526                     3.6954172 12.701     0.072 14.219  5.920
                      FeII-2382                                         3.6954172 $<$11.85      \\              
                      OI-1302                                           3.6954172 $<$12.81      \\

J1723+2243 z4.24710\\                                                                                           
                      MgII-2796.MgII-2803                               4.2470927 13.314     0.059 21.532  2.444
                      FeII-2382                                         4.2470927 12.785     0.188 21.532  2.444
                      OI-1302                                           4.2475195 13.701     0.044 21.532  2.444
                      SiII-1304.SiII-1526                               4.2475195 13.364     0.047 21.532  2.444
                      AlII-1670                                         4.2475195 12.252     0.084 21.532  2.444
                      CII-1334                                          4.2475195 14.252     0.158 21.532  2.444
                      CIV-1548.CIV-1550                                 4.2474747 13.802     0.019 27.054  1.997 
                      SiIV-1393.SiIV-1402                               4.2476752 13.403     0.020 27.054  1.997

J0042-1020 z3.62953\\                                                                                               
                      FeII-2344.FeII-2382.FeII-2586.FeII-2600           3.6285654 13.211     0.079  8.684  1.195
                      MgII-2796.MgII-2803                               3.6285848 13.950     0.332 12.783  2.424
                      AlIII-1854.AlIII-1862                             3.6286791 12.893     0.015 13.754  1.104
                      AlII-1670                                         3.6287102 12.768     0.040 13.608  1.217
                      OI-1302                                           3.6287126 13.430     0.034  6.941  3.512
                      SiII-1526                                         3.6287186 13.955     0.030 13.608  1.217
                      SiIV-1393.SiIV-1402                               3.6287378 13.856     0.018 17.287  0.594
                      CII-1334                                          3.6287736 $>$14.81   0.000 13.608  1.217
                      CIV-1548.CIV-1550                                 3.6287759 14.067     0.016 18.014  0.875
                      AlII-1670                                         3.6302644 12.454     0.017 11.496  1.172
                      SiIV-1393.SiIV-1402                               3.6302698 13.740     0.025 14.993  1.090
                      CIV-1548.CIV-1550                                 3.6302789 13.890     0.031 17.472  1.789
                      AlIII-1854.AlIII-1862                             3.6302843 12.571     0.034 22.862  2.895
                      CII-1334                                          3.6303109 14.146     0.037 20.012  2.445
                      MgII-2796.MgII-2803                               3.6303190 13.384     0.021 44.047  3.024
                      SiII-1526                                         3.6303271 13.230     0.041 13.485  3.564
                      CIV-1548.CIV-1550                                 3.6305685 14.098     0.023 40.420  5.060
                      SiIV-1393.SiIV-1402                               3.6306256 13.623     0.022 48.215  2.797
                      AlII-1670                                         3.6310256 12.037     0.031  6.000  0.000
                      CII-1334                                          3.6310670 14.023     0.431  6.000  0.000
                      SiII-1526                                         3.6310817 12.949     0.123  6.421  5.913

J0056-2808 z3.58045\\                                                                                           
                      MgII-2796.MgII-2803                               3.5800764 $>$14.77   0.000 15.472  2.208
                      FeII-1608.FeII-2344.FeII-2382.FeII-2586.FeII-2600 3.5802038 13.720     0.032 18.054  1.490
                      CIV-1548.CIV-1550                                 3.5802346 14.234     0.010 28.952  0.817
                      SiIV-1393.SiIV-1402                               3.5802681 14.025     0.028 23.233  0.994
                      CII-1334                                          3.5803508 $>$17.49   0.000 10.650  1.132
                      AlII-1670                                         3.5803860 14.598     0.492 10.650  1.132
                      AlIII-1854.AlIII-1862                             3.5804088 13.041     0.023 20.821  1.947
                      SiII-1304.SiII-1526                               3.5804422 14.570     0.140 13.910  1.529
                      OI-1302                                           3.5804535 15.016     0.233 10.650  1.132
                      SiIV-1393.SiIV-1402                               3.5815169 12.153     0.215  8.160  6.848
                      CIV-1548.CIV-1550                                 3.5815269 13.314     0.031 13.918  4.081
                      SiIV-1393.SiIV-1402                               3.5829358 12.844     0.125  7.498  5.723
                      CIV-1548.CIV-1550                                 3.5829358 13.781     0.015 30.399  1.897
                      CIV-1548.CIV-1550                                 3.5841343 13.252     0.040 10.209  4.796

J0100-2708 z3.24279\\                                                                                           
                      CIV-1548.CIV-1550                                 3.2426631 13.052     0.056 28.706  5.917
                      MgII-2796.MgII-2803                               3.2426905 13.482     0.380 10.388  5.163
                      OI-1302                                           3.2427368 14.587     0.071  9.000  0.000
                      CII-1334                                          3.2428929 14.244     0.215 10.360  2.369
                      SiII-1304.SiII-1526                               3.2429262 13.498     0.059 16.090  3.832
                      AlII-1670                                         3.2429444 12.371     0.179  9.341  9.112

J0124+0347 z3.67488\\                                                                                           
                      SiIV-1393.SiIV-1402                               3.6738359 12.972     0.109 13.968  4.363
                      CIV-1548.CIV-1550                                 3.6740549 13.774     0.028 46.233  4.839
                      MgII-2796.MgII-2803                               3.6747067 13.055     0.026 26.773  3.790
                      SiII-1260.SiII-1304.SiII-1526                     3.6748632 13.095     0.037 14.996  2.998
                      CIV-1548.CIV-1550                                 3.6748671 13.756     0.080 10.931  4.279
                      AlII-1670                                         3.6748803 11.900     0.131 14.712  6.443
                      SiIV-1393.SiIV-1402                               3.6748972 13.684     0.051 23.004  7.374
                      CII-1334                                          3.6749599 13.553     0.072 14.712  6.443
                      AlIII-1854.AlIII-1862                             3.6750259 12.529     0.085  9.013  2.167
                      SiIV-1393.SiIV-1402                               3.6754692 13.684     0.051  6.916  5.033
                      CIV-1548.CIV-1550                                 3.6754979 13.774     0.028 11.000  0.000
                      SiIV-1393.SiIV-1402                               3.6756513 13.684     0.051  5.466  7.413
                      CIV-1548.CIV-1550                                 3.6762452 13.363     0.150  6.951  7.326
                      FeII-2344                                         3.6748800 $<$11.73      \\

J0133+0400 z3.62025\\                                                                                           
                      CIV-1548.CIV-1550                                 3.6197033 13.920     0.028 45.863  8.606
                      SiIV-1393.SiIV-1402                               3.6199169 13.473     0.016 35.878  1.839
                      MgII-2796.MgII-2803                               3.6202142 13.126     0.025 40.742  3.443
                      AlII-1670                                         3.6202142 12.150     0.050 24.276  5.390
                      SiII-1526                                         3.6202884 12.785     0.139 20.038  6.902
                      CIV-1548.CIV-1550                                 3.6203266 13.920     0.028 47.714  9.061
                      SiIV-1393.SiIV-1402                               3.6210052 13.117     0.030 30.538  2.251
                      AlII-1670                                         3.6210417 11.724     0.080 10.682  5.352
                      SiII-1526                                         3.6211140 12.642     0.122 10.038  6.902

J0133+0400 z3.99668\\                                                                                           
                      SiIV-1393.SiIV-1402                               3.9942522 13.152     0.331  7.168  0.509
                      AlII-1670                                         3.9944956 12.154     0.050 17.191  4.019
                      CII-1334                                          3.9945067 13.796     0.025 15.263  1.948
                      SiIV-1393.SiIV-1402                               3.9946175 13.062     0.126 12.991  9.680
                      CIV-1548.CIV-1550                                 3.9948350 13.678     0.073 47.851  9.617
                      SiIV-1393.SiIV-1402                               3.9951623 12.573     0.193  7.000  0.000
                      FeII-1608.FeII-2344.FeII-2382.FeII-2586.FeII-2600 3.9952453 13.228     0.052 29.098  7.123
                      SiII-1304.SiII-1526                               3.9955386 13.743     0.074  9.740  3.307
                      CII-1334                                          3.9955601 14.326     0.164 10.914  2.928
                      AlII-1670                                         3.9955668 12.286     0.088  9.684  5.698
                      OI-1302                                           3.9956821 14.772     0.039 19.292  1.878
                      CII-1334                                          3.9960693 13.881     0.165  8.391  3.052
                      CIV-1548.CIV-1550                                 3.9963032 12.951     0.198 20.513  9.487
                      SiIV-1393.SiIV-1402                               3.9963388 13.128     0.058 41.099  9.442
                      FeII-1608.FeII-2344.FeII-2382.FeII-2586.FeII-2600 3.9963677 12.919     0.091 11.221  5.934
                      AlII-1670                                         3.9965784 12.062     0.067 24.374  6.393
                      OI-1302                                           3.9966150 14.549     0.149 10.620  3.228
                      CII-1334                                          3.9966732 14.236     0.377  8.391  3.052
                      SiII-1304.SiII-1526                               3.9967287 13.543     0.065 26.868  6.163
                      SiIV-1393.SiIV-1402                               3.9978337 12.034     0.265  7.000  0.000
                      CIV-1548.CIV-1550                                 3.9990006 13.816     0.018 42.191  2.071
                      SiIV-1393.SiIV-1402                               3.9993039 13.620     0.020 28.774  1.566

J0214-0517 z3.69203\\                                                                                           
                      SiIV-1393.SiIV-1402                               3.6907682 13.525     0.387 15.741  2.479
                      MgII-2796.MgII-2803                               3.6918349 13.280     0.048 21.327  2.961
                      SiIV-1393.SiIV-1402                               3.6919148 13.438     0.466 10.910  6.298
                      CIV-1548.CIV-1550                                 3.6919148 13.221     0.048 10.395  6.390
                      SiII-1304.SiII-1526                               3.6919224 13.402     0.043  7.872  0.967
                      AlII-1670                                         3.6919225 12.425     0.052  8.815  3.702
                      CII-1334                                          3.6919821 14.096     0.021 13.989  1.020
                      AlIII-1854.AlIII-1862                             3.6919894 12.458     0.066 28.261  7.036
                      SiIV-1393.SiIV-1402                               3.6925983 13.370     0.123 18.242  3.278
                      CIV-1548.CIV-1550                                 3.6925983 13.760     0.090  6.000  0.000
                      SiIV-1393.SiIV-1402                               3.6926660 13.507     0.244  5.518  2.630
                      AlII-1670                                         3.6927296 11.634     0.121  8.815  3.702
                      CII-1334                                          3.6927410 13.814     0.033 10.697  1.901
                      CIV-1548.CIV-1550                                 3.6939637 13.077     0.075  5.635  1.252

J0234-1806 z4.22817\\                                                                                           
                      CIV-1548.CIV-1550                                 4.2266931 13.318     0.167 21.324  9.892
                      SiIV-1393.SiIV-1402                               4.2267158 13.207     0.085 23.467  5.565
                      AlII-1670                                         4.2270742 13.040     0.196  7.000  0.000
                      CII-1334                                          4.2271099 14.484     0.085 24.274  3.488
                      SiII-1304.SiII-1526                               4.2271409 13.861     0.085 10.071  3.071
                      OI-1302                                           4.2271939 14.305     0.071 11.733  4.088
                      MgII-2796.MgII-2803                               4.2279385 $>$14.32   0.000 61.324  4.702
                      SiII-1304.SiII-1526                               4.2280106 14.688     0.454 10.053  2.585
                      AlII-1670                                         4.2280529 13.173     0.086 21.071  6.239
                      OI-1302                                           4.2281044 15.409     0.185 10.000  0.000
                      FeII-2382.FeII-1608                               4.2281057 13.944     0.056 32.748  4.044
                      SiIV-1393.SiIV-1402                               4.2281113 13.863     0.023 40.879  3.214
                      CIV-1548.CIV-1550                                 4.2281699 14.040     0.037 47.672  6.993
                      CII-1334                                          4.2285076 $>$16.60   0.000 24.274  3.488
                      SiII-1304.SiII-1526                               4.2289713 13.999     0.025 23.661  3.437
                      AlII-1670                                         4.2290203 12.912     0.074 24.256  7.128
                      OI-1302                                           4.2290874 14.100     0.045 21.691  5.621
                      MgII-2796.MgII-2803                               4.2301978 13.32      0.221 11.000  0.000
                      SiII-1304.SiII-1526                               4.2302193 13.334     0.060 13.066  7.993
                      CII-1334                                          4.2302629 14.189     0.057 24.274  3.488
                      OI-1302                                           4.2302639 13.650     0.084  7.000  0.000
                      AlII-1670                                         4.2303238 12.511     0.171  7.000  0.000
                      SiIV-1393.SiIV-1402                               4.2305821 12.467     0.122  8.146  5.922
                      CIV-1548.CIV-1550                                 4.2307156 13.565     0.042 11.000  0.000

J0307-4945 z4.21345\\                                                                                           
                      SiIV-1393.SiIV-1402                               4.2103114 13.596     0.132 19.647  5.248
                      AlIII-1854.AlIII-1862                             4.2104132 12.504     0.181 31.536  6.666
                      AlII-1670                                         4.2104959 12.229     0.066 30.459  7.130
                      MgII-2796.MgII-2803                               4.2107622 13.599     0.038 47.256  4.384
                      SiII-1526                                         4.2108996 13.610     0.104 30.459  7.130
                      CIV-1548.CIV-1550                                 4.2111338 $>$16.43   0.000 32.628  3.630
                      SiIV-1393.SiIV-1402                               4.2112301 14.159     0.038 48.115  3.255
                      AlIII-1854.AlIII-1862                             4.2116008 12.707     0.112 28.782  9.156
                      AlII-1670                                         4.2116708 12.447     0.035 19.387  3.114
                      SiII-1526                                         4.2118781 13.318     0.170 19.387  3.114
                      SiIV-1393.SiIV-1402                               4.2139348 12.939     0.056 27.995  5.737
                      CIV-1548.CIV-1550                                 4.2142054 $>$13.38   0.000 47.760  6.933
                      SiII-1526                                         4.2163124 13.334     0.053  7.125  2.745
                      AlIII-1854.AlIII-1862                             4.2163345 12.419     0.052  5.989  1.091
                      SiIV-1393.SiIV-1402                               4.2163975 13.614     0.050 12.973  1.760
                      MgII-2796.MgII-2803                               4.2164171 12.969     0.055 33.418  8.319
                      CIV-1548.CIV-1550                                 4.2164270 $>$13.91   0.000 18.685  3.210
                      AlII-1670                                         4.2165032 11.913     0.064  7.125  2.745
                      CIV-1548.CIV-1550                                 4.2172060 $>$13.98   0.000 17.291  3.591
                      SiIV-1393.SiIV-1402                               4.2172507 13.343     0.054 20.152  5.118
                      AlII-1670                                         4.2173493 11.905     0.066 13.921  3.574
                      SiII-1526                                         4.2173754 12.903     0.056 13.921  3.574
                      MgII-2796.MgII-2803                               4.2183106 12.873     0.219 12.310  9.341
                      SiIV-1393.SiIV-1402                               4.2184624 13.398     0.061 35.648  7.576
                      SiII-1526                                         4.2187059 13.108     0.078 13.921  3.574
                      AlII-1670                                         4.2187089 12.012     0.063 13.921  3.574
                      CIV-1548.CIV-1550                                 4.2192608 $>$14.49   0.000 48.692  5.026
                      MgII-2796.MgII-2803                               4.2195622 12.894     0.184 11.446  6.328
                      SiIV-1393.SiIV-1402                               4.2198246 13.389     0.038 26.580  3.353
                      SiII-1526                                         4.2199445 13.050     0.058  8.363  5.377
                      AlII-1670                                         4.2201987 12.088     0.078 45.674  6.686
                      SiII-1526                                         4.2206986 13.045     0.077 18.308  6.483
                      MgII-2796.MgII-2803                               4.4661569 14.120     0.036 48.803  4.042
                      FeII-2382                                         4.2134500 $<$11.47      \\              
                                                                                                                
J0529-3552 z3.71343\\                                                                                           
                      CIV-1548.CIV-1550                                 3.7119784 13.09      0.109  9.444  9.788
                      SiIV-1393.SiIV-1402                               3.7121394 12.833     0.078 11.060  7.446
                      MgII-2796.MgII-2803                               3.7130772 13.524     0.076 20.205  2.295
                      SiII-1526                                         3.7132260 13.218     0.245  7.728  1.575
                      CIV-1548.CIV-1550                                 3.7132658 13.180     0.242 10.691  9.522
                      AlII-1670                                         3.7132793 12.406     0.105  7.728  1.575
                      SiIV-1393.SiIV-1402                               3.7133445 13.393     0.059 13.559  3.921
                      AlII-1670                                         3.7138075 12.239     0.102  8.000  0.000
                      SiII-1526                                         3.7138368 12.922     0.291  8.000  0.000
                      CIV-1548.CIV-1550                                 3.7138818 13.120     0.322 21.620  9.137
                      SiIV-1393.SiIV-1402                               3.7140404 13.088     0.110 21.620  9.137
                                                                                                                
J0529-3552 z4.06561\\                                                                                           
                      FeII-2382.FeII-2600                               4.0651865 12.970     0.082 11.203  6.491
                      SiII-1260.SiII-1304.SiII-1526                     4.0656032 13.610     0.086  8.494  0.574
                      CII-1334                                          4.0656176 14.484     0.234  8.494  0.574
                      OI-1302                                           4.0656356 13.970     0.094  8.494  0.574
                      AlII-1670                                         4.0657064 12.364     0.096  8.494  0.574
                      CIV-1548.CIV-1550                                 4.0658631 12.889     0.235 17.480  6.459
                      SiIV-1393.SiIV-1402                               4.0658631 12.803     0.153 17.480  6.459
                      CIV-1548.CIV-1550                                 4.0667668 13.152     0.192 23.787  9.850
                      SiIV-1393.SiIV-1402                               4.0667668 13.114     0.112 23.787  9.850
                      CIV-1548.CIV-1550                                 4.0677568 13.760     0.042 35.791  4.596
                      SiIV-1393.SiIV-1402                               4.0679723 13.215     0.064 33.613  7.056

J0818+0958 z3.53141\\                                                                                           
                      CII-1334                                          3.5311415 13.383     0.105  6.000  0.000
                      SiIV-1393.SiIV-1402                               3.5313453 14.004     0.054 28.704  1.439
                      CIV-1548.CIV-1550                                 3.5313617 14.655     0.027 28.704  1.439
                      MgII-2796.MgII-2803                               3.5314230 13.143     0.102 12.294  4.706
                      AlIII-1854.AlIII-1862                             3.5316400 12.570     0.042 25.032  3.890
                      CII-1334                                          3.5316834 13.977     0.046 12.880  2.661
                      SiIV-1393.SiIV-1402                               3.5317187 13.800     0.119 10.275  3.315
                      SiII-1304.SiII-1526                               3.5317492 13.154     0.051 10.007  4.503
                      AlII-1670                                         3.5317492 12.184     0.039 14.657  3.161
                      FeII-2382                                         3.5314100 $<$12.03      \\

J1013+0650 z3.23534\\                                                                                           
                      SiII-1526                                         3.2348397 12.844     0.146 10.000  0.000
                      AlII-1670                                         3.2349365 11.783     0.134 10.757  3.755
                      MgII-2796.MgII-2803                               3.2351362 13.362     0.028 40.755  3.841
                      SiIV-1393.SiIV-1402                               3.2353287 13.262     0.033 44.178  4.164
                      CIV-1548.CIV-1550                                 3.2353967 13.876     0.026 44.426  3.199
                      SiII-1526                                         3.2355172 13.486     0.041  7.000  0.000
                      AlII-1670                                         3.2358547 12.267     0.056 20.757  3.755
                      SiII-1526                                         3.2359583 13.340     0.128  7.000  0.000
                      MgII-2796.MgII-2803                               3.2370035 12.447     0.124  8.512  1.929
                      SiIV-1393.SiIV-1402                               3.2371797 13.019     0.021 12.108  1.286
                      CIV-1548.CIV-1550                                 3.2372062 13.759     0.029 18.461  1.746

J1013+0650 z3.32076\\                                                                                           
                      MgII-2796.MgII-2803                               3.3189017 12.197     0.160 14.965  8.490
                      SiIV-1393.SiIV-1402                               3.3189509 13.241     0.051 34.209  4.266
                      CIV-1548.CIV-1550                                 3.3190820 14.002     0.040 36.558  3.770
                      MgII-2796.MgII-2803                               3.3202931 13.180     0.032 31.277  3.208
                      SiII-1526                                         3.3203198 13.099     0.048  9.012  3.084
                      SiIV-1393.SiIV-1402                               3.3203908 14.106     0.058 17.954  1.779
                      AlIII-1854.AlIII-1862                             3.3205171 12.492     0.071  7.016  1.227
                      AlII-1670                                         3.3205265 12.189     0.057 13.464  3.754
                      CIV-1548.CIV-1550                                 3.3205328 $>$14.63   0.000 46.304  1.940
                      SiIV-1393.SiIV-1402                               3.3205893 13.819     0.044 40.132  3.513
                      FeII-2600                                         3.3207600 $<$11.89      \\

J1024+1819 z3.18944\\                                                                                           
                      SiII-1526                                         3.1884923 13.276     0.064  8.119  4.660
                      MgII-2796.MgII-2803                               3.1885487 13.255     0.056 27.719  6.491
                      AlII-1670                                         3.1886657 12.330     0.048 34.821  5.458
                      CII-1334                                          3.1886657 14.087     0.048 32.131  5.727
                      SiIV-1393.SiIV-1402                               3.1887373 13.102     0.050 27.504  3.580
                      CIV-1548.CIV-1550                                 3.1890015 14.025     0.032 52.741  4.720
                      SiII-1526                                         3.1891796 13.091     0.069  8.119  4.660
                      SiIV-1393.SiIV-1402                               3.1897539 13.124     0.047 27.504  3.580
                      CIV-1548.CIV-1550                                 3.1898275 13.401     0.149 17.470  4.836
                      SiII-1526                                         3.1898536 12.823     0.100  9.874  2.513
                      CII-1334                                          3.1899126 13.893     0.078 37.638  4.534
                      AlII-1670                                         3.1899126 11.921     0.120 35.412  4.159

J1117+1311 z3.27522\\                                                                                           
                      CIV-1548.CIV-1550                                 3.2748012 14.076     0.044 26.623  3.237
                      CII-1334                                          3.2748455 12.922     0.132  7.000  0.000
                      AlII-1670                                         3.2748455 11.403     0.190  7.000  0.000
                      SiII-1526                                         3.2748455 12.272     0.269  7.000  0.000
                      SiIV-1393.SiIV-1402                               3.2748871 13.465     0.101 10.728  3.604
                      MgII-2796.MgII-2803                               3.2754055 13.350     0.112 14.833  2.389
                      SiIV-1393.SiIV-1402                               3.2754564 13.877     0.069 15.777  2.427
                      AlII-1670                                         3.2754740 12.584     0.058  7.000  0.000
                      CII-1334                                          3.2754815 14.800     0.187  7.000  0.000
                      CIV-1548.CIV-1550                                 3.2754889 14.106     0.044 19.138  3.405
                      SiII-1526                                         3.2755225 13.218     0.076  7.000  0.000
                      CIV-1548.CIV-1550                                 3.2763772 14.062     0.049 14.403  1.749
                      SiIV-1393.SiIV-1402                               3.2763865 12.621     0.697 14.403  1.749
                      FeII-2600                                         3.2752200 $<$12.03      \\

J1249-0159 z3.10265\\                                                                                           
                      SiIV-1393.SiIV-1402                               3.1020311 13.938     0.013 49.196  1.856
                      CIV-1548.CIV-1550                                 3.1020554 14.185     0.014 48.307  1.857
                      AlII-1670                                         3.1020901 12.795     0.039 31.089  3.765
                      SiII-1526                                         3.1021067 13.648     0.029 41.049  4.103
                      MgII-2796.MgII-2803                               3.1021632 $>$13.76   0.000 41.952  3.594
                      FeII-2344.FeII-2382.FeII-2586.FeII-2600           3.1022502 13.192     0.039 41.952  3.594
                      AlII-1670                                         3.1031869 12.443     0.068 11.085  3.765
                      SiIV-1393.SiIV-1402                               3.1033112 12.797     0.087 11.000  0.000
                      SiII-1526                                         3.1033606 13.311     0.090 11.089  3.765

J1304+0239 z3.21072\\                                                                                           
                      MgII-2796.MgII-2803                               3.2041280 $>$13.73   0.000 18.402  4.020
                      FeII-2344.FeII-2382.FeII-2586.FeII-2600.FeII-1608 3.2042416 13.197     0.053 10.592  2.861
                      CIV-1548.CIV-1550                                 3.2042500 $>$14.52   0.000 29.829  1.767
                      SiII-1526                                         3.2043086 13.878     0.048 15.125  2.470
                      AlII-1670                                         3.2043310 14.952     0.275  7.000  0.000
                      AlIII-1854.AlIII-1862                             3.2043665 12.728     0.039 18.733  4.322
                      SiIV-1393.SiIV-1402                               3.2043672 13.681     0.077 13.459  1.727
                      AlII-1670                                         3.2056974 12.014     0.159  7.000  0.000
                      SiIV-1393.SiIV-1402                               3.2057260 13.145     0.128  8.868  3.635
                      SiII-1526                                         3.2057283 13.212     0.101 15.125  2.470
                      CIV-1548.CIV-1550                                 3.2057569 $>$15.78   0.000  7.310  0.337
                      SiII-1526                                         3.2067001 13.466     0.070 15.949  0.000
                      AlII-1670                                         3.2067731 12.297     0.098  7.000  0.000
                      SiIV-1393.SiIV-1402                               3.2068420 13.354     0.033 20.703  3.509
                      CIV-1548.CIV-1550                                 3.2068514 $>$14.41   0.000 25.372  3.363
                      AlIII-1854.AlIII-1862                             3.2069064 12.995     0.077 10.660  6.825
                      MgII-2796.MgII-2803                               3.2073596 $>$13.88   0.000 43.286  4.079
                      AlII-1670                                         3.2075108 13.682     0.317  7.000  0.000
                      SiII-1526                                         3.2075571 13.852     0.065 15.125  2.470
                      FeII-2344.FeII-2382.FeII-2586.FeII-2600.FeII-1608 3.2076570 13.530     0.018 39.127  2.948
                      CIV-1548.CIV-1550                                 3.2080234 $>$14.24   0.000 38.330  6.250
                      CIV-1548.CIV-1550                                 3.2080449 $>$14.60   0.000 20.308  6.084
                      SiIV-1393.SiIV-1402                               3.2080727 13.803     0.026 25.414  1.778
                      AlIII-1854.AlIII-1862                             3.2080829 12.905     0.059 20.047  4.082
                      SiII-1526                                         3.2081135 13.953     0.065 15.125  2.470
                      AlII-1670                                         3.2081235 14.676     0.081  7.000  0.000
                      CIV-1548.CIV-1550                                 3.2100459 $>$14.38   0.000 25.334  3.894
                      AlII-1670                                         3.2101115 13.154     0.066 23.025  4.793
                      FeII-2344.FeII-2382.FeII-2586.FeII-2600.FeII-1608 3.2101171 13.664     0.034 29.370  3.158
                      SiII-1526                                         3.2101610 14.254     0.049 22.718  3.347
                      AlIII-1854.AlIII-1862                             3.2102046 12.992     0.034 30.809  4.194
                      SiIV-1393.SiIV-1402                               3.2105836 13.847     0.017 45.081  2.859
                      CIV-1548.CIV-1550                                 3.2106392 14.577     0.114  7.000  0.000
                      MgII-2796.MgII-2803                               3.2107154 $>$15.72   0.000 37.476  6.401
                      CIV-1548.CIV-1550                                 3.2110255 $>$14.57   0.000 40.052  6.178
                      AlII-1670                                         3.2111256 13.339     0.048 43.009  7.368
                      SiII-1526                                         3.2112383 14.122     0.062 47.314  9.664
                      FeII-2344.FeII-2382.FeII-2586.FeII-2600.FeII-1608 3.2112613 13.280     0.103  7.000  0.000
                      CIV-1548.CIV-1550                                 3.2121895 $>$13.69   0.000 39.931  6.900
                      SiII-1526                                         3.2124303 13.427     0.130 15.949  0.000
                      MgII-2796.MgII-2803                               3.2131044 $>$13.96   0.000 47.629  6.135
                      AlII-1670                                         3.2134025 12.993     0.025 44.982  3.325
                      SiII-1526                                         3.2134605 14.154     0.030 45.653  4.653
                      AlIII-1854.AlIII-1862                             3.2137245 12.827     0.047 26.444  5.220
                      SiIV-1393.SiIV-1402                               3.2137790 13.967     0.022 33.349  1.459
                      CIV-1548.CIV-1550                                 3.2137916 $>$15.14   0.000 24.138  3.144

J1352+1303 z3.00680\\                                                                                           
                      MgII-2796.MgII-2803                               3.0050038 $>$13.21   0.000 19.493  6.712
                      AlIII-1854.AlIII-1862                             3.0050286 12.670     0.029 11.148  3.441
                      SiII-1526                                         3.0050705 13.695     0.067  9.559  2.584
                      FeII-2344.FeII-2382                               3.0050816 12.700     0.031 17.762  3.358
                      AlII-1670                                         3.0051118 13.354     0.035  7.000  0.000
                      CIV-1548.CIV-1550                                 3.0051695 12.927     0.046  8.550  5.655
                      CIV-1548.CIV-1550                                 3.0062068 13.196     0.038 13.591  4.740
                      MgII-2796.MgII-2803                               3.0062103 $>$13.06   0.000 11.000  0.000
                      AlIII-1854.AlIII-1862                             3.0062264 12.068     0.053  7.000  0.000
                      AlII-1670                                         3.0062950 12.588     0.026 19.868  2.524
                      SiII-1526                                         3.0062975 13.655     0.020 20.428  2.326
                      FeII-2344.FeII-2382                               3.0064367 12.684     0.032 12.361  4.150
                      SiII-1526                                         3.0071065 13.050     0.109  7.000  0.000
                      MgII-2796.MgII-2803                               3.0072943 $>$13.82   0.000 35.596  5.785
                      CIV-1548.CIV-1550                                 3.0073385 13.996     0.014 34.759  1.846
                      AlIII-1854.AlIII-1862                             3.0073914 12.647     0.026 47.753  5.094
                      AlII-1670                                         3.0077267 12.970     0.020 39.494  3.877
                      FeII-2344.FeII-2382                               3.0078994 13.042     0.018 36.074  2.384
                      SiII-1526                                         3.0079694 13.891     0.019 26.608  2.962
                      MgII-2796.MgII-2803                               3.0081331 $>$14.70   0.000 11.000  0.000
                      CIV-1548.CIV-1550                                 3.0084926 13.574     0.031 33.724  3.440
                      AlIII-1854.AlIII-1862                             3.0085619 12.041     0.057  6.394  1.373
                      AlII-1670                                         3.0086085 12.423     0.098  7.000  0.000
                                                                                                                
J1517+0511 z3.26514\\                                                                                           
                      SiIV-1393.SiIV-1402                               3.2635871 12.973     0.067 44.423  6.778
                      CIV-1548.CIV-1550                                 3.2637100 13.436     0.027 39.963  3.914
                      MgII-2796.MgII-2803                               3.2647788 13.017     0.029 34.708  4.648
                      SiIV-1393.SiIV-1402                               3.2648783 13.389     0.027 20.587  2.503
                      SiII-1526                                         3.2648954 13.021     0.039  9.918  7.440
                      CIV-1548.CIV-1550                                 3.2649284 13.497     0.021 22.312  2.273
                      AlII-1670                                         3.2649303 11.956     0.010 11.121  1.373
                      CII-1334                                          3.2649804 13.638     0.027 18.862  2.865
                      SiIV-1393.SiIV-1402                               3.2663585 13.234     0.036 40.243  6.966
                      CIV-1548.CIV-1550                                 3.2666737 13.425     0.022 26.002  2.759

J1542+0955 z3.28223\\                                                                                           
                      AlII-1670                                         3.2815371 12.692     0.040 16.691  1.609
                      SiII-1526                                         3.2815813 13.515     0.045 11.059  1.498
                      FeII-2382.FeII-2600                               3.2816264 13.614     0.054 33.597  8.632
                      MgII-2796.MgII-2803                               3.2817909 $>$13.86   0.000 35.020  4.160
                      CIV-1548.CIV-1550                                 3.2819540 13.476     0.129 27.845  8.543
                      AlIII-1854.AlIII-1862                             3.2821696 12.814     0.025 25.711  2.395
                      AlII-1670                                         3.2822291 13.079     0.055 12.691  1.609
                      SiII-1526                                         3.2822663 14.006     0.074 11.059  1.498
                      CIV-1548.CIV-1550                                 3.2823478 13.155     0.291 10.000  0.000
                      SiII-1526                                         3.2830952 13.235     0.062 11.059  1.498
                      AlII-1670                                         3.2831544 12.430     0.101  9.525  5.089
                      MgII-2796.MgII-2803                               3.2840226 12.644     0.102  8.314  2.494
                      SiII-1526                                         3.2843213 12.775     0.153 11.059  1.498
                      AlII-1670                                         3.2843481 11.890     0.083  6.000  0.000

J1621-0042 z3.10570\\                                                                                           
                      MgII-2796.MgII-2803                               3.1040595 14.028     0.074 11.000  0.000
                      FeII-2344.FeII-2382.FeII-2600                     3.1041568 13.192     0.041  7.128  2.280
                      SiII-1526                                         3.1042204 13.72      0.066  7.000  0.000
                      AlII-1670                                         3.1042519 12.28      0.098  7.000  0.000
                      SiII-1526                                         3.1054372 13.99      0.103  7.000  0.000
                      MgII-2796.MgII-2803                               3.1057056 $>$15.34   0.000 24.562  3.540
                      FeII-2344.FeII-2382.FeII-2600                     3.1057353 13.659     0.023 29.927  2.280
                      AlII-1670                                         3.1058881 13.236     0.021 27.355  1.920
                      AlIII-1854.AlIII-1862                             3.1059537 13.142     0.045 42.164  6.870
                      SiII-1526                                         3.1060007 14.182     0.021 20.528  1.870
                      SiIV-1393.SiIV-1402                               3.1060483 14.084     0.012 40.048  1.010
                      CIV-1548.CIV-1550                                 3.1063037 14.383     0.020 48.082  2.700
                      MgII-2796.MgII-2803                               3.1073827 12.836     0.357  6.329  6.640
                      FeII-2344.FeII-2382.FeII-2600                     3.1074326 12.432     0.191 29.927  2.280
                      SiIV-1393.SiIV-1402                               3.1083556 13.270     0.065 54.474  6.200
                      CIV-1548.CIV-1550                                 3.1090149 14.415     0.023 39.545  2.000
                      MgII-2796.MgII-2803                               3.1090475 12.521     0.160  6.843  6.030
                      SiIV-1393.SiIV-1402                               3.1092942 13.443     0.101  9.757  2.480

J2215-1611 z3.70140\\                                                                                           
                      OI-1302                                           3.7012712 14.446     0.018 15.159  1.078
                      CII-1334                                          3.7012712 14.318     0.024 20.912  1.888
                      AlII-1670                                         3.7012712 12.195     0.116 18.393  4.164
                      SiII-1304.SiII-1526                               3.7013059 13.500     0.193 17.649  5.442
                      FeII-2344.FeII-2382.FeII-2586.FeII-2600           3.7013379 13.416     0.021 24.023  2.230
                      MgII-2796.MgII-2803                               3.7014862 $>$13.60   0.000 24.783  2.821
                      SiIV-1393.SiIV-1402                               3.7017166 13.271     0.129 29.035  4.820
                      AlII-1670                                         3.7018127 12.233     0.050 18.393  4.164
                      OI-1302                                           3.7019143 14.034     0.038 14.974  2.382
                      CII-1334                                          3.7019379 14.247     0.027 26.858  2.575
                      SiII-1304.SiII-1526                               3.7019406 13.628     0.152 23.772  7.833
                      CIV-1548.CIV-1550                                 3.7020800 13.905     0.018 64.323  3.526
                      MgII-2796.MgII-2803                               3.7023402 12.500     0.194 16.111  3.066
                      SiIV-1393.SiIV-1402                               3.7025072 13.493     0.077 32.823  4.197
                      CIV-1548                                          3.7092115 13.685     0.034 65.697  6.756

\end{document}